\documentclass[conference]{IEEEtran}
\ifCLASSINFOpdf
\else
\fi

\usepackage[absolute]{textpos}
\usepackage{tikz}
\usepackage{footmisc}
\usepackage{multirow}
\usepackage{booktabs}
\usepackage{graphicx}
\usepackage{float}
\usepackage{cite}
\usepackage{amsmath,amssymb,amsfonts}
\usepackage{textcomp}
\usepackage{algorithm}
\usepackage{algorithmicx}
\usepackage{algpseudocode}
\usepackage{datenumber}
\usepackage{tcolorbox}
\let\labelindent\relax
\usepackage{enumitem}% http://ctan.org/pkg/enumitem

\newcommand{\header}[1]{\noindent\textbf{#1}.}
\newcommand{\RQ}[1]{$\text{RQ}_#1$:\xspace}

\newcommand{\system}{\textsc{GraphCloak}\xspace}

\newcommand{\property}[1]{{\em \underline{#1}}.\xspace}

\let\proof\relax

\usepackage{xspace}

\usepackage{amsthm}
\theoremstyle{definition}

\usepackage{epsfig,amsmath,amsfonts,epsfig,multirow,makecell,caption,soul,csquotes,color,wrapfig,subcaption,mathtools,bm,spverbatim,booktabs,tcolorbox,diagbox,todonotes}
\usepackage[e]{esvect}

\usepackage{latexsym}
\usepackage{amsmath}
\usepackage{amssymb}
\usepackage{bm}
\usepackage{mathrsfs}
\usepackage{url}

\newcommand{\eb}{\mathbf{e}}

\newcommand{\wb}{\mathbf{w}}
\newcommand{\xb}{\mathbf{x}}

\newcommand{\zb}{\mathbf{z}}

\newcommand{\Dcal}{\mathcal{D}}

\newcommand{\Fcal}{\mathcal{F}}
\newcommand{\Gcal}{\mathcal{G}}

\newcommand{\Lcal}{\mathcal{L}}

\newcommand{\Ncal}{\mathcal{N}}

\newcommand{\Ycal}{\mathcal{Y}}

\newcommand{\RR}{\mathbb{R}} % Real numbers
\ifx\BlackBox\undefined
\newcommand{\BlackBox}{\rule{1.5ex}{1.5ex}}  % end of proof
\fi

\ifx\QED\undefined
\def\QED{~\rule[-1pt]{5pt}{5pt}\par\medskip}
\fi

\ifx\proof\undefined
\newenvironment{proof}{\par\noindent{\bf Proof\ }}{\hfill\BlackBox\\[2mm]}
\fi

\ifx\theorem\undefined
\newtheorem{theorem}{Theorem}

\fi

\ifx\axiom\undefined
\newtheorem{axiom}[theorem]{Axiom}
\fi

\def\etal{\emph{et al.}}

\newcommand{\ie}{\emph{i.e.}}

\renewcommand{\url}[1]{{\sffamily #1}}

\newcommand{\norm}[1]{\left\lVert #1 \right\rVert}

\newcommand{\ssup}[2]{{#1}^{{#2}}}
\newcommand{\ssub}[2]{{#1}_{{#2}}}

\usepackage{datenumber}
\usepackage{xcolor}

\newcounter{dateone}
\newcounter{datetwo}

\newcommand{\ddl}[5]{%
    \setmydatenumber{dateone}{#1}{#2}{#3}%
    \setmydatenumber{datetwo}{\the\year}{\the\month}{\the\day}%
    \addtocounter{datetwo}{-\thedateone}%
    \def\daystogo{\the\numexpr-\thedatetwo\relax}
    
    \begingroup
    \ifnum\daystogo<7
        \color{red}
    \else
        \ifnum\daystogo<14
            \color{orange}
        \else
            \ifnum\daystogo<30
                \color{pink}
            \fi
        \fi
    \fi
    \textbf{[#4] Time left:} 
    \daystogo\space day(s); Information: #5
    \endgroup
}

\begin{document}
%-------------------------------------------------------------------------------
%-------------------------------------------------------------------------------
% \begin{textblock}{15}(1,1)
% % To Appear in the 32nd USENIX Security Symposium, August 9–11, 2023.
% % \ddl{2023}{06}{06}{USENIX'24 }{13 main pages, unlimited ref. \& app. for v1 submission}
% % \ddl{2023}{10}{17}{USENIX'24 Fall Due}{13 main pages, unlimited ref. \& app. for v1 submission}
% \ddl{2023}{06}{29}{NDSS'24}{}
% \end{textblock}
%-------------------------------------------------------------------------------
% \date{}
\title{\system: Safeguarding Task-specific Knowledge within Graph-structured Data from Unauthorized Exploitation}
\vspace{-10pt}
\author{
{\rm Yixin Liu\textsuperscript{1,3,*}}\ \ \
{\rm Chenrui Fan\textsuperscript{2,*}}\ \ \
{\rm Xun Chen\textsuperscript{3}}\ \ \
{\rm Pan Zhou\textsuperscript{2}}\ \ \
{\rm Lichao Sun\textsuperscript{1}}\ \ \
\\
\\
\textsuperscript{1}\textit{Lehigh University} \ \ \ 
\textsuperscript{2}\textit{Huazhong University of Science and Technology} \ \ \ \\
\textsuperscript{3}\textit{Samsung Research American}
}

\maketitle
\begingroup\renewcommand\thefootnote{*}
\footnotetext{Equal contribution. Work done during Yixin Liu’s internship at Samsung Research American.}
\endgroup

%-------------------------------------------------------------------------------
\begin{abstract}
% \rtodo{need align}
As Graph Neural Networks (GNNs) become increasingly prevalent in a variety of fields, from social network analysis to protein-protein interaction studies, growing concerns have emerged regarding the unauthorized utilization of personal data. Recent studies have shown that imperceptible poisoning attacks are an effective method of protecting image data from such misuse. However, the efficacy of this approach in the graph domain remains unexplored. To bridge this gap, this paper introduces \system to safeguard against the unauthorized usage of graph data. Compared with prior work, \system offers unique significant innovations: (1) graph-oriented, the perturbations are applied to both topological structures and descriptive features of the graph; (2) effective and stealthy, our cloaking method can bypass various inspections while causing a significant performance drop in GNNs trained on the cloaked graphs; and (3) stable across settings, our methods consistently perform effectively under a range of practical settings with limited knowledge. To address the intractable bi-level optimization problem, we propose two error-minimizing-based poisoning methods that target perturbations on the structural and feature space, along with a subgraph injection poisoning method. Our comprehensive evaluation of these methods underscores their effectiveness, stealthiness, and stability. We also delve into potential countermeasures and provide analytical justification for their effectiveness, paving the way for intriguing future research.   
 %(3) out-of-domain utility, the cloaking process only targets safeguarding a specific task without affecting the utility of the graph for other purposes; 
\end{abstract}

%-------------------------------------------------------------------------------

\section{Introduction}
\label{sec. intro}

The abundance of data has led to the successful implementation of deep learning, which allows the integration of artificial intelligence (AI) into various domains, including computer vision~\cite{he2016deep}, speech recognition~\cite{hinton2012deep}, and natural language processing~\cite{devlin2018bert}. However, with the rise in the availability of publicly accessible data, there are growing concerns over unauthorized data exploitation. Numerous commercial AI models are trained using personal data inadvertently harvested from the internet. This raises serious questions about the potential misuse of this data for commercial or even illegal purposes, which poses a significant risk to individuals' privacy, security, and copyright~\cite{carlini2021extracting}. For example, a technology company has faced legal action due to allegations of unauthorized usage of portraits for training commercial facial recognition AI models~\cite{hill2020secretive}. A similar situation is observed with GitHub Copilot, which is trained on trillions of open-source code lines without user consent, leading to legal complications~\cite{sun2022coprotector}. %In addition, three well-known artists have filed a joint lawsuit against Stable Diffusion, alleging the unauthorized use of their copyrighted images in the company's training materials~\cite{Clark_2023, lang_2023}.
More worryingly, certain official entities have begun to retract copyright protection initially intended for AI training~\cite{Prime_2023}, thus making data protection from an individual's perspective more crucial than ever before.

A promising solution to this current dilemma is the development of innovative data privacy protection methods, i.e., ``Unlearnable Example''~\cite{huang2021unlearnable,fowl2021adversarial,fowl2021preventing}. These approaches aim to render the original data `unlearnable' by introducing subtle but deceptive perturbations to data samples, which subsequently result in a significant reduction in prediction accuracy when machine learning (ML) models are trained on this modified dataset. Specifically, during the training phase, these models are tricked into relying on these misleading and brittle perturbations. However, in the testing phase, once these injected perturbations are removed, these models are likely to generate erroneous predictions on clean test sets. Ultimately, this technique effectively safeguards the data from unauthorized usage by disrupting its utility for model training.

\begin{figure}[t]
    \centering
    \includegraphics[width=.95\linewidth]{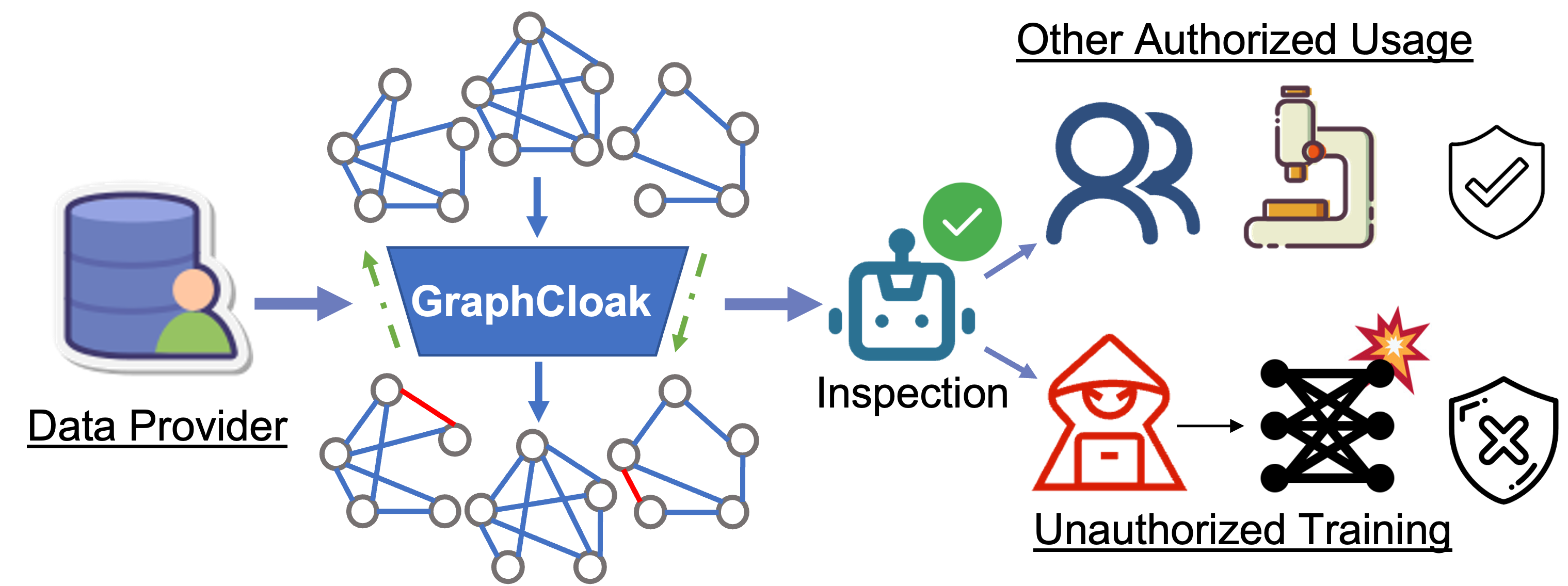}
    \caption{An illustration of \system. By cloaking sensitive graphs, data providers can prevent unauthorized exploitation. }
    \label{fig:motivation}
    % \vspace{-12pt}
\end{figure}

While current literature demonstrates significant potential for data protection, these focuses have been predominantly on visual data \cite{pmlr-v139-yuan21b,evtimov2021disrupting, huang2021unlearnable, jiang2023unlearnable,fu2022robust}. This leaves a substantial and relatively unexplored scope for research in the realm of graph data: whether methods like ``Unlearnable Examples'' can be effectively applied in this domain remains an open question. In recent years, graph-structured data, along with its associated graph neural networks (GNNs), are increasingly prevalent across various domains. Notable applications include malware analysis \cite{Wang:ijcai:2019}, memory forensics \cite{deepmem}, fraud detection \cite{fraud-detection}, and drug discovery \cite{drug-discovery}. %Given this trend, it becomes crucial to explore the potential for applying data poisoning techniques to prevent unauthorized exploitation of graph data. 
In this paper, we aim to bridge the gap by answering the following questions:
\begin{itemize}
\item \RQ{1} \textit{ Can we render graph data `unlearnable' with data poisoning technique?}\vspace{1mm}
        %\item \RQ{1} \textit{ Can we cloak graph data invisibly and render them `unlearnable' to prevent unauthorized exploitation?}
        
    \item \RQ{2} \textit{How effective are such defenses in practical setups, such as limited access to data or training setup?}\vspace{1mm}

    \item \RQ{3}\textit{Is there a way to bypass such defenses?}
    %\item \RQ{3}\textit{How robust is our proposed defense against potential counter measure What potential countermeasures exist against our proposed defenses, and how resistant are our methods in the face of these training setups?}
\end{itemize}

However, addressing these research questions within the graph domain is non-trivial due to two fundamental challenges: 1) Discreteness limits manipulation. The discrete nature of graph data renders the existing optimization-based, vision-oriented solutions infeasible. Existing approaches in vision domain exploit the extensive feature manipulation space of image data to craft deceptive noise that tricks ML models. However, structured graph data has limited manipulation combinations, making the adaptation of existing methods potentially sub-optimal and ineffective. 2) Stealthiness under inspections. The stealthiness of the cloaked graph is a greater concern as graph data is more likely to be reviewed by mechanical perception (e.g., an anomaly detection algorithm) rather than human visual observation, which is common in the image domain. Designing a graph cloaking algorithm capable of bypassing a wide range of inspections is a challenging task.

%the discrete nature of graph data. The discretization attribute makes the application of existing optimization-based, vision-oriented solutions infeasible. Therefore, compared to the vision domain, crafting a cloaked graph presents the following technical challenges: (1) \textbf{Discreteness limits manipulation}: existing approaches exploit the extensive feature manipulation space of image data to craft deceptive noise that tricks ML models. However, structured graph data has limited manipulation combinations, making the adaptation of existing methods potentially sub-optimal and ineffective; (2) \textbf{Stealthiness under inspections}: the stealthiness of the cloaked graph is a greater concern as graph data is more likely to be reviewed by mechanical perception (e.g., an anomaly detection algorithm) rather than human visual observation, which is common in the image domain. Designing a graph cloaking algorithm capable of bypassing a wide range of inspections is a challenging task. To address these unique challenges associated with graph data, we propose \system that is the first work for generating unlearnable graph-structured data. Our design ensures that \system possesses the following distinctive properties:

To deal with these unique challenges on graph data, we propose \system (a brief illustration shown in Fig.~\ref{fig:motivation}) that is the first work for generating unlearnable graph-structured data. Our design ensures that \system possesses the following distinctive properties:

\property{Effectiveness and stealthiness} \system can effectively secure graph data and the GNNs that trained on cloaked graphs will exhibit a significant decrease in clean testing accuracy. Furthermore, the modifications introduced by \system are not readily noticeable or detectable as outliers by various perceptual mechanisms (e.g., human vision, graph statistics, or anomaly detection algorithms).

%\property{Out-of-domain utility} \system is designed to target one specific predefined learning task without impairing the utility of the data for other normal uses. That is, \system can retain the utility of cloaked graphs for other non-target tasks, \eg, cloaked molecule graph used for toxicity prediction should not affect their utility for odor prediction.

\property{Stability across settings} Even under settings where a data owner may only have limited knowledge of the training settings (e.g., unknown GNN model architecture, training strategies, or data processing), \system remains effective.

Specifically, under different levels of knowledge settings, we propose three different methods, including two optimization-based methods under white-box setting: \underline{E}rror-\underline{min}imizing \underline{S}tructural Poisoning (EminS) and \underline{E}rror-\underline{min}imizing \underline{F}eature Poisoning (EminF), and a backdoor-based method under black-box setting: \underline{Sub}graph \underline{Inj}ection Poisoning (SubInj).
% In addition, we devise a handcrafted method known as \underline{Sub}graph \underline{Inj}ection Poisoning (SubInj) to achieve efficient yet effective graph cloaking. 
To validate the practicality of our methods, we apply them to a range of state-of-the-art GNNs and benchmark datasets, which yield the following intriguing findings.

\begin{itemize}
\item $\text{RA}_1$ -- We demonstrate that all of the proposed methods can lead the GNNs trained on cloaked graphs afflicted with large performance drops. Specifically, on the COLLAB dataset, the EMinF method can incur a 47.76\% accuracy drop for the GraphSage model with a small modification budget of $\beta=0.05$. %In addition, we use a case study to demonstrate that cloaking graph will not affect their out-of-domain utility. 
\vspace{1mm}
\item $\text{RA}_2$ -- Under the practical settings of limited budget and training setups misalignment, the proposed methods still remain to perform effectively. For instance, by only poisoning 80\% of data, the SubInj method can decrease the model performance from 78.33\% to 63.17\% on the IMDB-BINARY dataset. Moreover, under the black-box setting, two optimization-based models still perform well when using the misaligned surrogate model architecture. 
\vspace{1mm}
\item $\text{RA}_3$ -- We present three kinds of data compliance inspections (statistic analysis, anomaly detection, and visualization) and demonstrate that our cloaked graph can successfully bypass them. We also discuss two techniques, adversarial training, and the Robust GNN, as potential countermeasures. The results show that the effect of the proposed methods can be migrated but still effective. 

\end{itemize}

\textbf{Our contributions.} To our knowledge, we are the first to explore the potential of using imperceptible poisoning attacks to prevent unauthorized graph-structured data exploitation. Our contributions are summarized as follows: (i) We introduce the first imperceptible poisoning attack on graph data, termed as \system. This innovative attack exhibits key properties such as being effective and stealthy, and stable; (ii) we devised three methods to craft imperceptible cloaks for graph data based on different motivations and knowledge, named EminS, EminF, and SubInj. Each method exhibits its own distinct strengths in effectiveness, efficiency, and stealthiness; (iii) through our extensive empirical testing, we have verified the effectiveness of \system under various practical setups including potential countermeasures and black-box settings.

\section{Background}
\label{sec.back}

\subsection{Preliminary}
Let $\Gcal = \{G_i\}_{i=1}^{N}$ be a collection of graphs with $|\Gcal| = N$. Each graph $G_i = (V_i, E_i)$ consists of the node set $V_i = \{v^{(i)}_j\}_{j=1}^{|V_i|}$ and the edge set $E_i = \{\eb^{(i)}_j\}_{j=1}^{|E_i|}$. An edge $\eb^{(i)}_j = (\eb^{(i)}_{j, 1}, \eb^{(i)}_{j, 2}) \in V_i \times V_i$ connects the node $\eb^{(i)}_{j, 1}$ and the node $\eb^{(i)}_{j, 2}$. We assume the graphs are undirected, but our approach can be easily generalized to directed graphs. The nodes or edges may have features associated with them. We use $\xb(v^{(i)}_j) \in \RR^{d_{\text{node}}}$ and $\mathbf{w}(\eb^{(i)}_j) = \mathbf{w}(\eb^{(i)}_{j,1}, \eb^{(i)}_{j, 2})\in \RR^{d_{\text{edge}}}$ to denote the features of node $v^{(i)}_j$ and edge $\eb^{(i)}_j$, respectively.

\header{Graph neural network (GNN)} A GNN takes a graph $G$ as input, which has both topological structures and descriptive features and generates a representation (embedding) $z_{v}$ for each node $v$. We use $Z$ to denote the matrix form of the node embeddings. We consider GNNs that follow the neighborhood aggregation scheme \cite{xi2020graph}: $\ssup{Z}{(k)} = \mathsf{Aggregate}\left(A, \ssup{Z}{(k-1)}; \ssup{\theta}{(k)} \right)$, where $\ssup{Z}{(k)}$ is the node representation after the $k$-th iteration and also the ``message'' to be sent to neighboring nodes, and the {\em aggregation} function depends on the adjacency matrix $A$, the learnable parameters $\ssup{\theta}{(k)}$, and the node representations $\ssup{Z}{(k-1)}$ from the previous iteration. Usually $\ssup{Z}{(0)}$ is initialized as $G$'s node features. To get the graph representation $\ssub{z}{G}$, a global pooling can be applied over the node representations of the final layer. Formally, for each $k$-th layer, given aggregation function $h$ and the neighborhoods of a node $v$ represented by $\Ncal(v)$, we have
\begin{eqnarray}
	\zb_{v}^{(k)} = & h^{(k)}\big(\{\wb(u, v), \xb(u), \zb_{u}^{(k-1)}\}_{u \in \Ncal(v)}, \nonumber \\
& \xb(v), \zb_{v}^{(k-1)}\big), k \in \{1, 2, \ldots, K\}.
	\label{eq:gnn}
\end{eqnarray}
\header{Graph classification} 
In this work, we focus on graph-level classification, which is the most common task in graph learning \cite{wu2020comprehensive}. 
In graph classification, each graph $G_i$ is associated with a label $y_i \in \Ycal = {1, 2, \ldots, Y}$, with $Y$ denoting the number of categories. The dataset $\Dcal = {(G_i, y_i)}_{i=1}^N$ is comprised of graph instances and corresponding labels. 
% An example of this setting would be classifying drug molecule graphs according to their functionality. 
Given a standard cross-entropy loss function $L$, the classifier $f \in \Fcal: \Gcal \mapsto \Ycal $ is optimized to minimize the loss $\Lcal$ as defined by Eq.~\ref{eq:ind_loss}:
\begin{equation}
\Lcal = \frac{1}{N} \sum_{i=1}^N L(f(G_i), y_i).
\label{eq:ind_loss}
\end{equation}

% \begin{table}[t]
%     \centering
%     \caption{The defense settings. $\checkmark$ ($\times$) means the defender request(or not request) the knowledge/capability. }
%     \label{tab. thmodel}
% \resizebox{0.5\linewidth}{!}{
% \begin{tabular}{l|cc} 
% \hline
% Methods    & $\Mcal$      & $\Acal$       \\ 
% \hline

% grey-box-1 & $\times$     & $\checkmark$  \\
% grey-box-2 & $\checkmark$ & $\times$      \\
% white-box  & $\checkmark$ & $\checkmark$  \\
% black-box  & $\times$     & $\times$      \\
% \hline
% \end{tabular}
% }
% \end{table}

\header{Indiscriminate poisoning attack} By injecting imperceptible noises, indiscriminate poisoning attacks seek to prevent unauthorized exploitation of ``cloaked'' data for specific learning tasks. This attack works by exploiting the manipulate the learning process of ML models by poisoning the data and making them ``unlearnable''. Despite converging quickly during the training, the model trained on the cloaked data suffers from substantial performance drops on the clean testing data, which secures the knowledge inside data for specific learning tasks from being leaked to authorized third parties. Given the conflict between the growing need for data in deep learning across a broad spectrum of applications and the concerns of unauthorized data exploitation, data providers have strong incentives to leverage such an attack to secure their data before releasing it. The indiscriminate poisoning attack has not been explored in the graph domain. Next, we present the threat model and the formulation for crafting the unlearnable graph.

\section{Problem Formulation}
\label{sec. problem}

\subsection{Threat model}

% Following existing works \cite{huang2021unlearnable}, we assume a threat model as shown in Fig.~\ref{fig:motivation}. 
We follow the threat model considered in previous works on indiscriminate poisoning attack \cite{huang2021unlearnable, jiang2023unlearnable, fu2022robust}. The data owner is \textit{the defender} who seeks to cloak their graph data from being used by \textit{the attacker}, the unauthorized model trainers, who access and utilize the data owner's (cloaked) data to train AI models. %Specifically, we discuss in detail the defender's goals, capabilities, and background knowledge.
Specifically, we discuss in detail the goals and capabilities of both attacker and defender.

\header{Attacker's goal and capabilities} By its definition, the attacker aims to utilize the defender's data for some learning tasks, although it is unauthorized by the defender. The attacker might not have any knowledge about whether the graph data has been cloaked by the defender.  

Note that since the model trainer might collect graph data from different resources, the dataset constructed in the training process might contain some unperturbed clean data. We follow existing works \cite{huang2021unlearnable,jiang2023unlearnable, fu2022robust} and assume that the datasets considered in the graph cloaking and model training are restricted to be the same. 

\header{Defender's goal and capabilities} With full access and the modification ability to a portion of the graph, the data owner's goal is to secure their data from being learned for a specific learning task by injecting imperceptible perturbations. % (should bypass potential data inspections on the attack's side). 
%On the attackers' side, whether aware of the contaminated status of data or not, their goal is to leverage various training strategies for training models with better generalization ability. Since the model trainer might collect graph data from different resources, the dataset constructed in the training process might contain some unperturbed clean data. We follow existing works \cite{huang2021unlearnable,jiang2023unlearnable, fu2022robust} and assume that the datasets considered in the graph cloaking and model training are restricted to be the same.
% \header{Defender's capabilities} 
% \mathcal{D}
% \header{Defender's background knowledge} 

In addition, to craft cloaked graph, the defender might have additional background knowledge, such as model configuration and training strategies. The model configuation refers to the model architecture, network layers, and hidden size per layer, while training strategy refers to various optimizer types, the number of training epoch, data augmentation methods (e.g., adversarial training), and other training hyper-parameters. %On the defender side, leveraging the data collecting and training process knowledge is essential for crafting cloaked graphs. Here we decompose these knowledge $\Kcal$ into two main components $\mathcal{K}=\left( \mathcal{M}, \mathcal{A}\right)$:
% classify this knowledge into two categories:
%\begin{enumerate}[noitemsep,topsep=0pt]
    % \item {Data Source Control} $\mathcal{D}$: whether all the training graphs on the attacker's side are cloaked. 
%    \item {Model Configuration} $\mathcal{M}$: the model architecture, network layers, and hidden size per layer.
%    \item {Training Strategies} $\mathcal{A}$: various optimizer types, training \#epoch, data augmentation methods (e.g., adversarial training), and other training hyper-parameters. 
%\end{enumerate}
We then consider two defense settings: \textit{black-box} and \textit{white-box}.
% We study the effectiveness of \system in comprehensive settings. 
In the \textit{black-box} setting, the defender has no knowledge of either model configuration or training strategy. %While in the grey-box setting, the defender might have one of the two aspects from $\left(\mathcal{M}, \mathcal{A}\right)$, enabling adaptively crafting a cloaked graph specialized for target model or training strategies. 
In the \textit{white-box} protection setting, the defender has full knowledge of both model configuration and training strategy. 

\begin{figure*}[t]
    \centering
    \includegraphics[width=.95\linewidth]{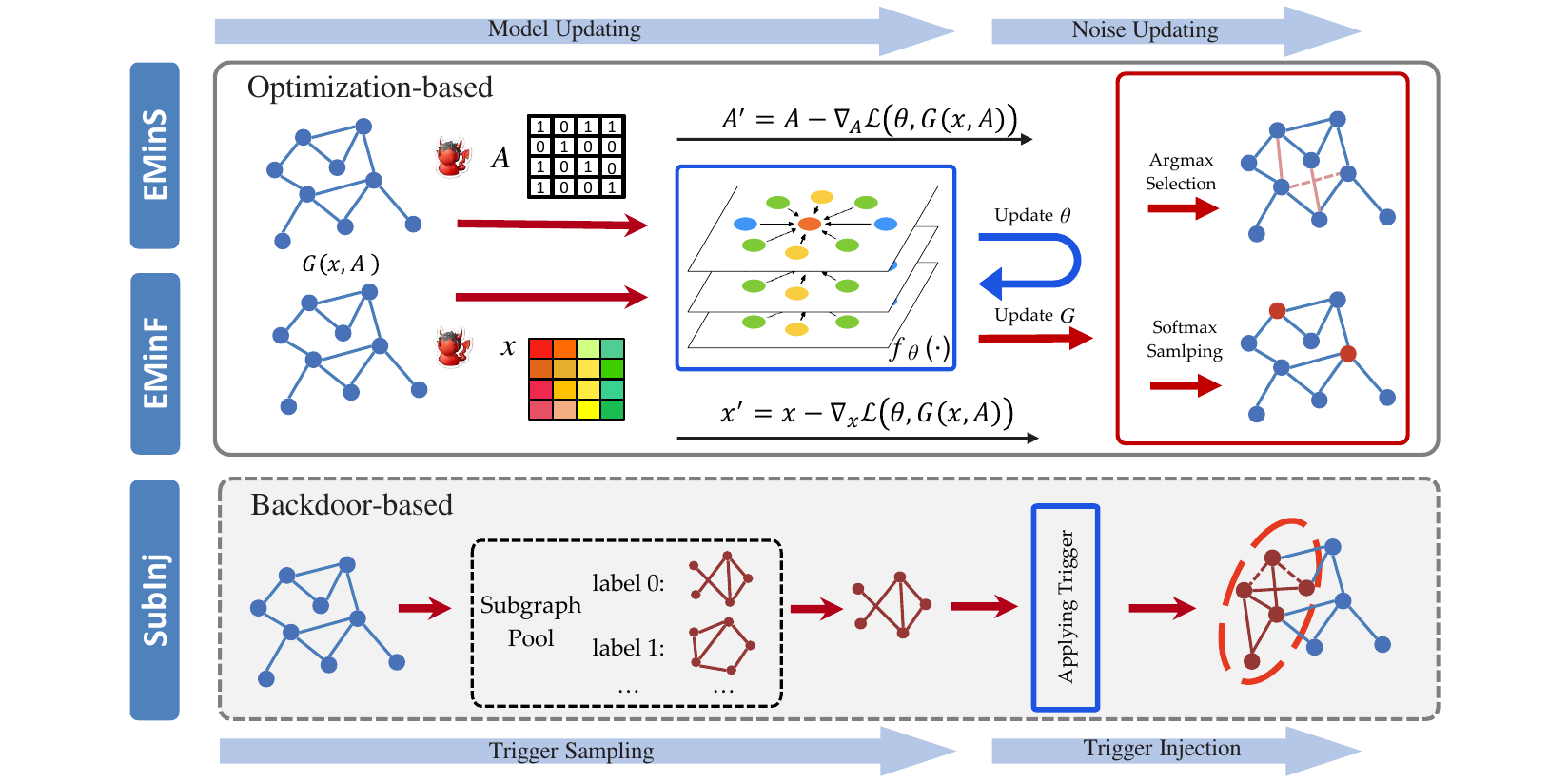}
    \caption{ 
    % \rtodo{optional}
    % We need a figure to visualize the main idea of the three methods.
    The framework of \system. We propose optimization-based (EMinS and EMinF) and backdoor-based (SubInj) poisoning methods for cloaking graphs from being able to be learned. The optimization-based methods aim to learn an optimal delusive perturbation (neither in the structure nor feature space) by solving the bi-level optimization problem. The backdoor-based method applies hand-craft triggers from predefined pools to create a false correlation between the injected trigger and class label. 
    }
    \label{fig:my_label}
\end{figure*}

\subsection{Design goal: bi-level optimization} Given a clean graph training dataset $\mathcal{G}_c= \{G_i,y_i\}_{i=1}^{N}$, our goal is to craft their cloaked counterpart $\mathcal{G}_u= \{\hat{G_i},y_i\}_{i=1}^{N}$ such that the models trained on the $\mathcal{G}_u$ have poor performance on the clean testing set $\mathcal{G}_t$. Note that the cloaked graph is obtained with $\hat{G_i}=G_i\oplus\delta_i$, where $\oplus$ denotes the application of perturbations/modifications of node features or topology structure on the original graph $G_i$. The magnitude of perturbations is bounded by certain constraints $\delta _i\preceq \mathrm{c}$ to ensure that the cloaked graph remains ``similar'' to the original ones in terms of some distance metrics that can bypass possible data compliance inspection where $\preceq$ represents the budget constraints relationship. 
In this study, we set $\delta = \Delta(G, \hat{G})$, where $\Delta$ is the edit distance \cite{gao2010survey} for structural manipulation and $l_p$ norm for graph feature perturbation. Given graph $G$, 
By adding imperceptible but delusive noise into data, we create misleading shortcut patterns that trick the model. 
Formally, the task can be formalized into the following bi-level optimization:
\begin{equation}
    \begin{gathered}
\max_{\delta _i\preceq \mathrm{c}} \underset{\left( \mathrm{G}_i,y_i \right) \sim \mathcal{G} _{\mathrm{t}}}{\mathbb{E}}\left[ \mathcal{L} \left( f_{\theta ^*}(\mathrm{G}_i),y \right) \right] ,\\
\,\,\mathrm{s}.\mathrm{t}. \;\; \theta ^*=\underset{\theta}{\mathrm{arg}\min}\sum_{\left( \mathrm{G}_i,y_i \right) \in \mathcal{G} _{\mathrm{t}}}{\left[ \mathcal{L} \left( f_{\theta}\left( \mathrm{G}_i\oplus \delta _i \right) ,y_i \right) \right]}.
\end{gathered}
\label{eq:bi-opt}
\end{equation}

To solve Eq.\ref{eq:bi-opt}, under different knowledge settings, we propose two cloaking methods in the following section: error-minimizing-based poisoning and subgraph injection poisoning. 
% we propose two kinds of cloaking methods . 

% --------------------------------------------
\section{\system: Design Details}
\label{sec. method}
At a high level, \system~cloaks the graph data in an imperceptible way to mislead the model into learning inaccurate knowledge for a specific task, which essentially attempts to optimize the bi-level problem. However, exactly solving Eq.~\ref{eq:bi-opt} is expensive since each update of $\delta$ require re-computing $\theta$ (\ie, re-training $f$ over $\mathcal{G}_u$). In this section, we propose two kinds of approximated solutions, including optimization-based poisoning and backdoor-based poisoning, with different levels of knowledge requirements. Under the white-box setting, we propose to solve another min-min optimization iteratively by training a surrogate perturbation generator. Depending on the updating component, the derived methods can be classified as \textit{structural poisoning} and \textit{feature poisoning}. Under the black-box setting, we propose a model-agnostic subgraph trigger injection to create misleading patterns for learning classifiers.  
% to inject a hand-craft invisible subgraph to create misleading patterns for learning. 

\subsection{Error-minimizing-based Poisoning}
To tackle the intractable bi-level problem in Eq.~\ref{eq:bi-opt}, an approximated \textit{min-min} optimization process \cite{huang2021unlearnable} is proposed first to learn a perturbation generator and leverage it to conduct perturbation generation. 
With the white-box assumptions, the defender can set the model configuration and training strategies of the perturbation generator to align the one that will be used later on the attacker side. 
The insight is to iteratively craft perturbation that can trick the models trained on the poisoned data. The problem is also a bi-level optimization problem with two levels of minimization. The inner level is a constrained optimization problem that finds the perturbation $\delta$ bounded by certain constraints and minimizes the model's classification loss. The outer level is another minimization problem that finds the parameters $\theta$ that minimize the model's classification loss.
\vspace{-3pt}
\begin{equation}\label{eq:min-min-obj}
\underset{\theta}{\mathrm{arg}\min}\,\mathrm{E}_{(\mathrm{G}_{\mathrm{i}},y)\sim \mathcal{G} _c}\left[ \underset{\delta}{\min}\mathcal{L} (f_{\theta}(\mathrm{G}_i\oplus \delta _i),y) \right] \; \mathrm{s}.\mathrm{t}.\; \delta _i\preceq \mathrm{c}.
\end{equation}
Note that the above optimization has two components that optimize the same objective. While the outer optimization of $\theta$ can be solved straightforwardly with the existing continual gradient descend methods \cite{mason1999boosting}, in the graph domain, the optimization of the constrained inner minimization for $\delta$ is more challenging due to the discrete property of graph $G$, making the vanilla Project Gradient Descend (PGD) not applicable to graph data.

\subsubsection{Error-minimizing Structural Poisoning}
\label{sec. emsp}
Solving the inner minimization in Eq. \ref{eq:min-min-obj} requires taking gradients with respect to discrete structures $G$, which is nontrivial with existing PGD methods for vision data. Drawing upon studies of PGD on graph \cite{dai2018adversarial, adv-graph-2}, in this paper, we adopt a variant of \textit{GradArgMax} to solve the inner minimization of Eq. \ref{eq:min-min-obj} by perturbing the graph structure (e.g., adding or removing edges). Our GradArgMax works by modifying the edges most likely to cause the maximum decrease in the objective based on the gradient. 

\header{Crafting delusive edges with \textit{GradArgMax}}
To elaborate on our method, we first recall the message-passing structure for graph neural networks. To get the embedding of node $v$ in the $k$-th layer, we leverage information from its neighbors:
\begin{equation}
    \small \zb_v^{(k)}=h^{(k)}\left(\left[\wb(u,v),\xb(u),\zb_u^{(k-1)}\right]_{u\in\mathcal{N}(v)},\xb(v),\zb_v^{(k-1)}\right).
\end{equation}
When the neighborhood information is encoded by an adjacent binary matrix $A$, the message-passing could then be as:
\begin{equation}
\small
\zb_{\mathrm{v}}^{(k)}=h^{(k)}\left( A_{(u,v)}\left[ \wb(u,v),\xb(u),\zb _{u}^{(k-1)} \right] _{u\in \mathrm{E}},\xb(v),\zb _{v}^{(k-1)} \right).
\end{equation}
The core of our method is to take gradients with respect to the adjacent matrix $A$ to obtain the gradient for any edge in the potential edge space ($\mathcal{V}\times\mathcal{V}$) no matter whether it exists or not. Specifically, for the element $\alpha_{u, v} \in A$, the gradient is computed with $\frac{\partial \mathcal{L}}{\partial \alpha_{u, v}}=\sum_{k=1}^K \frac{\partial \mathcal{L}^{\top}}{\mu_k} \cdot \frac{\partial \mu_k}{\partial \alpha_{u, v}}$. For one selected edge, we conduct the discrete version of gradient descend, \ie, $\alpha_{u, v} \leftarrow \alpha_{u, v}-\eta \frac{\partial \mathcal{L}}{\partial \alpha_{u, v}}$, by deleting existing edges ($\alpha_{u, v}=1$) with a positive gradient or adding non-exist edges ($\alpha_{u, v}=0$) with a negative gradient. We term such operation as flipping since the process is conducted by simply changing the state of edge with ${\alpha _{u,v}}^{\prime}=\mathbb{I} \left( \alpha _{u,v}=0 \right)$. To be more specific, given an edit distance of $c$ (the maximum modification of $c$ edges), we obtain a set of edges via a greedy selection,
\begin{equation}
    \left\{ u_t,v_t \right\} _{t=1}^{\mathrm{c}}=\underset{\left\{ u_t,v_t \right\} _{t=1}^{\mathrm{c}}}{\mathrm{arg}\max}\sum_{t=1}^{\mathrm{c}}{|\frac{\partial \mathcal{L}}{\partial \alpha _{u_t,v_t}}|}.
    \label{eq:top-c}
\end{equation}
After that, modifications are performed by sequentially modifying the edges of graph $\hat{G}^t$:
\begin{equation}
\hat{G}^{t+1}=\left\{ \begin{array}{l}
	\left( \hat{V}_t,\hat{E}_t\backslash \left( u_t,v_t \right) \right) :\frac{\partial \mathcal{L}}{\partial \alpha _{u_t,v_t}}>0\\
	\left( \hat{V}_t,\hat{E}_t\cup \left\{ \left( u_t,v_t \right) \right\} \right) :\frac{\partial \mathcal{L}}{\partial \alpha _{u_t,v_t}}<0\\
\end{array} \right. .
\label{eq:graph-upd}
\end{equation}
Note that we stop the modification process until we find all the gradients for existing edges are negative or the ones for non-exist edges are positive, in which case no perturbation for decreasing the objective is possible.

\begin{algorithm}[t]
\caption{Error-minimizing-based Poisoning}
\label{alg: emb}
\begin{algorithmic}[1]
\State {\textbf{Input}:} Source model $\theta$, clean training set $\mathcal{G}_c= \{G_i,y_i\}_{i=1}^{N}$, total steps $N_{\text{step}}$, cloaking algorithm $\mathcal{A}$, PGD step number $S$, and PGD step size $\alpha$
\State {\textbf{Output}:} cloaked graph data $\mathcal{G}_u= \{\hat{G_i},y_i\}_{i=1}^{N}$
\State Initialize $\mathcal{G}_u \leftarrow \mathcal{G}_c$
\For{$n$ {\bfseries in} $1 \cdots N_{\text{step}}$}
\State Sample a data batch $b,\{\hat{G_{b}}, y_b\} \sim \mathcal{G}_c$
\If{$\mathcal{A}$=EMinS}\Comment{Structural Cloaking}
% \If{$n\%M==0$}
    % \State // compute the budget
    % \State $\Delta_{\text{max}} = \min \left( r_VV^2,r_EE/5 \right) $ \Comment{Follow Eq. \ref{eq:cons}}
    
    \State // find the top-$C$ delusive edges and modify 
    % \State $ E^{|C|}_{G_i}=\underset{\left\{ u_t,v_t \right\} _{t=1}^{\mathrm{C}}}{\mathrm{arg}\max}\sum_{t=1}^{\mathrm{C}}{|\frac{\partial \mathcal{L}}{\partial \alpha _{u_t,v_t}}|}$\Comment{Follow Eq. \ref{eq:top-c}}
    % \State // conduct $C$-round update to modify graph $\hat{G_b}$
    \State $ E^{|C|}_{G_i}=\underset{\left\{ u_t,v_t \right\} _{t=1}^{\mathrm{C}}}{\mathrm{arg}\max}\sum_{t=1}^{\mathrm{C}}{|\frac{\partial \mathcal{L}}{\partial \alpha _{u_t,v_t}}|}$\Comment{Follow Eq. \ref{eq:top-c}}
    \For{$c$ {\bfseries in} 1 $\cdots C $}
    \State $\hat{G}_b^{c+1} \leftarrow \text{Modify}(\hat{G}_b^{c}, \{u_c, v_c\})$\Comment{Follow Eq. \ref{eq:graph-upd}}
    \EndFor
\ElsIf{$\mathcal{A}$=EMinF}\Comment{Feature Cloaking}
% \State //For loop all nodes and conduct PGD on feature X
% \State \rtodo{TODO}
\For{$s_i$ \textbf{in} $1$ to $S$} \Comment{Follow Sec.\ref{sec. EMinF}}
\State $\xb_b^{s_i} = \xb_b^{s_i-1} - \alpha \text{sign}(\xb_b^{s_i-1})$
% \State $\Xb_b^{s_i}$ = $P$($\Xb_b^{s_i}$) \Comment{Project into $\norm{\Xb_b^{s_i}-\Xb_b^{0}}_0 \leq 2$}
\EndFor
% \State // clip X using softmax sampling (one-hot feature)
% \State $\Xb_b^{s_i}$ = Softmax($\Xb_b^{s_i}, T$)
\State $\xb_b^{s_i}$ = Softmax($\xb_b^{s_i}, T$)
% \State \rtodo{TODO}
\EndIf
    % \State $\delta_i^\prime$ = Perturbation($x_i, y_i, \theta, \delta_i$) 
    % \State $x_i^\prime \leftarrow x_i+\delta_i$
% \EndIf
\State // Updating Source Model
\State $\theta ^{\prime}\gets \mathrm{Optimize(}\{\hat{G}_b,\mathrm{y}_b\},\theta )$
\EndFor
\end{algorithmic}
\end{algorithm}

\begin{algorithm}[t]
\caption{Subgraph Injection Poisoning}
\label{alg: subinj}
\begin{algorithmic}[1]
\State {\textbf{Input}:} Clean graph training dataset $\mathcal{G}_c= \{G_i,y_i\}_{i=1}^{N}$, class number $K$, trigger size $n$, graph density $\rho$
\State {\textbf{Output}:} cloaked graph data $\mathcal{G}_u= \{\hat{G_i},y_i\}_{i=1}^{N}$
\State Initialize $\mathcal{G}_u \leftarrow \mathcal{G}_c$

% \State \rtodo{TODO}
\For{k \textbf{in} 1 to $K$}
% for each class, craft subgraph g_t based on subgraph size and density p
% Get all graphs from this class and store 
\State // Create class-wise trigger
\State $g_k \gets \text{Erdos-Renyi}(n, p)$ \Comment{Follows Sec \ref{sec. subgraph}}
\State // Injecting trigger into graphs
% \State $\Gcal_c^{(k)} \gets \{G_i,y_i\}_{i=1}^{N}$, where $y_i = k$
\State $\Gcal_u^{(k)} \gets \{G_i \oplus g_k ,y_i\}_{i=1}^{N}$, where $y_i = k$
\EndFor
% \State //For loop all data, inject subgraph
% \State \rtodo{TODO}
\end{algorithmic}
\end{algorithm}

\subsubsection{Error-minimizing Feature Poisoning}
\label{sec. EMinF}
As a counterpart to the perturbation on the structural space, another intuitive idea is to take gradients through the feature of graph data and cloak the graph with node feature manipulation. 
Existing PGD methods can manipulate the continuous part of the node feature straightforwardly. 
However, the discreteness of node features space makes optimal error-minimizing feature poisoning harder. 
In this paper, we focus on the case that all the node features are discrete, \ie, $\xb(v^{(i)}_\cdot) \in \RR^{k\times d_k}$ is a one-hot embedding matrix, given $k$ is the number of discrete attributes and $d_k$ is the discrete class number for attribute $k$ (assumed aligned). For a single node, we term the cost of changing its discrete class label as ${\tt CostFeat}=1$. For a graph $G_i$ with $|V_i|$ nodes, the changing cost is bounded by $0\leq {\tt CostFeat} \leq |V_i|k$. We limit the ${\tt CostFeat} \leq c$ and will iteratively conduct feature modification before meeting the budget.

\header{Projected gradient descent with softmax sampling} 
To overcome this hurdle of discrete feature, inspired by \cite{xu2019topology}, we propose Projected Gradient Descent with Softmax Sampling, which augments the vanilla PGD with posthoc softmax protection to preserve the discreteness of the perturbed feature.
Compared to the vanilla PGD \cite{xu2019topology}, our approach only conducts one projection in the final round.
We apply standard PGD on the feature space at each update step, in the last round, we conduct a temperature-control softmax sampling as projections to the feasible region. Formally, let 
$\xb_v$ denote the original clean feature of node $v$, $\xb_{v,i}^t$ denote the $i$-th feature for nodes $v$ at step $t$ and $\alpha$ as step size, we adopt $\xb_{v,i}^{t+1} = \xb_{v,i}^t - \alpha \text{sign}(\frac{\partial \mathcal{L}}{\partial \xb_{v,i}^t})$. Once the final iteration of gradient ascent is completed, we employ a temperature-controlled softmax operation to calculate the probability vector $\mathbf{p_v}$, where $p_{v,i} = \frac{\exp{\xb_{v,i}\cdot T}}{\sum_{j=1}^k{\exp{\xb_{v,j}\cdot T}}}$. Then, we use the probability vector $\mathbf{p_v}$ to sample a new one-hot features $\xb_v ^{\prime}$. To constrain changing, we end the process at certain features of the node for a given graph once the total ${\tt CostFeat} > c$, \ie, in $l_0$-norm:
% \vspace{-3pt}
\begin{equation}
\sum_{i,v}{
\norm{{\xb_{v,i}}^{\prime}-\xb_{v,i}}_0
}\le c, \text{s.t.}\;\;\xb_{v,i}^{(\cdot)} \in e(d_k). 
\end{equation}
where $e(d_k)$ is the one-hot embedding space with $d_k$ classes.

\subsection{Subgraph Injection Poisoning}
\label{sec. subgraph}
The error-minimizing-based poisoning methods require some amount of model training, which might not be preferable and efficient sometimes for larger-scale datasets. Moreover, the white-box setting might be too strong in some realistic applications. Can we skip the possible surrogate model training part and conduct perturbation direction? Under the black-box setting, we give an affirmative answer in this section by proposing a training-free approach, sub-graph injection poisoning, to embed some highly regular class-wise sub-graphs. 
% However, such a setting requires too strong assumptions given that the data collection and model training process is usually confidential. 
Similar to the backdoor attack, which aims to create false connections between the injected trigger and the target class, the proposed \textit{SubInj} seek to link the delusive crafted subgraph to the \textit{original} correct class. During training, the model might highly rely on these injected subgraphs for classification, which can cause catastrophic performance degradation on the testing graph (with the subgraph removed). Following the convention, we term the injected subgraph as a trigger and set the trigger size to be $\text{size}(v)=E+V+F(v)$, where $F$ is the total feature number of node $v$. Such a setting of trigger size allows us to benchmark the modification budget across different methods. The class-wise subgraph triggers generation and embedding process are as follows.

\header{Class-wise trigger generation and injection} We first generate class-wise subgraphs based on the total number of nodes and the pre-defined graph density. Given $n$ nodes, there are $\frac{n(n-1)}{2}$ pairs of nodes, which is the maximum number of edges that a subgraph with $n$ nodes could have. We denote the graph density of a subgraph as the ratio between the number of edges in the subgraph and the number of node pairs in the subgraph. Formally, we have the graph density $\rho=\frac{2e}{n(n-1)}$ where $e$ denotes the number of edges in the subgraph. We then apply the Erdos-Renyi algorithm \cite{gilbert1959random} for the graph generation. Specifically, the algorithm takes as input the number of vertices in the graph and a probability value and generates a graph by iterating through all pairs of vertices and adding an edge between them with probability $\rho$. 
Following Zhang \cite{zhang2021backdoor}, we inject the class-wise subgraphs by randomly substituting $n$ nodes in the original graph. Different from their method, we replace both the node features and connectivity with the ones in subgraphs to improve poisoning efficiency. Upon injection, given the target subgraph as $g \in G$ and trigger $g_t$, the perturbation should consider both structural and feature change, which can be expressed as:
\begin{equation}
\Delta E\left( g_t,g \right) +\sum_{v_i=1}^{|V\left( g \right) |}
\norm{ \xb_{v_i\left( g \right)}-\xb_{v_i\left( g_t \right)} }_0
\le c, 
\label{eq: bug-subj}
\end{equation}
where the first term $\Delta E\left( g_t,g \right)$ is the edit distance of edge change, $\xb_{v_i\left( g \right)}$ denote the feature of node $v_i$ of sub-graph $g$, $V(g)$ is the number of nodes in sub-graph $g$, the second term is the  total modification in feature space between $g$ and $g_t$.

% % --------------------------------------------
\section{Evaluation}
\label{eval}
In this section, we carry out a comprehensive empirical investigation assessing the performance of our proposed methodologies within varied scenarios. Initially, we delineate the experimental configurations, followed by a detailed exposition of the results from our methodologies. These results are analyzed through four different lenses: i) Effectiveness and Efficiency; ii) Stealthiness; iii) Potential Counteractive Measures; and iv) Stability Analysis. We also present a case study of targeted protection on two medical datasets.
\textit{Note that}, due to space constraints, additional tables and figures are included in the appendix.

\subsection{Experiment Settings}\label{Exp.Settings}
\header{Dataset} We conduct experiments on six benchmark graph classification datasets. These datasets were chosen from two different domains: {bio-informatics and social networks}. The bio-informatics datasets \cite{debnath1991structure,borgwardt2005protein} included {MUTAG, ENZYMES, and PROTEINS}, and the social network datasets \cite{yanardag2015deep} are {IMDB-BINARY, IMDB-MULTI, and COLLAB}. We utilize different kinds of graph features with corresponding methods for perturbation generation and model evaluation. The statistics of the datasets can be found in Tab.~\ref{tab:data}. Our selection of datasets provides a representative set of graph classification tasks, as it encompasses an average graph size ranging from 17.93 to 74.49, and a varying number of graphs from 188 to 5000. To reduce the impact of random error, all the experiments are repeated \textit{three} times with the average and standard deviation of clean testing accuracy reported.

\header{Model architectures}
To evaluate the effectiveness of our proposed method, we conduct a thorough comparison using four commonly used graph classification models as our source models, including GCN \cite{kipf2017semisupervised}, GAT \cite{veličković2018graph}, GraphSage \cite{hamilton2018inductive}, and GIN \cite{xu2019powerful}. We set the number of convolution layers to be three and the hidden dimension to 32 across all GNNs, and the final graph embedding is then passed to an MLP for classification. 

\header{Baselines and metric} Since there are no existing works on graph cloaking defense, we adopt two baselines based on the proposed ones. The \textit{Random} baseline simply conducts structural perturbation in a random manner within a constrained budget. Similar to the EMinS, the EMaxS perturb the original graph in an opposite gradient, which is essentially an adversarial perturbation method on the structural graph space \cite{xu2019topology}. As for the evaluation metric, we evaluate the performance of the different cloaking methods by the clean testing accuracy, where a larger decrease in test accuracy indicates better effectiveness of our cloaking protection. 

% \rtodo{Need Rephrase}
\header{Perturbation budget}
% For the EMinS method, we adopt the budget setting in section \ref{budget}. Based on previous studies in adversarial examples, we choose the perturbation ratio $r_V=0.05$ to keep the perturbation imperceptible in our experiment unless otherwise explicitly stated. We further restrict that $r_E\leq0.2$ for a single update to avoid over-modification that results in high sparsity.
% For the EMinF method, we adopt $\epsilon=0.025$ for a single-step PGD update and $T=5$ as the sampling temperature to obtain a one-hot node encoding. 
% For the SubInj method, we take 5 as the node number of the injected subgraphs and 0.6 as the density of injected subgraphs.  
% To implement the EMinS method, we follow the budget configuration outlined in Section \ref{budget}. Considering prior research on adversarial examples, we set the perturbation ratio $r_V=0.05$ to ensure the imperceptibility of perturbations in our experiment, unless explicitly stated otherwise. Additionally, to prevent excessive modifications leading to high sparsity, we impose the constraint $r_E\leq0.1$ for the update. Regarding the EMinF method, we utilize $\epsilon=0.025$ for a single-step PGD update and set the sampling temperature $T=5$ to generate a one-hot node encoding. As for the SubInj method, we specify a node count of 5 for the injected subgraphs and a density of 0.6 for the injected subgraphs.
Since different graph datasets possess different densities, applying a fixed modification budget $c$ might be problematic. To overcome it, we leverage adaptive budget $c(E, V)=\lfloor \min \left(E/10, \beta V^2\right)\rfloor$ considering both the effect of \#\textit{edge} and \#\textit{node}, where we set $\beta=0.05$ unless explicitly stated in the following sections. For the SubInj method, we set the density as 0.6 and compute the \#size $n$ of injected subgraph as $n=\lfloor (-1+\sqrt{1+4c})/2) \rfloor$ by solving $\max (\Delta E + {\tt CostFeat}) = n(n-1) + 2n =c$ from Eq~\ref{eq: bug-subj}. In practice, we found that the SubInj method can generalize well across different settings with a small number of $n=5$. 
% \header{Metrics}
% To evaluate the performance of the generated perturbations, we allocated 20\% of each dataset as a clean test set, while the remaining data was subjected to perturbations. 
% We train two separate models on the perturbed and clean training sets and evaluate their performance on the clean test set. The effectiveness of the unlearnable examples is determined by the difference in accuracy between the two models, with a larger loss indicating better effectiveness.

\begin{table}[t]
\caption{Summary of graph classification datasets.}
\label{tab:data}
\centering
\resizebox{0.9\linewidth}{!}{
\begin{tabular}{cccc}
\hline
Dataset         & \#Graphs & \#Classes & Avg\#Graph Size \\ \hline
MUTAG    & 188      & 2         & 17.93           \\
ENZYMES  & 600  & 6       & 32.63    \\
PROTEINS & 1,113     & 2         & 39.06           \\
IMDB-B   & 1,000     & 2         & 19.77           \\
IMDB-M &1,500	&3	 			& 13.00						\\
COLLAB   & 5,000     & 3         & 74.49           
\\ \hline
\end{tabular}
}
\end{table}

\begin{table*}[!thbp]
\centering
\caption{
% Test accuracies (\%) of models trained by clean, random, Error-Maximizing Structural noise (EMaxS) or Error-Maximizing Structural noise (EMinS) perturbed datasets. \textbf{Bold} indicates the best (lowest) performance of GNN models trained on the various sets of generated unlearnable graphs.
The main results across models and datasets for different methods. 
% \rtodo{ Step-wise setting and epoch-wise setting still need to be aligned. std for Eminf and subinj. } 
}
\label{tab:main}
\renewcommand{\arraystretch}{0.9}
\resizebox{0.9\linewidth}{!}{
\begin{tabular}{cccccccc} 
    \toprule
    \multirow{2}{*}{Models} & \multirow{2}{*}{Methods} & \multicolumn{6}{c}{Dataset}                                                                                                                                                                                                                                                                                                                                                        \\ 
    \cmidrule{3-8}
                            &                          & MUTAG                                                       & ENZYMES                                                     & PROTEINS                                                    & IMDB-BINARY                                                 & IMDB-MULTI                                                  & COLLAB                                                       \\ 
    \midrule
    \multirow{6}{*}{GCN}    & Clean                    & 83.33 $\pm$ 3.04                           & 22.50 $\pm$ 6.34                           & 71.45 $\pm$ 0.52                           & 78.33 $\pm$ 3.88                           & 57.44 $\pm$ 3.22                           & 84.73 $\pm$ 2.65                            \\
                            & Random                   & 84.21 $\pm$ 0.00                           & 23.61 $\pm$ 2.68                           & 68.76 $\pm$ 3.24                           & 73.00 $\pm$ 0.87                           & 54.33 $\pm$ 1.20                           & 79.23 $\pm$ 0.38                            \\
                            & EMaxS                    & \textbf{34.21 $\pm$ 0.00} & 25.56 $\pm$ 4.81                           & 70.10 $\pm$ 2.99                           & 72.17 $\pm$ 2.75                           & 53.11 $\pm$ 2.52                           & 79.63 $\pm$ 2.36                            \\
                            & EMinS                    & 80.70 $\pm$ 1.52                           & 18.61 $\pm$ 3.37                           & 66.67 $\pm$ 6.09                           & 69.67 $\pm$ 1.76                           & 51.22 $\pm$ 0.38                           & 77.50 $\pm$ 2.01                            \\
                            & EMinF                    & 83.33 $\pm$ 1.52                           & 20.56 $\pm$ 3.76                           & 68.91 $\pm$ 2.02                           & 68.50 $\pm$ 1.00                           & 42.89 $\pm$ 3.34                           & 66.90 $\pm$ 3.65                            \\
                            & SubInj                   & 35.96 $\pm$ 3.04                           & \textbf{16.11 $\pm$ 1.92} & \textbf{40.96 $\pm$ 0.26} & \textbf{61.83 $\pm$ 7.32} & \textbf{41.33 $\pm$ 5.51} & \textbf{61.60 $\pm$ 3.00}  \\ 
    \midrule
    \multirow{6}{*}{GAT}    & Clean                    & 80.70 $\pm$ 3.04                           & 26.50 $\pm$ 3.20                           & 67.41 $\pm$ 2.74                           & 77.00 $\pm$ 2.78                           & 56.47 $\pm$ 4.86                           & 83.43 $\pm$ 4.19                            \\
                            & Random                   & 81.58 $\pm$ 2.63                           & 22.78 $\pm$ 4.81                           & 59.79 $\pm$ 4.17                           & 72.50 $\pm$ 2.18                           & 52.33 $\pm$ 1.20                           & 78.43 $\pm$ 0.98                            \\
                            & EMaxS                    & \textbf{34.21 $\pm$ 0.00} & 23.89 $\pm$ 0.48                           & 68.61 $\pm$ 1.79                           & 68.67 $\pm$ 1.04                           & 51.89 $\pm$ 0.38                           & 80.53 $\pm$ 0.51                            \\
                            & EMinS                    & 77.19 $\pm$ 7.60                           & 21.67 $\pm$ 0.83                           & \textbf{50.37 $\pm$ 9.03} & 69.00 $\pm$ 1.80                           & 51.67 $\pm$ 0.88                           & 78.73 $\pm$ 2.20                            \\
                            & EMinF                    & 83.33 $\pm$ 1.52                           & 22.22 $\pm$ 4.19                           & 68.31 $\pm$ 3.18                           & 67.00 $\pm$ 1.50                           & 47.89 $\pm$ 0.51                           & 67.10 $\pm$ 3.90                            \\ 
                            & SubInj                   & 76.32 $\pm$ 0.00                           & \textbf{20.28 $\pm$ 3.47} & 52.62 $\pm$ 2.12                           & \textbf{60.50 $\pm$ 8.23} & \textbf{38.00 $\pm$ 0.67} & \textbf{51.13 $\pm$ 9.98}  \\ 
    \midrule
    \multirow{6}{*}{GIN}    & Clean                    & 87.72 $\pm$ 3.04                           & 28.02 $\pm$ 7.03                           & 69.06 $\pm$ 4.23                           & 73.50 $\pm$ 3.12                           & 50.20 $\pm$ 2.43                           & 73.80 $\pm$ 5.12                            \\
                            & Random                   & 65.79 $\pm$ 0.00                           & 22.50 $\pm$ 2.20                           & 65.02 $\pm$ 4.68                           & 67.33 $\pm$ 2.08                           & 49.33 $\pm$ 4.70                           & 73.00 $\pm$ 4.19                            \\
                            & EMaxS                    & 79.82 $\pm$ 3.04                           & 25.56 $\pm$ 3.94                           & 71.00 $\pm$ 0.93                           & 69.50 $\pm$ 2.29                           & 48.11 $\pm$ 3.10                           & 75.10 $\pm$ 0.66                            \\
                            & EMinS                    & 74.56 $\pm$ 1.52                           & 23.61 $\pm$ 4.74                           & 69.81 $\pm$ 1.37                           & 66.17 $\pm$ 2.36                           & 50.33 $\pm$ 1.76                           & 73.26 $\pm$ 4.65                       \\
                            & EMinF                    & 85.09 $\pm$ 3.04                           & 22.22 $\pm$ 3.94                           & 70.40 $\pm$ 2.69                           & \textbf{65.83 $\pm$ 1.15} & 45.89 $\pm$ 1.58                           & 61.80 $\pm$ 3.80                            \\
                            & SubInj                   & \textbf{35.96 $\pm$ 3.04} & \textbf{20.56 $\pm$ 1.73} & \textbf{45.14 $\pm$ 1.29} & 67.33 $\pm$ 3.62                           & \textbf{37.78 $\pm$ 3.60} & \textbf{47.20 $\pm$ 6.55}  \\ 
    \midrule
    \multirow{6}{*}{Sage}   & Clean                    & 79.82 $\pm$ 4.02                           & 21.67 $\pm$ 4.82                           & 63.00 $\pm$ 1.77                          & 75.00 $\pm$ 4.58                           & 53.27 $\pm$ 4.97                           & 77.83 $\pm$ 2.90                            \\
                            & Random                   & 81.58 $\pm$ 4.56                           & 20.83 $\pm$ 5.20                           & 54.86 $\pm$ 4.14                           & 71.67 $\pm$ 1.26                           & 50.00 $\pm$ 2.65                           & 65.73 $\pm$ 1.10                            \\
                            & EMaxS                    & \textbf{73.68 $\pm$ 4.56} & \textbf{16.67 $\pm$ 0.00} & 65.92 $\pm$ 1.62                           & 70.67 $\pm$ 3.55                           & 52.78 $\pm$ 1.07                           & 70.00 $\pm$ 2.05                            \\
                            & EMinS                    & 78.07 $\pm$ 5.48                           & 22.78 $\pm$ 5.85                           & 60.39 $\pm$ 1.70                           & 70.67 $\pm$ 2.31                           & 51.11 $\pm$ 0.96                           & 63.00 $\pm$ 4.03                            \\
                            & EMinF                    & 83.33 $\pm$ 4.02                           & 16.67 $\pm$ 0.00                           & 55.46 $\pm$ 6.50                           & 64.50 $\pm$ 5.41                           & 37.22 $\pm$ 6.74                           & \textbf{30.07 $\pm$ 1.75}  \\
                            & SubInj                   & 73.68 $\pm$ 0.00                           & 17.78 $\pm$ 0.48                           & \textbf{43.05 $\pm$ 4.66} & \textbf{51.50 $\pm$ 2.65} & \textbf{35.67 $\pm$ 4.06} & 52.17 $\pm$ 2.24                            \\
    \bottomrule
    \end{tabular}
    
}

\end{table*}

\header{Training setting and hyper-parameter}
In our study, we utilized the Adam optimizer with a weight decay of 1e-4 for both the generation of noise and the training of our model and selected the cross-entropy as the loss function. The learning rate was scheduled using a learning rate scheduler that automatically reduces the learning rate when the optimization process reaches a plateau with patience \#epoch set to 20. We study various training settings, including standard training, adversarial training, and training with the early-stop method. 
% To enhance the training procedure, we applied the early-stop mechanism that halts the training if the loss on the validation set does not decrease for 20 epochs. 
Further details on the training approach and its hyper-parameter configurations can be found in the appendix.
% \header{Hyper-parameter for unlearnable methods}
% To solve the bi-level optimization problem for EMinS and EMinF, we use the alternating optimization procedure outlined in algorithm \ref{alg:ada-gradargmin}. 
As for the hyper-parameter in EMinS and EMinF, we set the maximum number of training steps $N_{\text{step}}$ as 5000 for intermediate-scale datasets, including MUTAG, ENZYMES, IMDB-BINARY, and IMDB-MULTI. We set $N_{\text{step}}$ to 500 for larger-scale datasets due to constrained computation resources in practice. For the EMinF, we set the temperature $T=5$, the PGD step size $\alpha=0.025$, and the step number $S=4$. 

% Each iteration includes $M$ steps for the model updating, followed by the perturbation updating proposed in Section \ref{sec. emsp}. The maximum number of training steps is set to $K$, and the budget is denoted as $\Delta$, with default values of 5000 for $K$ and 1 for $M$. We adopt 0.05 as the perturbation budget for structural perturbation and 0.025 as the perturbation for feature perturbation. For the SubInj method, we generate class-wise random subgraphs of 5 nodes with 0.6 as the graph density. We do not optimize for the structure of subgraphs and the position to inject them.

% \header{Hyperparameter for other benchmarks}
% To generate different types of noise, we use either random or maximization operations to replace the inner minimization in the EMinS method. For random noise, we randomly alter elements in the adjacency matrix according to the budget. For creating error-maximizing structural noise, we employ the proposed \textit{GradArgMax} approach, which involves flipping an existing edge with the smallest gradient or a non-existing edge with the maximum gradient based on its absolute value.

\subsection{Effectiveness and Efficiency}
\label{sec. eff}
% We first present the effectiveness of our methods on the simple white-box setting. Then we discuss the effectiveness of our methods in different defense settings. 
% We present the effectiveness of our methods
% Lastly, we analyze the generation efficiency and time complexity of our methods. 

\begin{figure*}[!htbp]
    \centering
    \begin{minipage}{0.30\linewidth}
    \centering
    \includegraphics[width=\textwidth]{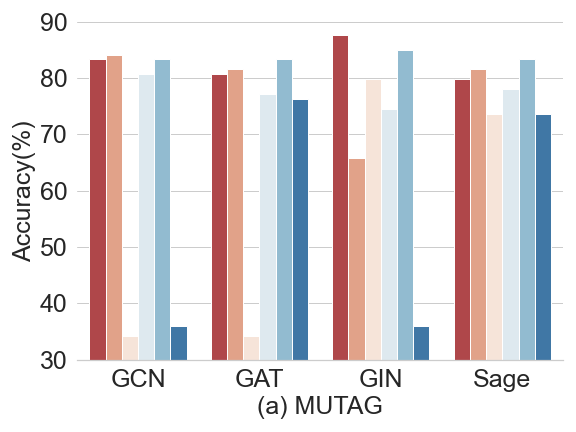}
    \end{minipage}
    \begin{minipage}{0.30\linewidth}
    \centering
    \includegraphics[width=\textwidth]{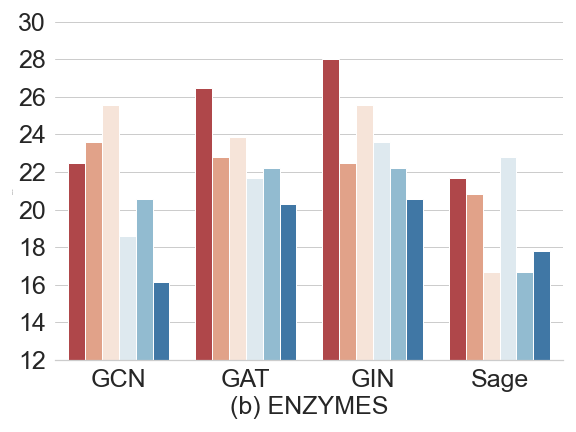}
    \end{minipage}
    \begin{minipage}{0.30\linewidth} 
    \includegraphics[width=\textwidth]{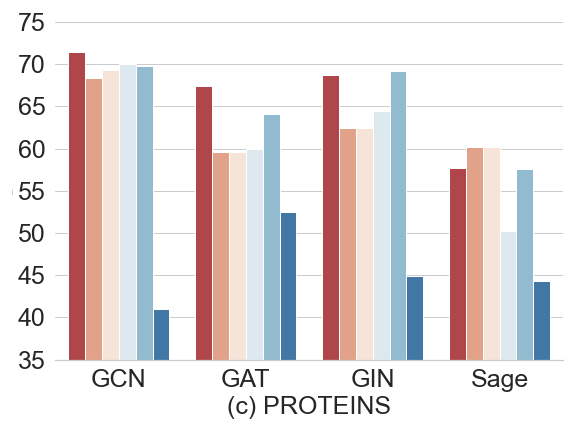}
    \end{minipage}\\
    \begin{minipage}{0.30\linewidth}
    \centering
    \includegraphics[width=\textwidth]{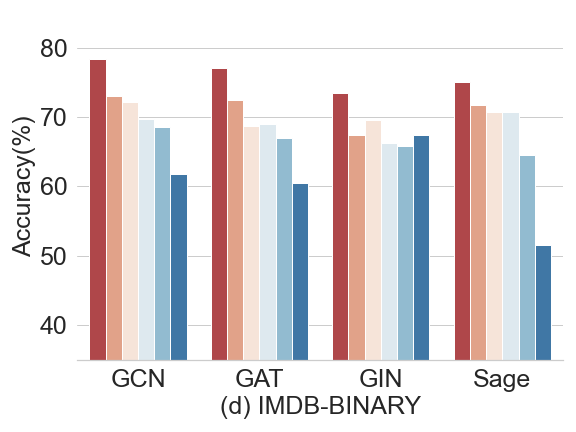}
    \end{minipage}
    \begin{minipage}{0.30\linewidth}
    \centering
    \includegraphics[width=\textwidth]{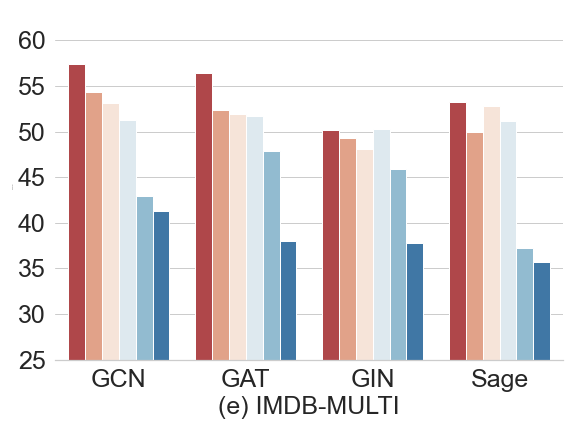}
    \end{minipage}
    \begin{minipage}{0.30\linewidth}
    \centering
    \includegraphics[width=\textwidth]{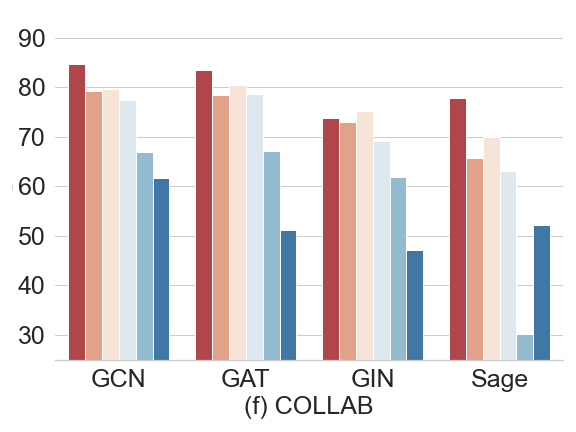}
    \end{minipage}\\
    \begin{minipage}{0.70\linewidth}
    \centering
    \includegraphics[width=\textwidth]{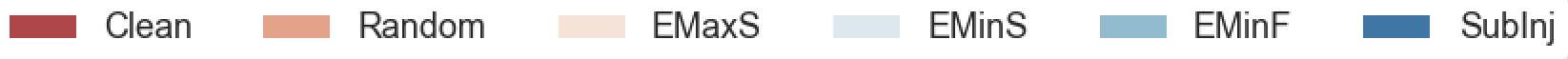}
    \end{minipage}
    \caption{A comparison among different methods on six datasets. }
    \label{appfig:result}
\end{figure*}

% \input{table/main-table}
\header{Effectiveness} 
As observed in Tab.~\ref{tab:main}, both structural and feature perturbations applied to graphs exhibit a notable impact on the accuracy of trained models. Specifically, when considering the COLLAB dataset, the EMinS, EMinF, and SubInj methods proposed in this study successfully decrease the classification accuracy from 84.73\% to 77.50\%, 66.90\%, and 61.60\%, respectively. Meanwhile, it's worth noting that the majority of the trained models are tricked into having a random-guess-level performance on clean data by more than one of our methods, which shows the effectiveness of our cloaked graph in protecting data from unauthorized training. The results in Fig.~\ref{appfig:result} suggest that the proposed methods can significantly decrease the model accuracy compared to the two baseline methods that conduct Random and EMaxS, which conduct random and structural-error-maximizing perturbations. 

% Notably, the SubInj method is the most effective method across different architectures and datasets. The test accuracy on all tested datasets significantly drops after we insert the class-wise subgraphs. Also, compared to the Error-Maximizing Structural (EMaxS) noises, our EMinS method outperforms them on the majority of datasets, which aligns with findings in the previous works on the vision domain \cite{huang2021unlearnable}. Nevertheless, we do observe that EMaxS noise performs more effectively in the MUTAG dataset across most of the source models. We speculate that EMaxS noise might have some advantages over the EMinS noise on smaller datasets. Further studies are required to demonstrate under what circumstances one of these cloaking methods works better on graph data.

The SubInj method stands out as the most effective technique across different architectures and datasets. Notably, when we incorporate class-wise subgraphs using this method, the test accuracy experiences a significant decline across all tested datasets. Furthermore, in comparison to the Error-Maximizing Structural (EMaxS) noises, our EMinS method consistently outperforms them on the majority of datasets, aligning with the findings from previous works in the vision domain \cite{huang2021unlearnable}. However, it is worth noting that the EMaxS noise demonstrates superior performance on the MUTAG dataset across most of the source models. This observation leads us to speculate that EMaxS noise may possess certain advantages over EMinS noise when applied to smaller datasets. Further investigation is necessary to ascertain the specific circumstances in which each of these cloaking methods excels in graph data settings.

In addition, we have observed that models trained on perturbed datasets occasionally outperform those trained on clean data. For example, on the ENZYMES dataset, the Sage model trained on the EMinS perturbed dataset outperforms the one trained on clean data by 1.11\%. We speculate that in these cases, small perturbations can act as a form of data augmentation, enhancing the model's capacity to learn and generalize.
%Consequently, the model trained on the augmented data exhibits improved performance compared to models trained on clean data. 
However, we have also discovered through experiments that this phenomenon diminishes as we continue to increase the perturbation budget.

In order to provide a comprehensive demonstration of the effectiveness of the EMinS, EMinF, and SubInj methods, we conduct an extensive analysis of the results across diverse models and datasets, as illustrated in Fig.~\ref{appfig:result}. The findings unequivocally affirm the efficacy of our proposed methods across various models and datasets. Notably, when applied to bio-information datasets (MUTAG, ENZYMES, and PROTEINS), the structural perturbations exhibit a notably superior effectiveness compared to feature perturbation. However, an intriguing pattern emerges when examining social network datasets (IMDB-BINARY, IMDB-MULTI, and COLLAB), where the EMinF method surpasses the EMinS approach. This distinction can be attributed to the substantially higher graph density observed in social networks, as supported by the data presented in Tab.~\ref{tab:statics_all}. As a consequence, non-robust node features \cite{ilyas2019adversarial} are more prone to being captured by the graph classifier in such cases. Remarkably, the SubInj method yields commendable performance across all datasets and architectures. Even for large-scale social networks, the modification of only 5 nodes and the edges connecting them proves sufficient to prevent the graph classifier from extracting knowledge from graph data. 

% , revealing the non-robust nature \cite{zhang2022unsupervised} of the GNN classifier.

% Note that since the three proposed methods do not share the same budget setting, one cannot simply compare their performance in the table. However, they could be applied in different scenarios and are all proven effective by our experiments.

% \header{Performance across Different Architectures}
% From Tab.~\ref{tab:main}, we can see that our EMinS noise achieves good performance under the settings with different models. Compared to other models, GAT turns out to be more susceptible under our attack since the trained GAT models get the most accuracy drop. On the opposite, although the GCN model has lower accuracy on many classification tasks, it is less sensitive to perturbations than the other three classes of models, which might be due to the fact that the GCN can be seen as a low-pass filter when extracting features from graph data making them less sensitive to our high-frequency perturbation.

\begin{figure}
    \centering
    % \includegraphics[width=0.5\linewidth]{figs/time.pdf}
    % \caption{The time for training noise generator on MUTAG with a GCN source model. }
    \begin{subfigure}{0.48\linewidth}
        \centering
        \includegraphics[width=\linewidth]{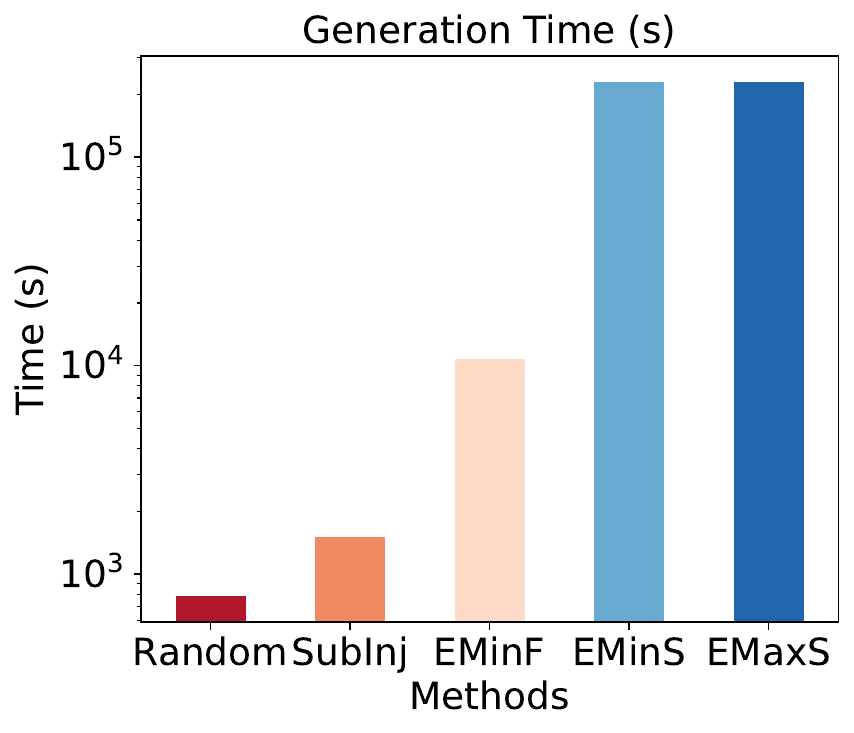}
        \caption{Generation time comparison. }
    \end{subfigure}
    \begin{subfigure}{0.48\linewidth}
        \centering
        \includegraphics[width=\linewidth]{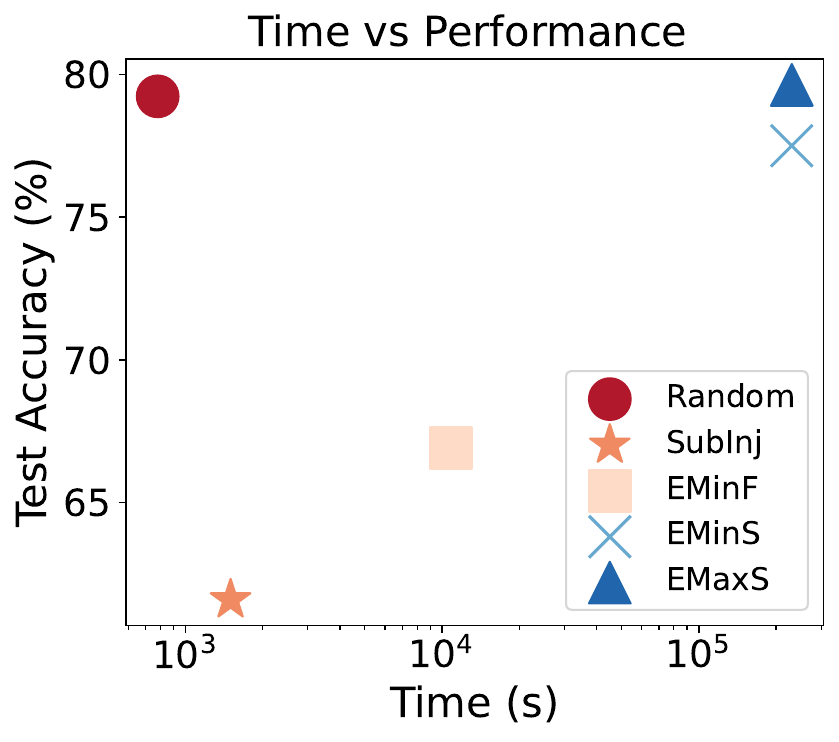}
        \caption{Time-performance analysis. }
    \end{subfigure}
    \caption{The generation efficiency analysis across different poisoning methods on COLLAB dataset with GCN model. }
    \label{fig: time}
\end{figure}

\vspace{-5pt}
\header{Efficiency} To execute the \textit{GradArgMax} operation for a single graph, it is necessary to consider the gradient between all pairs of nodes in the graph. As a result, the computational cost for modifying an edge in the graph is at least $O(|V|^2)$, where $|V|$ represents the number of nodes in the graph. However, the EMinF method proves to be more efficient as it only requires performing Sampling PGD on the node embedding space. Among the proposed methods, the SubInj method stands out as the most efficient, as it solely involves inserting class-wise subgraphs into the clean ones. 

Regarding the generation time, Tab.~\ref{tab: gen-time} presents the timing results of different methods on various datasets for the GCN model. Additionally, Fig.~\ref{fig: time} illustrates the generation time and performance on the COLLAB dataset. These findings align with our analysis, as the EMinS and EMaxS methods take over $10^5$ seconds for noise generation, the EMinF method takes approximately $10^4$ seconds, while the SubInj method requires only $10^3$ seconds to generate the most effective perturbations.

\begin{tcolorbox}[colback=white,colframe=black,boxsep=1pt]
\textbf{Takeaway}: The SubInj shows better effectiveness compared to other baselines across various datasets and GNN models. However, it's worth noting that other proposed models are quite competitive, and under certain conditions, they can potentially outperform SubInj.
\end{tcolorbox}

% The results in Fig.~\ref{fig: time} show that on the COLLAB dataset, the SubInj method outperforms others in both efficiency and effectiveness. 

% Fig.~\ref{fig: time} presents the results obtained for the COLLAB dataset, and the generation time aligns with our analysis. For a comprehensive overview, please refer to Tab.~\ref{tab: gen-time} for the complete results.

% \input{table/threat-exp-main}

% \begin{figure*}[t]
%     \centering
%     \begin{minipage}{0.30\linewidth}
%     \centering
%     \includegraphics[width=\textwidth]{figs/main/mutag-RdBu.png}
%     \end{minipage}
%     \begin{minipage}{0.30\linewidth}
%     \centering
%     \includegraphics[width=\textwidth]{figs/main/enzymes-RdBu.png}
%     \end{minipage}
%     \begin{minipage}{0.30\linewidth} 
%     \includegraphics[width=\textwidth]{figs/main/proteins-RdBu.png}
%     \end{minipage}\\
%     \begin{minipage}{0.30\linewidth}
%     \centering
%     \includegraphics[width=\textwidth]{figs/main/imdb-binary-RdBu.png}
%     \end{minipage}
%     \begin{minipage}{0.30\linewidth}
%     \centering
%     \includegraphics[width=\textwidth]{figs/main/imdb-multi-RdBu.png}
%     \end{minipage}
%     \begin{minipage}{0.30\linewidth}
%     \centering
%     \includegraphics[width=\textwidth]{figs/main/collab-RdBu.png}
%     \end{minipage}\\
%     \begin{minipage}{0.70\linewidth}
%     \centering
%     \includegraphics[width=\textwidth]{figs/main/legend.png}
%     \end{minipage}
%     \caption{A comparison among different methods on six datasets. }
%     \label{appfig:result}
% \end{figure*}

\subsection{Stealthiness}

\header{Statistics of the perturbed datasets}
To verify the stealthiness of cloaked graph datasets, we first employ two important graph statistics, the average number of edges and graph density as metrics to assess the imperceptibility of our perturbations. The average graph density of a dataset is defined as $\rho = \frac{1}{n}\sum_{i=1}^n\frac{|E_i|}{|V_i|^2}$, where $|E|$ and $|V|$ represent the number of edges and nodes in a graph, respectively. From Tab.~\ref{tab:statics}, we observe that the modified datasets exhibit only minor discrepancies in their statistical characteristics compared to the original clean datasets. There are no significant changes in the number of edges or graph density. This finding assures us that our defense mechanism cannot be easily circumvented through simple graph statistics analysis.
% todo: too simple

% As we can see from Tab.\ref{tab:statics}, our method is less perceptible on graphs of social networks (IMDB-B, IMDB-M, COLLAB) than graphs concerning bio-information (MUTAG, ENZYMES, PROTEINS). This is related to the fact that the average density of graphs in social network datasets is much higher than that of bio-information datasets. 
% Our perturbation budget is mainly based on the potential edge space, which is the size of the adjacency matrix, so the perturbation is more perceptible on sparser graphs. This suggests that developing methods for crafting less perceptible noise on sparse graphs could be a promising area for future research.

% ,ding2019deep,xu2022contrastive
% \usepackage{multirow}
\begin{table}[t]
\caption{The statistical changes in average edge number and average density of cloaked graphs compared to originals.}
\label{tab:statics}
\renewcommand{\arraystretch}{1.1}
\resizebox{\linewidth}{!}{
\setlength{\tabcolsep}{0.75mm}{
\begin{tabular}{cccccccccc}
\toprule
\multirow{2}{*}{Models} &
  \multirow{2}{*}{Methods} &
  \multicolumn{4}{c}{PROTEINS} &
  \multicolumn{4}{c}{IMDB-BINARY} \\
                       & & Avg E & $\Delta E(\%)\uparrow$  & Avg $\rho$ &$\Delta \rho(\%)\uparrow$  & Avg E &$\Delta E(\%)\uparrow$ & Avg $\rho$ &$\Delta \rho(\%)\uparrow$ \\ \hline
\multirow{3}{*}{GCN}  &  Clean / EMinF   & 36.044 & - & 0.049 & - & 97.123 & - & 0.245 & - \\
                      & EMinS  & 36.183 & +{0.38\%} & 0.047 & -{2.93\%} & 96.211 & -{0.94\%} & 0.242 & -{1.38\%}\\
                      % & EMinF    & 36.044 & - & 0.049 & - & 97.123 & - & 0.245 & -\\
                      &SubInj & 37.775 & +{4.80\%} & 0.049 & +{0.08\%} & 97.030 & -{0.10\%} & 0.244 & -{0.54\%}\\ \hline 
\multirow{3}{*}{GAT}  &  Clean / EMinF  & 36.044 & - & 0.049 & - & 97.123 & - & 0.245 & - \\
                      &  EMinS  & 36.183 & +{0.38\%} & 0.047 & -{2.93\%} & 96.257& -{0.89\%} & 0.242 & -{1.25\%}\\
                      % &  EMinF  & 36.044 & - & 0.049 & - & 97.123 & - & 0.245 & - \\
                      &  SubInj & 37.775 & +{4.80\%} & 0.049 & +{0.08\%} & 97.030 & -{0.10\%} & 0.244 & -{0.54\%} \\ \hline                       
\multirow{3}{*}{GIN}  &  Clean / EMinF   & 36.044 & - & 0.049 & - & 97.123 & - & 0.245 & -\\
                      &  EMinS  & 35.811 &-{0.65\%} & 0.048 & -{1.66\%} & 96.349 & -{0.80\%} & 0.242  & -{1.11\%}\\
                      % &  EMinF  & 36.044 & - & 0.049 & - & 97.123 & - & 0.245  & -\\
                      &  SubInj  & 37.775 & +{4.80\%} & 0.049 & +{0.08\%} & 97.030 & -{0.10\%}& 0.244 & -{0.54\%}\\ \hline
\multirow{3}{*}{Sage} &  Clean / EMinF  & 36.044 & - & 0.049 & - &97.123 & - &0.245 & - \\
                      &  EMinS  & 35.484 & -{1.55\%} & 0.046 & -{5.13\%} &96.349 & -{1.03\%} &0.241 & -{0.54\%}  \\
                      % &  EminF  & 36.044 & - & 0.049 & - & 96.123 & -&0.245  & - \\
                      &  SubInj  & 37.775 & +{4.80\%} &0.049 & +{0.08\%} & 97.030 & -{0.10\%} & 0.244 & -{0.54\%}\\ \bottomrule
\end{tabular}

}
}

\end{table}

\header{Anomaly dectection} 
Numerous techniques \cite{bandyopadhyay2020outlier,fan2020anomalydae,peng2018anomalous,xu2022contrastive} have been developed to detect anomalies in graphs. 
% In the work of \cite{bandyopadhyay2020outlier,fan2020anomalydae}, attribute and structure autoencoders were employed to compute outlier scores. \cite{peng2018anomalous} utilized CUR decomposition and residual analysis for anomaly detection. \cite{liu2021anomaly} proposed a contrastive self-supervised learning approach that incorporated random neighbor sampling. \cite{ding2019deep} suggested the use of a shared graph convolutional encoder, structure reconstruction decoder, and attribute reconstruction decoder for anomaly detection, while \cite{xu2022contrastive} incorporated contrastive loss during model training. These works showcase a variety of approaches and techniques that have been explored to detect anomalies in graphs, utilizing methods such as autoencoders, decomposition, self-supervised learning, and contrastive loss.
% \cite{wang2021oneclass} employed multiple layers of GCN for graph embedding and measured the anomaly's distance to the centroid.
We leveraged seven graph anomaly detection methods
% We employed a diverse array of anomaly detection methods 
implemented in \textit{PyGOD} \cite{liu2022pygod,liu2022bond} and present the detection graph-level outlier score of the cloaked dataset. We select the GCN source model and the IMDB-BINARY as our dataset. Despite the distinct scoring strategies adopted by different detectors, higher scores always indicate a higher degree of graph anomalousness. As we can see from Tab.~\ref{tab:detection}, among all the tested detectors, \textit{CONAD} and \textit{DOMINANT} demonstrate the best performance, as evidenced by a slight increase in the computed outlier scores. However, overall, most of the detectors are ineffective in identifying modified graphs. Surprisingly, we found that the EMinF method tricks most of the detectors into believing that the modified graphs are less anomalous than the original ones. This may be because our method adds confusing features to the data, making the data ``easier'' to learn and thus indirectly reducing the anomaly score of the cloaked data. In a nutshell, the findings presented in Tab.~\ref{tab:detection} clearly illustrate that the majority of existing off-the-shelf graph anomaly detectors were unable to effectively detect our cloaked graphs, verifying the stealthiness of our methods. See Tab.~\ref{tab:det-1} for more results. 

% One intriguing observation from Tab.~\ref{tab:detection} is that the graphs perturbed by our methods often appear even less anomalous than the graphs in the original datasets. This indicates that our methods are deceptive enough to make the detectors believe that the data becomes more concentrated after perturbation. We have introduced features that make classification tasks easier, thereby reducing the anomalousness of certain nodes or edges that were originally outliers. As a result, our methods can effectively bypass the majority of existing detectors.

\begin{table}[t]
\renewcommand{\arraystretch}{1.0}

\caption{The result of the GCN model on the IMDB-BINARY dataset, the number represents the outlier score computed by each detector. }
\label{tab:detection}
\resizebox{\linewidth}{!}{
% \begin{tabular}{ccccc}
% \hline
% Detector           & Clean & EMinS & EMinF & SubInj \\ \hline
% % AdONE      &  0.0001     & 0.0001      & 0.0001      & 0.0001       \\ 
% ANOMALOUS  &  0.5041     & 0.5269      & 0.3251      & 0.3389       \\ 
% AnomalyDAE & 35.937      & 32.240      & 32.510      & 32.356       \\ 
% CoLA       & -6.235      & -6.361      & -3.852      & -4.211      \\ 
% CONAD      & 3.1452      & 3.1373      & 3.3146      & 3.2454       \\ 
% DOMINANT   & 3.1448      & 3.1242      & 3.3116      & 3.2464       \\ 
% % DONE       & 0.0001      & 0.0001      & 0.0001      & 0.0001       \\ 
% OCGNN       & -0.041      & -0.049      & -0.030      & -0.045       \\ 
% % ONE       & 0.0001      & 0.0001      & 0.0001      & 0.0001       \\ 
% Radar       & 0.4850      & 0.4921      & 0.3265      & 0.3366       \\ \hline

% \end{tabular}
\begin{tabular}{c|c|cc|cc|cc}
\toprule
Detector    & Clean   & EMinS   &  $\Delta$(\%) & EMinF   &$\Delta$(\%) & SubInj  & $\Delta$(\%) \\ \hline
ANOMALOUS   &  0.5041 & 0.5269  & 4.52              & 0.3251  & \textbf{-35.51}             & 0.3389  & -32.78 \\
AnomalyDAE  & 35.937  & 32.240  & \textbf{-10.28}            & 32.510  & -9.53              & 32.356  & -9.96 \\
CoLA        & -6.235  & -6.361  & 2.02              & -3.852  & \textbf{-38.22}             & -4.211  & -32.46 \\
CONAD       & 3.1452  & 3.1373  & \textbf{-0.25}             & 3.3146  & 5.38               & 3.2454  & 3.19 \\
DOMINANT    & 3.1448  & 3.1242  & \textbf{-0.66}             & 3.3116  & 5.30               & 3.2464  & 3.23 \\
OCGNN       & -0.041  & -0.049  & 19.51             & -0.030  & \textbf{-26.83}             & -0.045  & 9.76 \\
Radar       & 0.4850  & 0.4921  & 1.46              & 0.3265  & \textbf{-32.68}             & 0.3366  & -30.60 \\
\bottomrule
\end{tabular}
}

\end{table}

% \header{More Visualizations}
% We visualize the perturbed graphs with different perturbation methods. As we can see from Fig.~\ref{fig:morevisual}, more modifications have been applied to graphs as the budget rises, and our default $r_V=0.05$ is a reasonable budget for the majority of the datasets. Also, it is aligned with the statistics in Tab.~\ref{apptab:data} that our EMinS noise is less perceptible on social network datasets such as IMDB-B and IMDB-M as the graphs in these datasets are denser than those in bio-informatics datasets. It could also be observed from \ref{fig:morevisual} that although we apply the flipping operation to modify edges, our method tends to add edges rather than remove them in a graph.

% Meanwhile, as many symmetric structures could be observed in graphs of bio-information, such as molecules, symmetrical edges would share the same gradient when operating by the GradArgMax method. Further work could be done to specify the edge selection strategy when encountering edges of the same gradient. Ideally, a good distribution of budget among symmetrical edge selections would further improve the stealthiness of modifications.

\header{Perturbation visualization} Despite the good performance of our methods in degrading the models' test accuracy, one might wonder how is the original graph and modified graph different under visual inspections. In order to address this question, we present visualizations of the graphs before and after perturbation in Tab.~\ref{fig:visual}, utilizing colors to indicate different one-hot node features. Upon examining the visualizations, we observe that, for the majority of graphs, there are no noticeable visual discrepancies between the original and modified versions. 
% However, for less dense graphs, the sparsity change is visible in some cases, which indicates a good direction for future studies.

\begin{tcolorbox}[colback=white,colframe=black,boxsep=1pt]
\textbf{Takeaway}: Three common inspection approaches all fail to effectively identify the cloaked graphs. 
\end{tcolorbox}

% ,dolatabadi2023devils,liu2021going
\subsection{Potential Countermeasures from Attacker}
As our methods achieve good performance in various conditions, we discuss the potential countermeasures that unauthorized exploiters might adopt to bypass our privacy protection. The research of unlearnable examples in the vision domain \cite{jiang2023unlearnable,qin2023learning} has demonstrated the vulnerability of unlearnable approaches. Several early methods could be circumvented by robust training schemes such as adversarial training. Also, \cite{fu2022robust,wang2021fooling} has proposed several methods to enhance the robustness of unlearnable examples effectively. As we are the first team to study unlearnable examples of graph data, we adopt some common techniques of adversarial literature to evaluate our methods with possible defense. We inspect our methods under potential countermeasures from the data exploiters in two perspectives: robust training method and inherently robust model structure. They have been proven to enhance the model's robustness under various adversarial scenarios effectively.

\header{Adversarial training}
Adversarial training is a widely used technique in machine learning to enhance the robustness of models against adversarial attacks. By incorporating both regular training examples and adversarial examples, which are carefully crafted to deceive the model, the training process enables the model to better understand and defend against potential attacks. In the context of our research on unlearnable examples of graph data, we draw inspiration from the adversarial literature \cite{adv-graph-2} and apply common techniques of adversarial training to evaluate the effectiveness of our methods against possible defense strategies.

We design adversarial training strategies as adaptive countermeasures against our methods. We start by iteratively crafting adversarial examples based on our unlearnable data and corresponding unlearnable methods. For the EMinS method, we generate the adversarial example by conducting the topology adversarial attack on graphs. As for the EMinF method, we adopt PGD attack to the node features. We then train the graph classification model with these examples. Intuitively, training with such examples would help the models capture more robust features rather than non-robust ones, thus improving the models' performance on clean examples. 

\begin{table}[t]
\caption{Graph visualization on IMDB-MULTI and PROTEINS (the first row are clean graphs, and the following rows are the graphs we generated with the proposed EMinS, EMinF, and SubInj method). }
    \label{fig:visual}
    \centering
    \resizebox{1.01\linewidth}{!}{
    \begin{tabular}{c|cc}
        % \toprule
        &{\large IMDB-M}& {\large PROTEINS}\\
        \midrule
        \rotatebox{90}{ \large Clean}&  \includegraphics[width=0.8\linewidth]{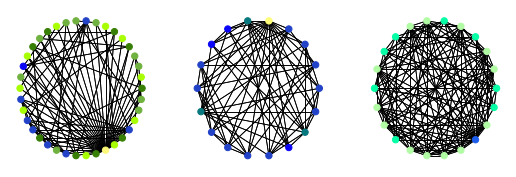}&\includegraphics[width=0.8\linewidth]{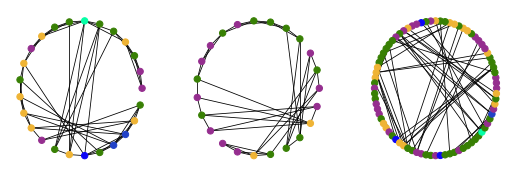}\\
        \rotatebox{90}{\large EMinS} &  \includegraphics[width=0.8\linewidth]{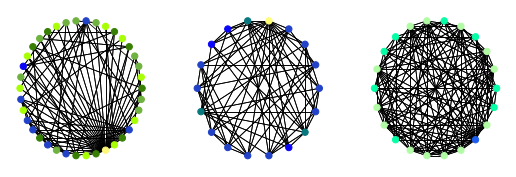}&\includegraphics[width=0.8\linewidth]{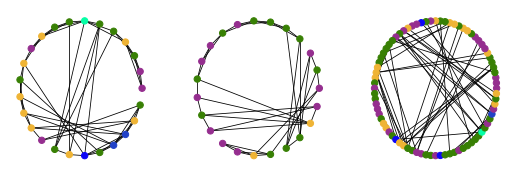}\\
        \rotatebox{90}{\large EMinF} &  \includegraphics[width=0.8\linewidth]{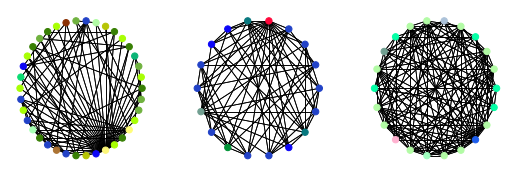}&\includegraphics[width=0.8\linewidth]{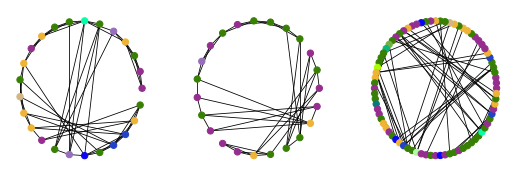}\\
        \rotatebox{90}{\large SubInj} &  \includegraphics[width=0.8\linewidth]{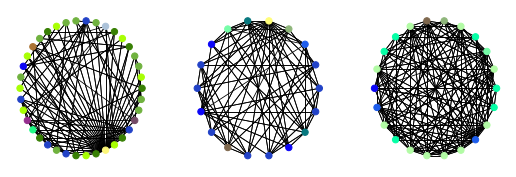}&\includegraphics[width=0.8\linewidth]{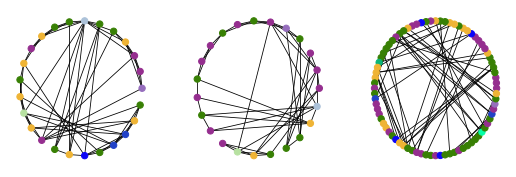}
    % \bottomrule
    \end{tabular}
    }
    
\end{table}

\begin{table}[h]
\centering
\renewcommand{\arraystretch}{0.8}
\caption{The test accuracy of the model with adversarial training on poisoned MUTAG graph. }
\label{tab:adv_train}
\resizebox{1.0\linewidth}{!}{
\begin{tabular}{ccccccc} 
\toprule
\multirow{2}{*}{Methods} & \multicolumn{2}{c}{GCN}         & \multicolumn{2}{c}{GIN} & \multicolumn{2}{c}{Sage}         \\ 
\cmidrule(l){2-7}
                         & before         & after          & before & after          & before         & after           \\ 
\midrule
Clean                    & 83.33          & \textbf{86.84} & \textbf{87.72}  & 86.84 & \textbf{79.82} & 34.21           \\
Random                   & \textbf{84.21}          & 73.68 & 65.79  & \textbf{84.21} & \textbf{81.58} & 34.21           \\
EMaxS                    & \textbf{34.21} & 34.21          & 79.82  & \textbf{81.58} & \textbf{73.68}         & 68.42  \\
EMinS                    & \textbf{80.70} & 68.42          & 74.56  & \textbf{81.58} & \textbf{78.07}          & 71.05  \\
EMinF                    & \textbf{83.33}         & 78.95 & \textbf{85.09}  & 78.95 & \textbf{83.33} & 55.26           \\
SubInj                   & \textbf{35.96} & 34.21          & 35.96  & \textbf{65.79} & \textbf{73.68} & 57.89           \\
\bottomrule
\end{tabular}
}
\end{table}

The results of adversarial training are shown in Tab.~\ref{tab:adv_train}. Surprisingly, instead of enhancing the model's performance, adversarial learning often results in its degradation, thus reinforcing the effectiveness of our data protection methods. In fact, it often negatively impacts the model's performance. Specifically, we noted an average performance drop of 13.25\% on the GCN and Sage models. While the efficacy of our methods, particularly on GIN models, is only slightly mitigated by adversarial training, but this mitigation is not substantial when compared to the clean model (e.g., 21.93\% performance drop with SubInj). These findings suggest that relying solely on adversarial training may be insufficient to completely mitigate the impact of our methods. 

\begin{figure*}[t]
\centering  
    \includegraphics[width=\linewidth]{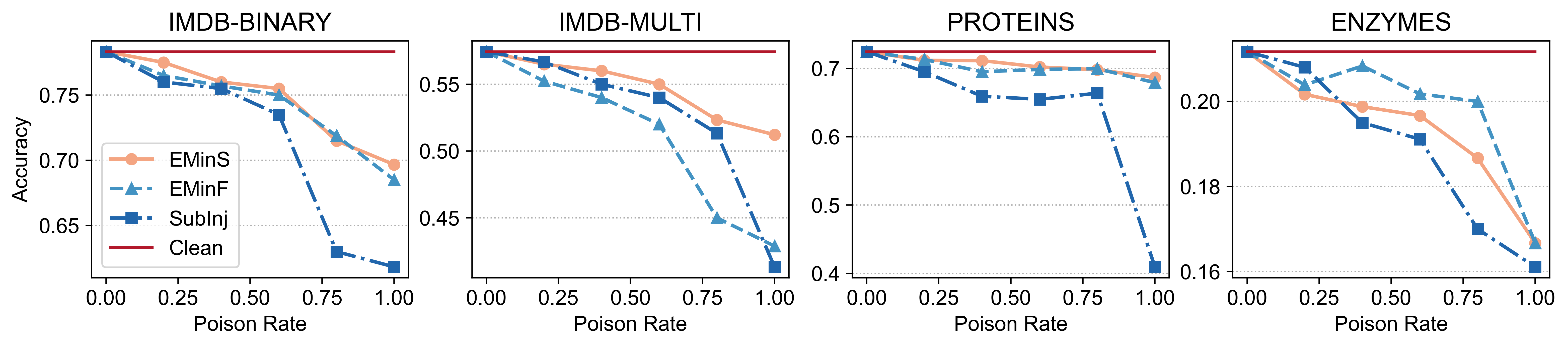}
    \caption{The test result of GCN model on datasets with different poison rates. }
    \label{fig:stab}
\end{figure*}

\header{Robust GNN}
In addition to developing robust training methods, researchers are currently engaged in the development of inherently robust GNN architectures. In the context of node classification tasks, \cite{zhu2021shiftrobust} introduced the Shift-Robust GNN, which addresses distributional disparities between biased training data and the true inference distribution of the graph. Similarly, \cite{Zhuang_2022} combined Bayesian label transition and topology-based label propagation to enhance the robustness of GNNs against topological disturbances. Additionally, \cite{geisler2023robustness} proposed a robust aggregation function that serves as an effective defense mechanism across all scales.

\begin{table}[h]
\caption{The test accuracy of RGNN and the comparison with vanilla GCN trained with poisoned graph.}
\label{tab:robust-gnn}
\centering
\resizebox{\linewidth}{!}{
\begin{tabular}{ccccccc} 
\toprule
\multicolumn{1}{l}{\multirow{2}{*}{Methods}} & \multicolumn{2}{c}{ENZYMES}               & \multicolumn{2}{c}{IMDB-MULTI}           & \multicolumn{2}{c}{PROTEINS}              \\ 
\cmidrule{2-7}
\multicolumn{1}{l}{}                             & RGNN & $\Delta (\uparrow)$            & RGNN & $\Delta (\uparrow)$           & RGNN & $\Delta (\uparrow)$        \\ 
\midrule
Clean                                         & 22.00 & \textbf{+0.83}  & 48.00 & -4.67  & 64.00 & -4.16  \\
EMinS                                         & 23.05 & \textbf{+2.22}  & 48.33 & -3.00  & 57.54 & -3.00 \\
EMinF                                        & 27.50  & \textbf{+6.94}  & 47.11 & \textbf{+2.11}  & 72.07 & \textbf{+2.11}  \\
SubInj                                        & 22.23 & \textbf{+3.06}  & 36.01 & -2.66  & 38.15 & -2.66  \\
\bottomrule
\end{tabular}
}

\end{table}

To evaluate the effectiveness of our privacy protection under robust GNN, we replace the aggregation function of the tested graph classifier as the \textit{Soft Median} function proposed by \cite{geisler2023robustness}. Intuitively, the \textit{Soft Median} is a weighted mean where the weight for each instance is determined based on the distance to the dimension-wise median $\Bar{x}$. This way, instances far from the dimension-wise median are filtered out. The \textit{Soft Median} function is defined as:
\begin{equation}
    \mu_{SoftMedian}(X)=\text{softmax}(-c/T\sqrt{d})^\mathrm{T}X
\end{equation}
with the distance $c_v=\Vert\bar{x}-X_{v,:}\Vert$, temperature $T$, and number of dimension $d$. The temperature $T$ controls the steepness of weight distribution between the neighbors. In extreme case, as $T\rightarrow0$ we recover the instance which is closest to the dimension-wise Median, and when $T\rightarrow\infty$, the \textit{Soft Median} is equivalent to the sample mean. This aggregation function is proved to be more robust than common aggregations (e.g. sum or mean).  

 % \rtodo{Results}

The results presented in Tab.~\ref{tab:robust-gnn} demonstrate the effectiveness of Robust-GNN in mitigating the impact of our approaches. When compared to the GCN model, Robust-GNN consistently shows improvements across all methods for the ENZYMES dataset. In other cases, the accuracy either increases or decreases, but overall, among the three proposed methods, Robust-GNN exhibits the highest effectiveness in countering the EMinF method.

\begin{tcolorbox}[colback=white,colframe=black,boxsep=1pt]
\textbf{Takeaway}: In most datasets and under various GNN model conditions, neither adversarial learning nor robust GNN strategies are able to bypass our data protection methods. Robust GNN shows better performance than adversarial learning in general.
\end{tcolorbox}
% \subsection{Generation Efficiency}

\begin{figure*}[t]
\centering  
    \begin{minipage}{0.3\linewidth}
    \centering
    \includegraphics[width=\textwidth]{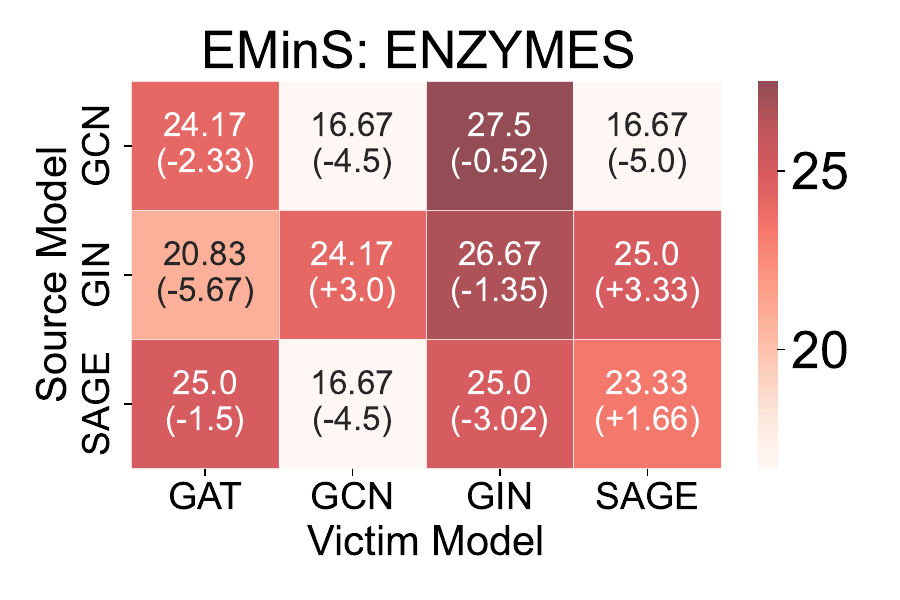}
    \end{minipage}
    \begin{minipage}{0.3\linewidth}
    \centering
    \includegraphics[width=\textwidth]{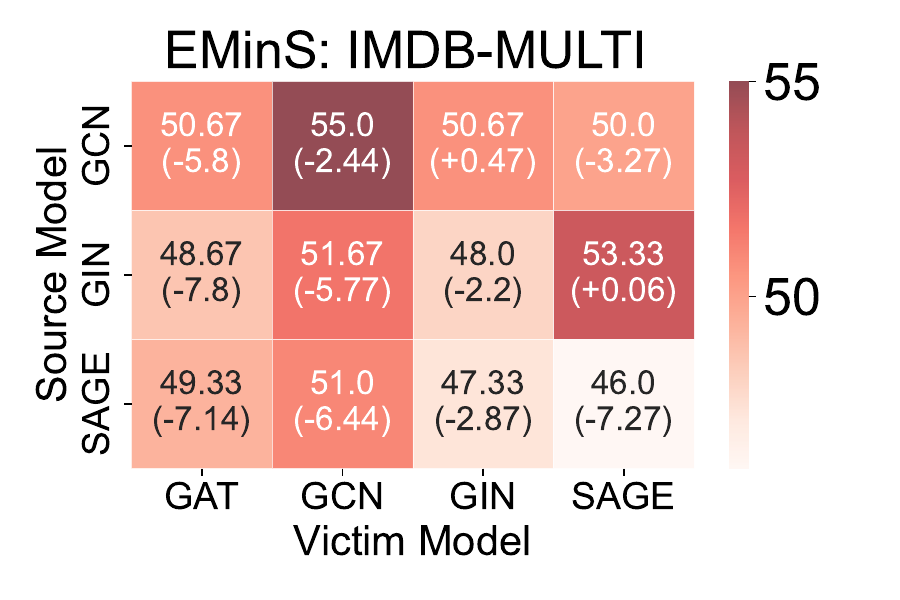}
    \end{minipage}
    \begin{minipage}{0.3\linewidth}
    \centering
    \includegraphics[width=\textwidth]{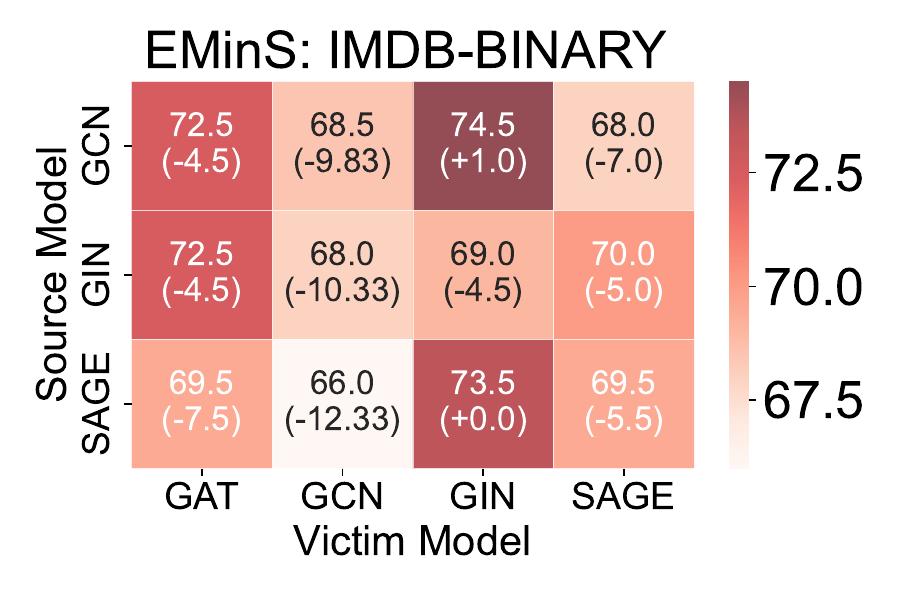}
    \end{minipage}    \\
    \begin{minipage}{0.3\linewidth}
    \centering
    \includegraphics[width=\textwidth]{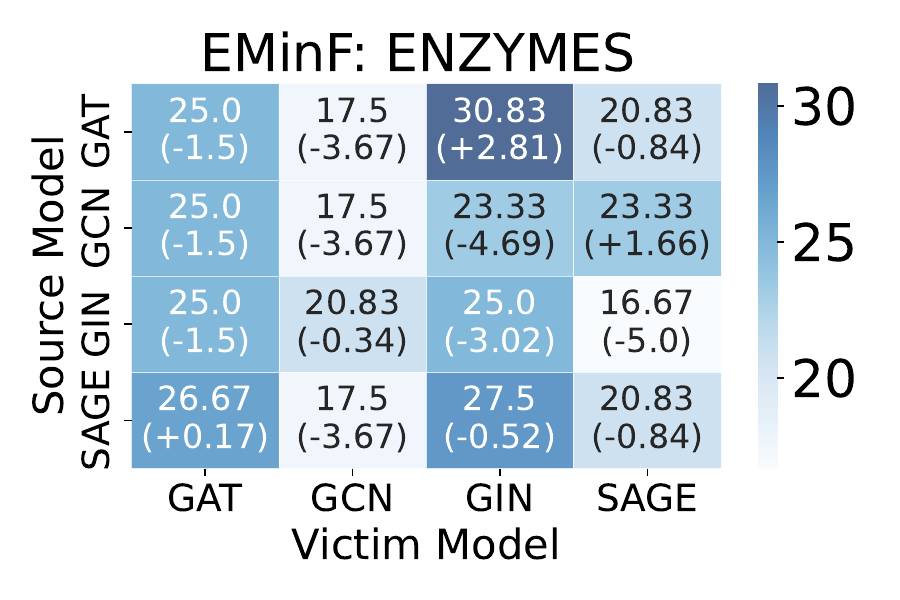}
    \end{minipage}
    \begin{minipage}{0.3\linewidth}
    \centering
    \includegraphics[width=\textwidth]{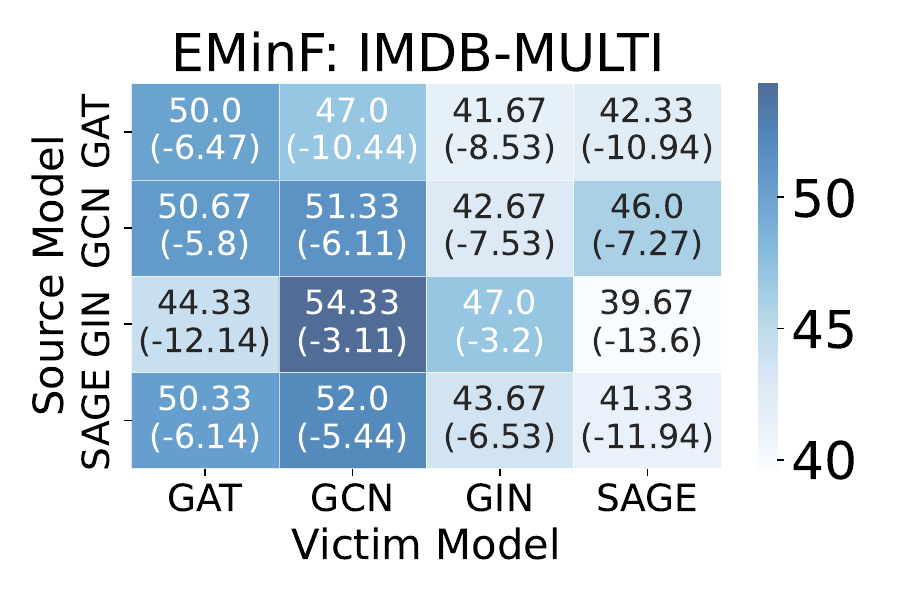}
    \end{minipage}
    \begin{minipage}{0.3\linewidth}
    \centering
    \includegraphics[width=\textwidth]{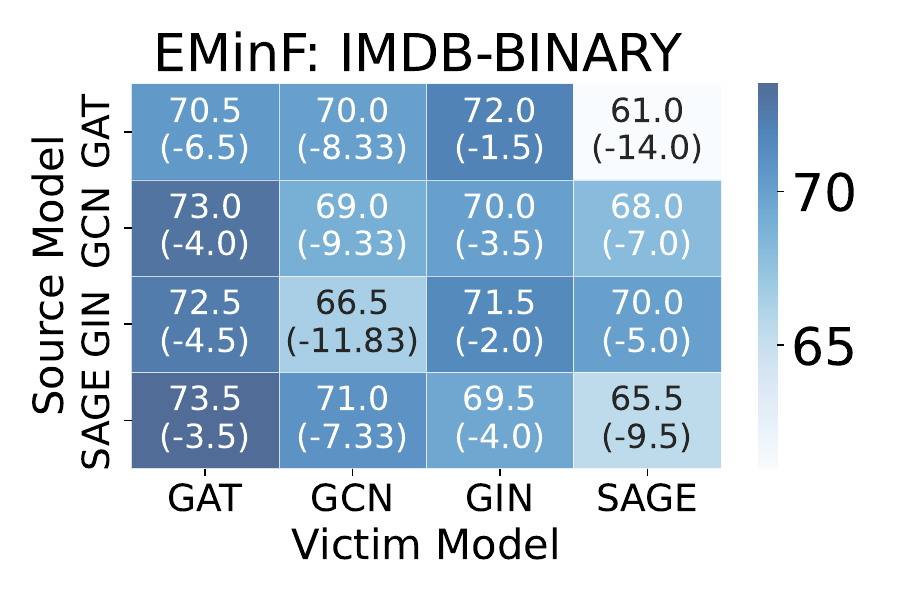}
    \end{minipage}
    \caption{The transferability test of our proposed \textcolor{red}{EMinS} and \textcolor{blue}{EMinF} across models. The clean testing accuracy of different victim models trained on cloaked data crafted by various source models. To compare with the results of clean training, we also report the accuracy change inside the brackets. The results show that our methods transfer well.
    % As \textit{dgl} currently does not support auto-grad for edges in GAT, we do not use it as a source for the EMinS method.
    }
    \label{fig:trans}
\end{figure*}

\subsection{Stability Analysis}
\header{Poison rate}
It is often the case that not all data are processed by the owner who wants to apply our technique and make their data unexploitable. To handle this real-life scenario, we examine the effectiveness of our methods when only a given proportion of data can be modified in the data set. We follow the definition of unlearnable percentage in \cite{https://doi.org/10.48550/arxiv.2101.04898} and conduct experiments when perturbation could only be added to a percentage of graphs in the training set. 

The result in Fig.~\ref{fig:stab} is consistent with the stability study in the unlearnable vision domain. Once the unlearnable noise is not applied to the whole training set, the models could capture the true correlations between the graph and the labels and the test accuracy of the trained model would sharply increase on clean data. Out of the three proposed methods, the SubInj is the most stable one with respect to the poisoning rate. The data perturbed by the SubInj method could still be unlearnable when only a portion of graphs is modified. The results suggest that generating unlearnable graphs with partial modification constraints is imperative to further make the application of this technique more flexible.

% \begin{figure*}[t]
% \centering  
%     \includegraphics[width=\linewidth]{figs/poison rate-v2.png}
%     \caption{The test result of GCN model on datasets with different poison rates. }
%     \label{fig:stab}
% \end{figure*}

\begin{table}[thbp]
\caption{The transferability of EMinS perturbation crafted by GCN.}
\label{tab:trans}
\renewcommand{\arraystretch}{0.9}
\centering
\resizebox{1.0\linewidth}{!}{
\begin{tabular}{ccccc} 
\toprule
\multirow{2}{*}{Models} & \multirow{2}{*}{Methods} & \multicolumn{3}{c}{Dataset}                                                        \\ 
\cmidrule{3-5}
                        &                          & IMDB-MULTI                & ENZYMES                   & IMDB-BINARY                \\ 
\midrule
\multirow{6}{*}{GAT}    & Clean                    & 56.58 $\pm$ 4.55          & 26.77 $\pm$ 2.97          & 77.00 $\pm$ 2.49           \\
                        & Random                   & 52.33 $\pm$ 1.20          & 22.78 $\pm$ 4.81          & 72.50 $\pm$ 2.18           \\
                        & EMaxS                    & 51.89 $\pm$ 0.38          & 23.89 $\pm$ 0.48          & 68.67 $\pm$ 1.04           \\
                        & EMinS                    & 51.67 $\pm$ 0.88          & 21.67 $\pm$ 0.83          & 69.00 $\pm$ 1.80           \\
                        & EMinF                    & 47.89 $\pm$ 0.51          & 22.22 $\pm$ 4.19          & 67.00 $\pm$ 1.50           \\
                        & SubInj                   & \textbf{38.00 $\pm$ 0.67} & \textbf{20.28 $\pm$ 3.47} & \textbf{60.50 $\pm$ 8.23}  \\ 
\midrule
\multirow{6}{*}{GIN}    & Clean                    & 51.25 $\pm$ 2.67          & 28.18 $\pm$ 6.73          & 72.67 $\pm$ 2.44           \\
                        & Random                   & 48.78 $\pm$ 1.64          & 22.50 $\pm$ 2.20          & 68.00 $\pm$ 0.87           \\
                        & EMaxS                    & 49.56 $\pm$ 1.95          & 25.56 $\pm$ 3.94          & \textbf{66.83 $\pm$ 5.01}  \\
                        & EMinS                    & 46.89 $\pm$ 1.92          & 23.61 $\pm$ 4.74          & 69.50 $\pm$ 0.00           \\
                        & EMinF                    & 46.22 $\pm$ 0.19          & 22.22 $\pm$ 3.94          & 69.17 $\pm$ 2.47           \\
                        & SubInj                   & \textbf{37.44 $\pm$ 2.14} & \textbf{20.56 $\pm$ 1.73} & \textbf{66.83} $\pm$ 0.58           \\ 
% \midrule
% \multirow{6}{*}{Sage}   & Clean                    & 53.83 $\pm$ 4.51          & 22.08 $\pm$ 3.93          & 75.75 $\pm$ 3.60           \\
%                         & Random                   & 52.33 $\pm$ 0.88          & 23.89 $\pm$ 6.74          & 71.50 $\pm$ 1.32           \\
%                         & EMaxS                    & 51.56 $\pm$ 3.02          & \textbf{16.67 $\pm$ 0.00} & 71.83 $\pm$ 0.29           \\
%                         & EMinS                    & 52.00 $\pm$ 0.67          & 19.17 $\pm$ 4.33          & 69.50 $\pm$ 2.18           \\
%                         & EMinF                    & 38.33 $\pm$ 4.58          & 18.61 $\pm$ 3.37          & 59.50 $\pm$ 7.86           \\
%                         & SubInj                   & \textbf{33.44 $\pm$ 3.42} & 17.78 $\pm$ 2.55          & \textbf{56.33 $\pm$ 5.06}  \\
\bottomrule
\end{tabular}
}

\end{table}
\header{Transferability analysis}
% We conduct transferability experiments on graph classification datasets and the results are presented in Fig.~\ref{fig: trans}. The EMinS noise demonstrates good transferability across source models on all datasets tested. There is no clear conclusion about which source model generates the best transferable noise. However, as the transferability studies on all datasets report satisfactory results, it may be a promising direction to design an unlearnable graph generation algorithm in a black-box setting in future work.
As the models used by potential data exploiters are unknown to private data owners, the privacy-defending approach must have a certain transferable capacity (i.e., the perturbations generated using one source model could be able to fool other kinds of models). To demonstrate that our approach has such ability, we generate the perturbations with GCN as the source model and train other kinds of models with the perturbed dataset. 
The results presented in Tab.~\ref{tab:trans} indicate that SubInj offers superior transferability protection across various GNN models. Notably, it leads to improvements of 9.89\%, 1.94\%, and 6.5\% for GAT model. Comparable results were observed for the GIN model as well.

% The results presented in Tab.~\ref{tab:trans} demonstrate that our methods exhibit transferability to various models across different datasets. Although the effect of the transferred perturbations may not be highly significant in some cases, we attribute this to the fact that different graph classification models learn distinct data representations and are sensitive to different structural components within graphs. To further verify the transferability of our unlearnable examples, we conduct a comprehensive experiment to examine the datasets crafted by different methods from various source models.  As the SubInj is a black box approach regardless of the source model, we plot the result of the EMinS and EMinF and the corresponding accuracy drop in Fig.~\ref{fig: trans}. As \textit{dgl} currently does not support auto-grad for edges in GAT, we do not use it as a source for the EMinS method. 

From Fig.~\ref{fig:trans}, it is evident that most victim models experience a decline in accuracy when subjected to transferable noise. Specifically, the Sage model generates the most effective noise for the EMinS methods, while the GIN model proves to be the most effective perturbation generator for the EMinF methods. Based on these observations, we recommend that private data owners utilize the Sage model for defense against potential unauthorized data exploiters using EMinS methods, and the GIN model for defense against exploiters employing EMinF methods. By leveraging these source models, data owners can enhance their protection against data exploitations.

\begin{tcolorbox}[colback=white,colframe=black,boxsep=1pt]
\textbf{Takeaway}: SubInj is the most stable one with respect to both poison rate and transferability across GNN models. 
\end{tcolorbox}

% % --------------------------------------------

% \input{no_odd_utility-v1}
\subsection{Case study on Targeted Protection}
% \rtodo{Too many dataset descriptions and processing details}
% In order to showcase the effectiveness of the modified graph for other usages,
To investigate whether the cloaking process affects the utility of modified graphs for other usages, 
we perform the following case study on two multi-label molecule graph datasets, namely Tox21 and Odor. Specifically, we apply the cloaking process targeting one of the label fields, termed the \textit{target task}. 
% intentionally introduce modifications to the graphs by altering one of the multiple labels to render them unlearnable. 
Subsequently, we train a GNN classifier on the cloaked graph for another \textit{non-target task} with different label filed. 
% By successfully rendering the targeted task unlearnable while maintaining high accuracy for other classification tasks, our methods demonstrate their ability to effectively preserve certain general features of the graphs, thereby offering potential applications for other purposes. 
Ideally, the cloaked graph targeting one specific label field should not affect its utility for other non-target label fields. Specifically, the model trained on cloaked graphs for non-target tasks should not decrease a lot in test accuracy. 
In the rest of this section, we first provide an overview of these datasets, followed by our experimental setup and the results of our case study. 

% \subsubsection{Tox21}
\begin{table*}[t]
\caption{The results for the case study. 
% \rtodo{need more descriptions,like what's target and non-target? }
}
\label{tab:case_study}
\centering
\resizebox{0.9\linewidth}{!}{

\begin{tabular}{cc|ccc|ccc|ccc|ccc} 
\toprule
\multirow{3}{*}{Models} & \multirow{3}{*}{Methods} & \multicolumn{6}{c}{Tox21}                                                                        & \multicolumn{6}{c}{Odor}                                                                  \\ 
\cline{3-14}
                        &                          & \multicolumn{3}{c}{Targeted}                    & \multicolumn{3}{c|}{Non-targeted}               & \multicolumn{3}{c}{Targeted}                    & \multicolumn{3}{c}{Non-targeted}        \\ 
\cline{3-14}
                        &                          & before                 & after & drop           & before                 & after & drop          & before                 & after & drop           & before                 & after & drop   \\ 
\hline
\multirow{3}{*}{GCN}    & EMinS                    & \multirow{3}{*}{76.00} & 67.00 & \textbf{9.00}  & \multirow{3}{*}{62.33} & 56.00 & 6.33          & \multirow{3}{*}{80.86} & 69.44 & \textbf{11.42} & \multirow{3}{*}{72.84} & 65.97 & 6.87   \\
                        & EMinF                    &                        & 60.33 & \textbf{15.67} &                        & 52.67 & 9.66          &                        & 64.19 & \textbf{16.67} &                        & 61.23 & 11.61  \\
                        & SubInj                   &                        & 37.00 & \textbf{39.00} &                        & 53.00 & 9.33          &                        & 44.44 & \textbf{36.42} &                        & 63.28 & 9.56   \\ 
\hline
\multirow{3}{*}{GAT}    & EMinS                    & \multirow{3}{*}{76.33} & 72.33 & 4.00           & \multirow{3}{*}{57.67} & 49.67 & \textbf{8.00} & \multirow{3}{*}{76.54} & 57.58 & \textbf{18.96} & \multirow{3}{*}{75.30} & 68.26 & 7.04   \\
                        & EMinF                    &                        & 71.67 & \textbf{4.66}  &                        & 53.33 & 4.34          &                        & 62.53 & \textbf{14.01} &                        & 65.37 & 9.93   \\
                        & SubInj                   &                        & 37.00 & \textbf{39.33} &                        & 51.67 & 6.00          &                        & 44.13 & \textbf{32.14} &                        & 63.32 & 11.98  \\ 
\hline
\multirow{3}{*}{GIN}    & EMinS                    & \multirow{3}{*}{71.67} & 41.67 & \textbf{30.00} & \multirow{3}{*}{59.00} & 53.00 & 6.00          & \multirow{3}{*}{72.84} & 61.16 & \textbf{11.68} & \multirow{3}{*}{71.60} & 62.93 & 8.67   \\
                        & EMinF                    &                        & 41.67 & \textbf{30.00} &                        & 52.33 & 6.67          &                        & 61.60 & \textbf{11.24} &                        & 61.97 & 9.63   \\
                        & SubInj                   &                        & 36.67 & \textbf{35.00} &                        & 51.67 & 7.33          &                        & 54.19 & \textbf{19.65} &                        & 62.52 & 9.08   \\ 
\hline
\multirow{3}{*}{Sage}   & EMinS                    & \multirow{3}{*}{79.33} & 56.67 & \textbf{22.66} & \multirow{3}{*}{65.33} & 56.33 & 9.00          & \multirow{3}{*}{75.93} & 63.89 & \textbf{12.14} & \multirow{3}{*}{73.14} & 64.52 & 8.62   \\
                        & EMinF                    &                        & 65.33 & \textbf{14.00} &                        & 56.00 & 9.33          &                        & 64.13 & \textbf{11.80} &                        & 63.09 & 10.09  \\
                        & SubInj                   &                        & 40.33 & \textbf{39.00} &                        & 56.67 & 8.66          &                        & 55.86 & \textbf{20.07} &                        & 64.94 & 8.20   \\
\bottomrule
\end{tabular}
}

\end{table*}

\header{Tox21 dataset}
The Tox21 dataset \cite{mayr2016deeptox,huang2016tox21challenge} is a machine learning dataset created as part of the \textit{Toxicology in the 21st Century} initiative, aiming to aid scientists in comprehending the potential of chemicals and compounds under examination. It assists in evaluating the capacity of these substances to interfere with biological pathways, potentially leading to toxic effects. The dataset encompasses 12,060 training samples and 647 test samples representing various chemical compounds. It includes 801 dense features and 272,776 sparse features that characterize chemical substructures. Each sample is associated with 12 binary labels, indicating the results of distinct toxicological experiments. It is important to note that the label matrix may contain missing values, and we employ masks to signify these missing entries.

\header{Data processing of Tox21}
In order to demonstrate the effectiveness of modified graphs, we exclude the chemical features available in the dataset and solely utilize the graph structure of chemical compounds for graph classification. We employ the official API provided by \textit{dgllife} to convert the SMILES string representations of chemical compounds into their corresponding graph representations. To maintain consistency with our previous experiments, we simplify the bond type by considering uniform edge connections and encoding node features using the canonical atom featurizer. Additionally, we incorporate self-loops into the graphs, as it is a common practice when training graph classification models.

To assess the utility of modified graphs, we select a specific label as the target and another label as the non-target from the 12 available binary labels provided by the Tox21 dataset. We compare the test accuracy of the model trained on the clean graph and cloaked graph (crafted for the target task) for both the target and the non-target classification task. 
% The experimental procedure involves the following steps: (1) Initially, we evaluate the accuracy of two graph classification models trained on the original, unmodified graphs and their associated labels. (2) Subsequently, we conduct cloaking on the graphs based on the target label, aiming to secure the data from being trained for the target task. (3) Finally, we repeat step (1), but this time using the modified graphs, in order to observe any decrease in accuracy for both the target and non-target labels. 

To address the challenges posed by missing values and the sparsity of the label matrix, we have implemented preprocessing steps to ensure reliable and balanced results, thereby avoiding potential biases caused by imbalanced datasets. Initially, we calculate the distribution of 0s and 1s for each of the 12 binary labels. From this analysis, we identify the top 4 labels that exhibit the most balanced proportions. These selected labels will be considered as candidates for the target and non-target labels in our subsequent analysis. Next, we compute the covariance matrix among these four labels and select label pairs that demonstrate minimal covariance. This step is crucial to avoid correlations between the chosen labels. Our objective is to capture distinct perspectives of the graphs, and if the target and non-target labels were closely related, it would make the cloaked graph inevitably impact its utility for the other learning task.
Furthermore, we filter out data instances that have at least one label missing, ensuring that our analysis focuses on instances with complete label information. 
% , thus mitigating any potential biases arising from missing data. 
% By performing these preprocessing steps, we aim to enhance the reliability and accuracy of our subsequent analysis of the dataset.

Since the dataset remains imbalanced even after these preprocessing steps, we conduct down-sampling by limiting the maximum number of instances with majority labels. As a result, we obtain a relatively balanced graph classification dataset devoid of missing values. The budget settings and training scheme employed are consistent with what we described in Section \ref{Exp.Settings}.

% \header{Results}
% We evaluate the performance of the EMinS, EMinF, and SubInj methods using the Tox21 datasets, and document the classification accuracy as well as the accuracy drop for both targeted and non-targeted labels in Table \ref{tab:case_study}.

% % \subsubsection{Odor}

% As demonstrated in Table \ref{tab:case_study}, our methods effectively reduce the classification accuracy of trained models on the targeted labels. However, the accuracy drop for non-targeted labels is considerably smaller in the majority of cases. These findings indicate that our methods can be successfully employed to prevent unauthorized exploitation of specific properties within the data, while still maintaining the data's utility for other purposes.

% \subsubsection{Odor}
\header{Odor dataset}
The Leffingwell Odor dataset \cite{sanchez2019machine} is a machine learning dataset created for predicting diverse properties of molecules. This area of research has gained significant attention in machine learning, especially as models capable of learning from graph-valued inputs have advanced in complexity and robustness. To promote research in the field of Structure-Odor Relationships (SOR), \cite{sanchez2019machine} assembled and refined a dataset comprising 3,523 molecules. These molecules are associated with expert-labeled odor descriptors sourced from the \textit{Leffingwell PMP 2001} database. The dataset also provides various featurizations for all molecules, including bit-based and count-based fingerprints, Morded molecular descriptors, and embeddings generated from a trained GNN model. The dataset encompasses 113 binary labels, which describe the odor properties of the chemical compounds.

\header{Data processing of Odor} In order to showcase the effectiveness of perturbed graphs generated by our methods, we focus solely on the graph structure of molecules and select two binary labels from the Odor dataset to serve as the target and non-target labels for evaluating our models. For node features, we employ the \textit{rdkit} \footnote{https://rdkit.org/} library to convert different atoms into one-hot embeddings. Additionally, we utilize edges to represent chemical bond connections. To align with our previous experiments, we simplify the problem by employing uniform bonds to denote different connections. Given that the majority of the labels in the dataset exhibit sparsity and imbalance, we perform the same preprocessing procedure as we did with the Tox21 dataset to create a balanced dataset. The budget settings and training scheme employed in this study remain consistent with the ones used in Tox21.

\header{Results and analysis}
We evaluate the performance of the EMinS, EMinF, and SubInj methods on Tox21 and Odor datasets and document the classification accuracy as well as the accuracy drop for both targeted and non-targeted labels in Table \ref{tab:case_study}. As demonstrated in Table \ref{tab:case_study}, on the target task, our methods effectively decrease the classification accuracy of trained models task by $\sim$10\% to $\sim$40\% across different models. On the non-target task, although there is a slight drop in accuracy, the magnitude of the drop is within 10\%, which is relatively small compared to the drop on the target task.  
% However, the accuracy drop for the non-targeted task is considerably smaller in the majority of cases. 
These findings indicate that our methods can be successfully employed to prevent unauthorized exploitation of specific properties within the data, while still maintaining the data's utility for other purposes.
\begin{tcolorbox}[colback=white,colframe=black,boxsep=1pt]
\textbf{Takeaway}: Results on multi-label datasets prove that our methods well preserve the graph's utility for any targeted tasks.
\end{tcolorbox}
% \rtodo{this part should have more analysis}

% % --------------------------------------------

% \input{no_ood_utility}

% % --------------------------------------------

\section{Discussion, Limitations and Future Work}
\label{sec:discussion}
In this section, we will discuss some intriguing results and promising feature directions revealed by the experimental results in Section \ref{eval}. 

% Unlearnable or easier to learn?
\header{Determining budget threshold}
Users of our methods should be cautioned against using excessively small budgets. The results presented in Fig.~\ref{appfig:result} demonstrate that perturbations can occasionally enhance the model's performance on clean data. We hypothesize that these small perturbations act as a form of data augmentation, improving the model's learning and generalization capabilities. However, we have observed that this phenomenon diminishes as the perturbation budget increases. There exists a threshold for the budget, beyond which the data becomes effectively unlearnable. When budgets fall below this threshold, the cloaked graph is easier to learn, and the cloaking process actually accelerates knowledge leakage from the graph. Therefore, it is crucial for data owners who employ our methods to select a budget above the threshold to ensure effective data protection. It is important to note, however, that an efficient algorithm to determine this threshold has not yet been derived. Further research is necessary to develop methods that can accurately determine the optimal budget threshold to achieve the desired level of data protection.
% The results in Fig.~\ref{appfig:result} demonstrate that small perturbations can occasionally improve the model's performance on clean data, acting as a form of data augmentation. However, this effect diminishes as the perturbation budget increases. There exists a threshold beyond which the data becomes effectively unlearnable, and budgets below this threshold can accelerate knowledge leakage from the graph. To ensure effective data protection, data owners should use a budget above the threshold. Determining this threshold efficiently requires further research and the development of optimal methods.

\header{Effectiveness under low poison rate}
The performance of our methods under low poison rate is not satisfactory.
As depicted in Fig.~\ref{fig:stab}, the model's performance exhibits a significant improvement when the dataset contains a portion of clean data rather than being entirely poisoned. This observation implies that data exploiters can substantially degrade the effectiveness of our methods by introducing a small fraction of clean data. It is evident that future studies should focus on developing more effective methods that can perform well even under low poison rates, mitigating the impact of such adversarial attempts.
% Fig.~\ref{fig:stab} shows that including a portion of clean data in the dataset significantly improves the model's performance compared to when the dataset is entirely poisoned. This highlights how data exploiters can undermine our methods by introducing even a small fraction of clean data. Future research should prioritize the development of more effective methods that can perform well under low poison rates, thus mitigating the impact of such adversarial attempts.

\header{Improving transferability}
No models have demonstrated exceptional effectiveness as source models in our transferability test. While the majority of victim models experience a decline in accuracy as illustrated in Fig.~\ref{fig:trans}, no definitive conclusion can be drawn regarding which source model produces the most transferable noise across various methods and datasets. This finding indicates that the crafted features are not sufficiently deceptive to consistently deceive all graph classification models, as they sometimes manage to capture other features crucial for accurate predictions. As part of our future work, we will explore alternative approaches to identify a source model capable of generating highly effective noise with significant transferability across different models, methods, and datasets.

\vspace{-5pt}
\section{Related Work}
\label{sec. related work}
% \rtodo{polish}
\header{Graph poisoning attack}
The body of work on poisoning attacks specifically tailored to graph learning tasks remains comparatively small. Initial pioneering works Sun \etal \cite{sun2018data} and Bojcheski \etal \cite{bojcheski2018adversarial} both propose data poisoning attacks with a bi-level optimization formulation targeting factorization-based embedding models on homogeneous graphs, where the former leverages the eigenvalue perturbation theory~\cite{stewart1990matrix} to handle the problem and the latter directly apply the iterative gradient method~\cite{carlini2017towards} to find a solution. Zhang \etal \cite{zhang2019towards} further extends the settings to heterogeneous graphs and proposes a collection of data poisoning attack strategies on attacking knowledge graph embedding models. To generalize the setting, \cite{liu2019unified} propose a general framework for data poisoning attacks to graph-based semi-supervised learning, where the regression and classification problems are studied as two important cases. Recently, Zhang \etal \cite{zhang2022unsupervised} propose Contrastive Loss
Gradient Attack (CLGA), a graph-poisoning attack against graph-contrastive learning. CLGA aims to damage contrastive learning by flipping the edges with the largest absolute gradients using contrastive loss as guidance. However, the poisoning settings and techniques studied in CLGA only are relatively limited, \ie, they only focus on structural manipulation, and it simply takes the absolute value of gradients when making the gradient descent. In this paper, we turn our attention to supervised graph-level classification settings and introduce a variety of poisoning techniques specifically designed for these contexts. We aim to explore the performance of structure-oriented and feature-oriented poisoning under various settings.

% tran2018spectral

\header{Graph backdoor attack}
Neural networks for vision data have been demonstrated to be susceptible to backdoor attacks~\cite{gu2017badnets,chen2017targeted,liu2017trojaning,clements2018hardware,yao2019latent,salem2020dynamic}. Specifically, a backdoored neural network classifier produces attacker-desired behaviors when a trigger is injected into a testing example. On the graph domain, studies \cite{xi2020graph, zhang2021projective, yang2022transferable} indicated that GNNs are also vulnerable to backdoor attacks. Zhang et al. \cite{zhang2021projective} propose a simple subgraph-based backdoor attack to GNN for graph classification and demonstrate the effectiveness of the proposed method under randomized smoothing defense. Xi et al. \cite{xi2020graph} present Graph Trojaning Attack (GTA) with more advanced optimization-based trigger generation with a bi-level optimization framework. Different from the existing backdoor attack on the graph, our attack can be viewed as a variant of them where we focus on the clean-label and triggerless setting. In the design of our SubInj method, we brought some insights from the backdoor attack methods. We also study how effective our techniques can be when applying existing defenses \cite{liu2022pygod, liu2022bond} against backdoor attacks with anomaly detection algorithms. 

\header{Indiscriminate poisoning attack}
Huang et al. \cite{huang2021unlearnable} introduced a form of error-minimizing noise that proves to be potent in both sample-wise and class-wise scenarios. This method forces deep learning models to learn irrelevant features, thereby deteriorating their performance on clean data. The noise crafting process employs Projected Gradient Descent (PGD) to address a bi-level min-min optimization challenge. This concept has spurred further research into how Unlearnable Examples can be adapted to various learning paradigms and data structures. In the context of different learning paradigms, He et al. \cite{he2022indiscriminate} expanded upon unlearnable examples by delving into unsupervised learning tasks. Moreover, Ren et al. \cite{ren2022transferable} introduced Transferable Unlearnable Examples, an approach that harmonizes data protection for both supervised and unsupervised tasks. When it comes to different data structures, Sun et al. \cite{sun2022coprotector} proposed CoProtector, a method exploring how to shield open-source code from unauthorized training using data poisoning techniques. Similarly, Gan et al. \cite{gan2021triggerless} developed the clean-label backdoor attack, a strategy designed to misguide the model into making wrong predictions, consequently reducing its accuracy. 
% To the best of our knowledge, there is no study discussing how to conduct such an attack on graph-structured data. 

\vspace{-10pt}
% % --------------------------------------------
\section{Conclusion}
\label{sec8:conclusion}
In this paper, we propose \system, a novel framework consisting of two optimization-based poisoning methods and a backdoor-based poisoning method. Our methods explore how to conduct imperceptible perturbation on graph data to prevent GNN models from exploiting graph data freely. We verify our method by conducting experiments on six benchmark graph datasets, and the extensive experimental results show that our method can be applied effectively to various GNN architectures under different settings. This study represents an essential first step in safeguarding personal graph data from being exploited by GNN models.

\clearpage
% % --------------------------------------------
\bibliographystyle{ieeetr}
\bibliography{ref}
% % --------------------------------------------

% \newpage

\appendix
\subsection{Implementation Details}

\header{Implementation} We implement \texttt{Unlearnable Graph} using the \texttt{pytorch} and the \texttt{DGL} package. 
We use three graph convolution layers of each kind of model with 32 hidden neurons and an MLP for the final classification. 
We use ReLU as the activation function and a final dropout rate of 0.5 for the GIN model. Note that since \texttt{DGL} does not support the gradient computation on edge weight for the model of GAT, we simply use a surrogate GCN model for noise generation for GAT in the EMinS approach. We use the learning rate scheduler \texttt{ReduceLROnPlateau} which automatically reduces the learning rate by 75\% when the validation loss stop reducing for 20 epochs. To avoid overfitting, we adopt the early-stop mechanism to terminate model training when the validation loss stop reducing for 50 epochs.

% For larger datasets, conducting the GradArgMax operation might take a long time without offering any loss decrease. We stopped searching early when we encountered no loss decrease within 5 seconds. The code of our method and the baseline approaches is attached to the supplement.

\header{Evaluation}
We use test accuracy as a measure of the data privacy protection ability of the generated perturbations on graphs. A lower test accuracy indicates that the model learns less from the training data, which implies that the noise is more effective in protecting the knowledge contained in the graphs.

\header{Generating time}
We have recorded the time required for noise generation on different datasets for the GCN model in \ref{tab: gen-time}. Based on the recorded data, we can observe that the time required for noise generation on different datasets for the GCN model aligns with our previous analysis. 
% The EMinS/EMaxS method takes the longest time, which is expected since modifying an edge in a graph incurs a computational cost of at least $O(|V|^2)$. The EminF method follows as it involves PGD computation on the node embedding space. The SubInj method proves to be the most efficient, as it directly modifies the graphs.

\begin{table}[htbp]
\caption{The time for training noise generator on six different datasets with a GCN source model. 
% \rtodo{fill all the cases}
}
\label{tab: gen-time}
\centering
\resizebox{\linewidth}{!}{
% \begin{tabular}{c|ccccccc}
% \toprule
%      Datasets   & Random & EMaxS      & EMinS      & EMinF    & SubInj  \\
% \midrule
% MUTAG   & 29s    & 23h31m24s  & 23h21m26s  & 3h13m42s & 46s     \\
% ENZYMES & 1m47s  & 78h45m22s  & 82h50m20s  & 4h49m47s & 2m17s   \\
% PROTEINS& 2m10s  & 5h32m3s    & 6h4m57s    & 29m33s   & 38s     \\
% IMDB-BINARY & 2m11s & 6h30m7s & 6h59m3s  & 2h14m8s  & 1m2s     \\
% IMDB-MULTI  & 3m15s & 21h27m23s & 19h42m4s & 4h12m53s & 7m53s    \\
% COLLAB  & 13m4s  & 64h48m29s  & 164h47m52s & 3h9m11s  & 25m     \\
% \bottomrule
% \end{tabular}
\begin{tabular}{c|cccccc}
\toprule
Datasets & Random & SubInj & EMinF & EMinS & EMaxS \\
\midrule
MUTAG & $\sim$29s & $\sim$46s & $\sim$3h & $\sim$23h & $\sim$23h \\
ENZYMES & $\sim$1m & $\sim$2m & $\sim$4h & $\sim$82h & $\sim$78h \\
PROTEINS & $\sim$2m & $\sim$38s & $\sim$29m & $\sim$6h & $\sim$5h \\
IMDB-BINARY & $\sim$2m & $\sim$1m & $\sim$2h & $\sim$6h & $\sim$6h \\
IMDB-MULTI & $\sim$3m & $\sim$7m & $\sim$4h & $\sim$19h & $\sim$21h \\
COLLAB & $\sim$13m & $\sim$25m & $\sim$3h & $\sim$64h & $\sim$64h \\
\bottomrule
\end{tabular}
}

\end{table}

% \begin{figure}[h]
%     \begin{minipage}{0.23\textwidth}
%     \centering
%     \includegraphics[width=\textwidth]{figs/enzymes.png}
%     \end{minipage}
%     \begin{minipage}{0.23\textwidth}
%     \centering
%     \includegraphics[width=\textwidth]{figs/proteins.png}
%     \end{minipage}
%     \begin{minipage}{0.23\textwidth}
%     \centering
%     \includegraphics[width=\textwidth]{figs/imdb-b.png}
%     \end{minipage}
%     \begin{minipage}{0.23\textwidth}
%     \centering
%     \includegraphics[width=\textwidth]{figs/IMDB-M.png}
%     \end{minipage}
%     \caption{A comparison among different methods on ENZYMES, PROTEINS, IMDB-BINARY, and IMDB-MULTI datasets. \rtodo{move to app.}}
%     \label{fig:result}
% \end{figure}

% \subsection{Potential countermeasures}
% \header{Adv. Training}
% \input{table/app/adv_more}

\header{Stealthiness} i) Statistics: we adopt average edge number and graph density as the basic statistics of graph data and make a comparison between them before and after modifications. The comprehensive results are shown in Tab.~\ref{tab:statics_all}.
% \usepackage{multirow}
\begin{table*}[!htbp]
\caption{The average edge number and average density of graphs before and after modification.}
\label{tab:statics_all}
\resizebox{\linewidth}{!}{
\begin{tabular}{cccccccccccccc}
\hline
\multirow{2}{*}{Models} &
  \multirow{2}{*}{Methods} &
  \multicolumn{2}{c}{MUTAG} &
  \multicolumn{2}{c}{ENZYMES} &
  \multicolumn{2}{c}{PROTEINS} &
  \multicolumn{2}{c}{IMDB-BINARY} &
  \multicolumn{2}{c}{IMDB-MULTI} &
  \multicolumn{2}{c}{COLLAB} \\ \cline{3-14}
                       & & Avg E  & Avg $\rho$  & Avg E  & Avg $\rho$ & Avg E  & Avg $\rho$ & Avg E &Avg $\rho$ & Avg E  & Avg $\rho$  & Avg E  & Avg $\rho$ \\ \hline
\multirow{4}{*}{GCN}  &  Clean  & 9.877  & 0.032  & 30.831  & 0.038  & 36.044  & 0.049  & 97.123  & 0.245  & 66.023  & 0.346  & 2411.66  & 0.249   \\
                      & EMinS  & 10.003  & 0.033  & 30.908  & 0.038  & 36.183  & 0.047  & 96.211  & 0.242  & 65.535  & 0.341  &  2410.49 & 0.249    \\
                      & EMinF  & 9.877  & 0.033  & 30.831  & 0.038  & 36.044  & 0.049  & 97.123  & 0.245  & 66.023  & 0.346  & 2411.66  & 0.249   \\
                       &SubInj & 11.340  & 0.039  & 32.799  & 0.042  & 37.775  & 0.049  & 97.030  & 0.244  & 64.478  & 0.324   & 2411.26
  & 0.249\\ \hline 
\multirow{4}{*}{GAT}  &  Clean  & 9.877  & 0.032  & 30.831  & 0.038  & 36.044  & 0.049  & 97.123  & 0.245  & 66.023  & 0.346  & 2411.66  & 0.249 \\
                      &  EMinS  & 10.003  & 0.033  & 30.908  &  0.038 & 36.183  & 0.047  & 96.257  & 0.242  & 65.467 & 0.340  & 2410.47  & 0.249  \\
                      &  EMinF  & 9.877  & 0.032  & 30.831  & 0.038  & 36.044  & 0.049  & 97.123  & 0.245  & 66.023  & 0.346  & 2411.66  & 0.249  \\
                      &  SubInj  & 11.34  & 0.039  & 32.799  & 0.042  & 37.775  & 0.049  & 97.030  & 0.244  & 64.478  & 0.324  & 2411.26
  & 0.249 \\ \hline 
\multirow{4}{*}{GIN}  &  Clean   & 9.877  & 0.032  & 30.831  & 0.038  & 36.044  & 0.049  & 97.123  & 0.245  & 66.023  & 0.346  & 2411.66  & 0.249\\
                      &  EMinS  & 10.27  & 0.034  & 30.889  & 0.037  & 35.811  & 0.048  & 96.349  & 0.242  & 65.335  & 0.339  &  2410.28 & 0.249  \\
                      &  EMinF  & 9.877  & 0.032  & 30.831  & 0.038  & 36.044  & 0.049  & 97.123  & 0.245  & 66.023  & 0.346  & 2411.66  & 0.249  \\
                      &  SubInj  & 11.34  & 0.039  & 32.799  & 0.042  & 37.775  & 0.049  & 97.030  & 0.244  & 64.478  & 0.324  & 2411.26
  & 0.249\\ \hline
\multirow{4}{*}{Sage} &  Clean  & 9.877  & 0.032  & 30.831  & 0.038  & 36.044  & 0.049  &97.123   &0.245   & 66.023  & 0.346  & 2411.66  & 0.249 \\
                      &  EMinS  & 9.377  & 0.031  & 30.333  & 0.036  & 35.484  & 0.046  &97.123   &0.241   & 65.023  & 0.335  & 2409.74  &  0.249 \\
                      &  EminF  & 9.877  & 0.032  & 30.831  & 0.038  & 36.044  & 0.049  & 96.123  &0.245   & 66.023  & 0.346  & 2411.66  & 0.249 \\
                      &  SubInj  & 11.34  & 0.039  & 32.799  &0.042   & 37.775  &0.049   & 97.030  & 0.244  & 64.478  & 0.324  & 2411.26
  & 0.249 \\ \hline
\end{tabular}
}

\end{table*}; ii){Detections}: considering that existing graph anomaly detectors commonly operate on large graphs used for node classification, while our graphs are smaller and used for graph-level classification, we need to adapt the input format accordingly. To address this, we utilize the \textit{dgl.batch} function to form a single graph input. Importantly, this process does not alter the anomaly at the node level. Subsequently, we apply the detectors implemented by the \textit{PyGod} \cite{liu2022pygod} library to obtain outlier scores for each node. These scores are then summed and averaged to derive a graph-level anomalous score. The comprehensive results obtained from this process are reported in Tab.~\ref{tab:statics_all}, providing a detailed overview of the performance of the detectors on our graph-level classification task.

% \usepackage{multirow}
\begin{table*}[]
\caption{The outlier score computed by 10 off-the-shelf graph anomaly detectors. Different detectors apply different kinds of scoring strategies. A higher outlier score implies more anomalous graphs.}
\label{tab:det-1}
\resizebox{\linewidth}{!}{
\begin{tabular}{cccccccccccccc}
\hline
\multirow{2}{*}{Models} & \multirow{2}{*}{Detectors} & \multicolumn{4}{c}{MUTAG}      & \multicolumn{4}{c}{ENZYMES}    & \multicolumn{4}{c}{PROTEINS}   \\ \cline{3-14} 
                        &                            & Clean & EMinS & EMinF & SubInj & Clean & EMinS & EMinF & SubInj & Clean & EMinS & EMinF & SubInj \\ \hline
\multirow{10}{*}{GCN}  & AdONE      & 0.0003 & 0.0004 & 0.0004 & 0.0004 & 0.0001 & 0.0001 & 0.0001 & 0.0001 & 0.0001 & 0.0001 & 0.0001 & 0.0001 \\
                       & ANOMALOUS  & 0.5582 & 0.5963 & 0.5598 & 0.5102 & 0.5011 & 0.5825 & 0.5068 & 0.4626 & 0.4915 & 0.4925 & 0.5048 & 0.4872 \\
                       & AnomalyDAE & 18.0889 & 21.8547 & 22.0866 & 16.9439 & 37.9476 & 34.3155 & 36.3776 & 33.7460 & 25.6802 & 24.9090 & 24.8486 & 24.8782 \\
                       & CoLA       & -3.9178 & -3.3082 & -3.9178 & -2.4449 & -2.3860 & -2.3369 & -2.3429 & -2.1735 & -2.3368 & -2.5814 & -2.2203 & -1.9556 \\
                       & CONAD      & 0.9793 & 0.9979 & 0.9872 & 1.2150 & 1.2887 & 1.2905 & 1.2997 & 1.3288 & 1.2834 & 1.2849 & 1.3096 & 1.3393 \\
                       & DOMINANT   & 0.9793 & 1.0054 & 0.9904 & 1.7557 & 1.2913 & 1.2897 & 1.3036 & 1.3264 & 1.2836 & 1.2959 & 1.3144 & 1.3241 \\
                       & DONE       & 0.0003 & 0.0004 & 0.0004 & 0.0004 & 0.0001 & 0.0001 & 0.0001 & 0.0001 & 0.0001 & 0.0001 & 0.0001 & 0.0001 \\
                       & Radar      & 0.5533 & 0.5547 & 0.5518 & 0.5153 & 0.4914 & 0.4904 & 0.4981 & 0.4693 & 0.4860 & 0.4804 & 0.5216 & 0.4857 \\
                       & OCGNN      & -0.0538 & -0.0495 & -0.0353 & -0.0019 & -0.0419 & -0.0419 & -0.0341 & -0.0324 & -0.0381 & -0.0342 & -0.0349 & -0.0373 \\
                       & ONE        & 0.0003 & 0.0004 & 0.0004 & 0.0004 & 0.0001 & 0.0001 & 0.0001 & 0.0001 & 0.0000 & 0.0000 & 0.0000 & 0.0000 \\ \hline
\multirow{10}{*}{GAT}  & AdONE      & 0.0003 & 0.0004 & 0.0004 & 0.0004 & 0.0001 & 0.0001 & 0.0001 & 0.0001 & 0.0001 & 0.0001 & 0.0001 & 0.0001 \\
                       & ANOMALOUS  & 0.5555 & 0.6233 & 0.6453 & 0.5108 & 0.5135 & 0.5184 & 0.4980 & 0.4703 & 0.5297 & 0.5060 & 0.5265 & 0.4773 \\
                       & AnomalyDAE & 17.3728 & 22.2619 & 15.4488 & 24.4705 & 38.6345 & 37.5032 & 33.5046 & 34.3672 & 24.4819 & 23.7747 & 23.7199 & 23.6933 \\
                       & CoLA       & -4.3048 & -4.5010 & -4.0047 & -2.1679 & -2.2632 & -2.1439 & -2.1713 & -2.3409 & -2.2603 & -2.3092 & -2.3665 & -2.3298 \\
                       & CONAD      & 0.9793 & 0.9955 & 0.9915 & 1.2461 & 1.2896 & 1.2933 & 1.2998 & 1.3292 & 1.2854 & 1.2884 & 1.3119 & 1.3292 \\
                       & DOMINANT   & 0.9787 & 0.9986 & 0.9901 & 1.3546 & 1.2876 & 1.2861 & 1.3003 & 1.3318 & 1.2803 & 1.2923 & 1.3124 & 1.3267 \\
                       & DONE       & 0.0003 & 0.0004 & 0.0004 & 0.0004 & 0.0001 & 0.0001 & 0.0001 & 0.0001 & 0.0001 & 0.0001 & 0.0001 & 0.0001 \\
                       & Radar      & 0.5533 & 0.5547 & 0.5524 & 0.5153 & 0.4914 & 0.4904 & 0.4985 & 0.4693 & 0.4860 & 0.4804 & 0.5221 & 0.4857 \\
                       & OCGNN      & -0.394 & -0.0229 & -0.0384 & -0.0041 & -0.0343 & -0.0343 & -0.0397 & -0.0353 & -0.0489 & -0.0302 & -0.0261 & -0.0434 \\
                       & ONE        & 0.0003 & 0.0004 & 0.0004 & 0.0004 & 0.0001 & 0.0001 & 0.0001 & 0.0001 & 0.0000 & 0.0000 & 0.0000 & 0.0000 \\ \hline
\multirow{10}{*}{GIN}  & AdONE      & 0.0003 & 0.0004 & 0.0004 & 0.0004 & 0.0001 & 0.0001 & 0.0001 & 0.0001 & 0.0001 & 0.0001 & 0.0001 & 0.0001 \\
                       & ANOMALOUS  & 0.5805 & 0.5453 & 0.5946 & 0.5193 & 0.5143 & 0.5067 & 0.4932 & 0.4605 & 0.5137 & 0.4779 & 0.4960 & 0.4957 \\
                       & AnomalyDAE & 19.6741 & 19.8067 & 19.6436 & 17.7185 & 39.2516 & 34.9576 & 35.8857 & 35.7971 & 24.9304 & 24.2677 & 24.2269 & 24.2163 \\
                       & CoLA       & -3.9649 & -3.7701 & -3.8522 & -2.0902 & -2.3059 & -2.2453 & -2.1954 & -2.2135 & -2.3611 & -2.3379 & -2.2793 & -2.1407 \\
                       & CONAD      & 0.9791 & 1.0084 & 0.9930 & 1.3971 & 1.2880 & 1.2890 & 1.3003 & 1.3275 & 1.2830 & 1.2828 & 1.3085 & 1.3182 \\
                       & DOMINANT   & 0.9828 & 1.0114 & 0.9871 & 1.5776 & 1.2921 & 1.2911 & 1.3021 & 1.3295 & 1.2824 & 1.2892 & 1.3166 & 1.3295 \\
                       & DONE       & 0.0003 & 0.0004 & 0.0004 & 0.0004 & 0.0001 & 0.0001 & 0.0001 & 0.0001 & 0.0001 & 0.0001 & 0.0001 & 0.0001 \\
                       & Radar      & 0.5533 & 0.5376 & 0.5516 & 0.5153 & 0.4914 & 0.4898 & 0.4992 & 0.4693 & 0.4860 & 0.4847 & 0.5223 & 0.4857 \\
                       & OCGNN      & -0.0411 & -0.0307 & -0.0405 & -0.0017 & -0.0422 & -0.0414 & -0.0489 & -0.0410 & -0.0467 & -0.0371 & -0.0450 & -0.0369 \\
                       & ONE        & 0.0003 & 0.0004 & 0.0004 & 0.0004 & 0.0001 & 0.0001 & 0.0001 & 0.0001 & 0.0000 & 0.0000 & 0.0000 & 0.0000 \\ \hline
\multirow{10}{*}{Sage} & AdONE      & 0.0003 & 0.0004 & 0.0004 & 0.0004 & 0.0001 & 0.0001 & 0.0001 & 0.0001 & 0.0001 & 0.0001 & 0.0001 & 0.0001 \\
                       & ANOMALOUS  & 0.6054 & 0.6226 & 0.5467 & 0.5100 & 0.5179 & 0.5218 & 0.5073 & 0.4688 & 0.5087 & 0.4825 & 0.5058 & 0.4717 \\
                       & AnomalyDAE & 18.9142 & 17.0185 & 16.8544 & 18.6348 & 39.0539 & 34.5416 & 36.1084 & 35.7745 & 24.7217 & 23.9871 & 23.9729 & 23.9295 \\
                       & CoLA       & -4.1667 & -3.8499 & -4.4659 & -2.8213 & -2.2961 & -2.2044 & -2.2813 & -2.1284 & -2.3880 & -2.3796 & -2.3360 & -2.3135 \\
                       & CONAD      & 0.9774 & 0.9660 & 0.9859 & 1.9448 & 1.2904 & 1.2795 & 1.2971 & 1.3314 & 1.2874 & 1.2780 & 1.3122 & 1.3399 \\
                       & DOMINANT   & 0.9807 & 0.9672 & 0.9878 & 2.6526 & 1.2949 & 1.2840 & 1.2975 & 1.3320 & 1.2832 & 1.2854 & 1.3155 & 1.3246 \\
                       & DONE       & 0.0003 & 0.0004 & 0.0004 & 0.0004 & 0.0001 & 0.0001 & 0.0001 & 0.0001 & 0.0001 & 0.0001 & 0.0001 & 0.0001 \\
                       & Radar      & 0.5533 & 0.5594 & 0.5520 & 0.5153 & 0.4914 & 0.4978 & 0.4986 & 0.4693 & 0.4860 & 0.4861 & 0.5232 & 0.4857 \\
                       & OCGNN      & -0.0373 & -0.0314 & -0.0371 & 0.0014 & -0.0409 & -0.0408 & -0.0387 & -0.0517 & -0.0558 & -0.0559 & -0.0380 & -0.0301 \\
                       & ONE        & 0.0004 & 0.0003 & 0.0003 & 0.0003 & 0.0001 & 0.0001 & 0.0001 & 0.0001 & 0.0000 & 0.0000 & 0.0000 & 0.0000 \\ \hline
\end{tabular}
}

\end{table*}

\header{More visualizations}
We visualize the perturbed graphs with different perturbation budgets. As we can see from Fig.~\ref{fig:morevisual}, more modifications have been applied to graphs as the budget rises, and our default $\beta=0.05$ and $\alpha=0.025$ are the reasonable budget for the EMinS and EMinF methods on the majority of the datasets.

\begin{table*}[htbp]
\caption{An illustration of the SubInj attack. A number of nodes(N=5) are chosen, and the features and connections between them are cleared and then replaced with the class-wise subgraph generated by the Erdos-Renyi method.}
\centering
\resizebox{\linewidth}{!}{
\begin{tabular}{cccc}
\toprule
\multicolumn{2}{c}{Class: 0} & \multicolumn{2}{c}{Class: 1} \\ \hline
Subgraph           & Original Graphs & Subgraph           & Original Graphs \\ \hline
\multirow{3}{*}{\includegraphics[width=0.1\linewidth]{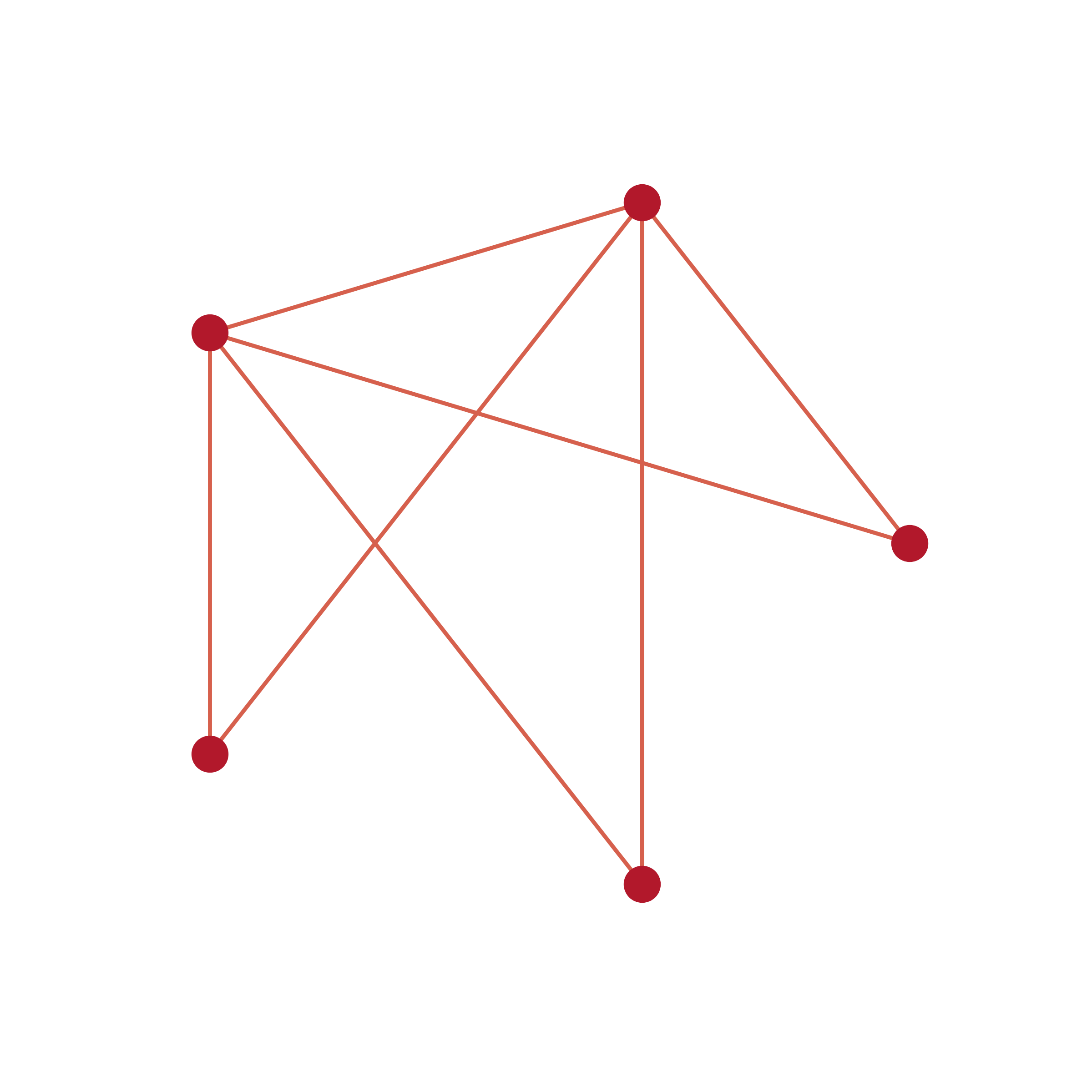}}  & \includegraphics[width=0.45\linewidth]{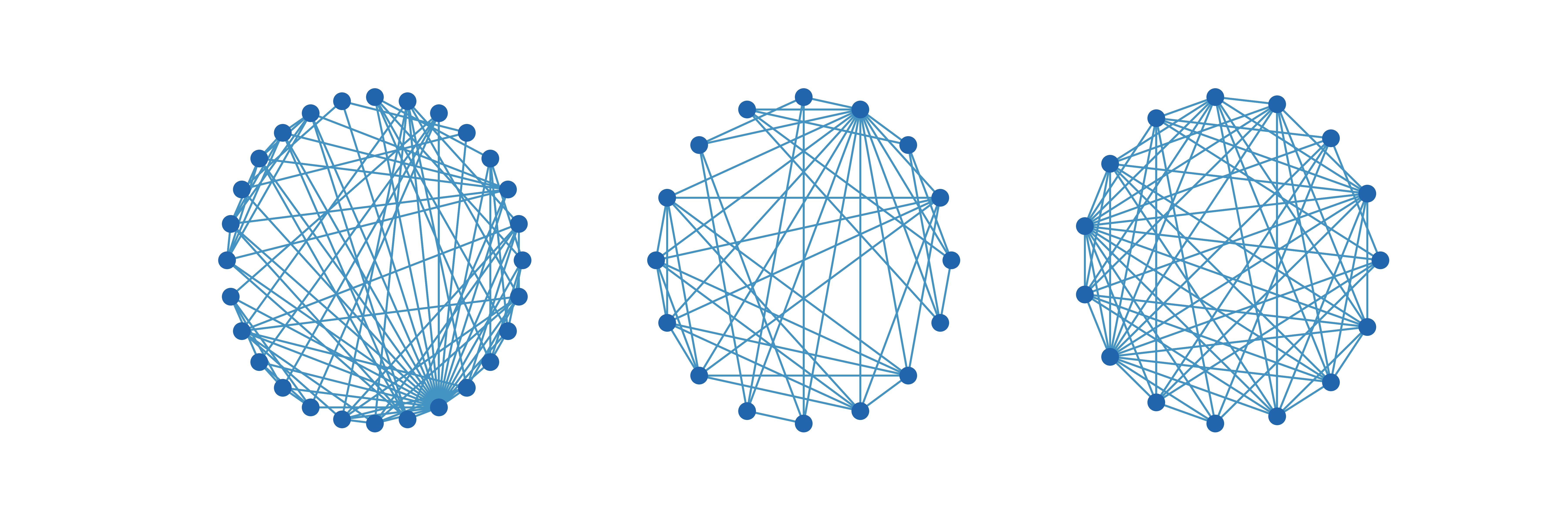}        & \multirow{3}{*}{\includegraphics[width=0.1\linewidth]{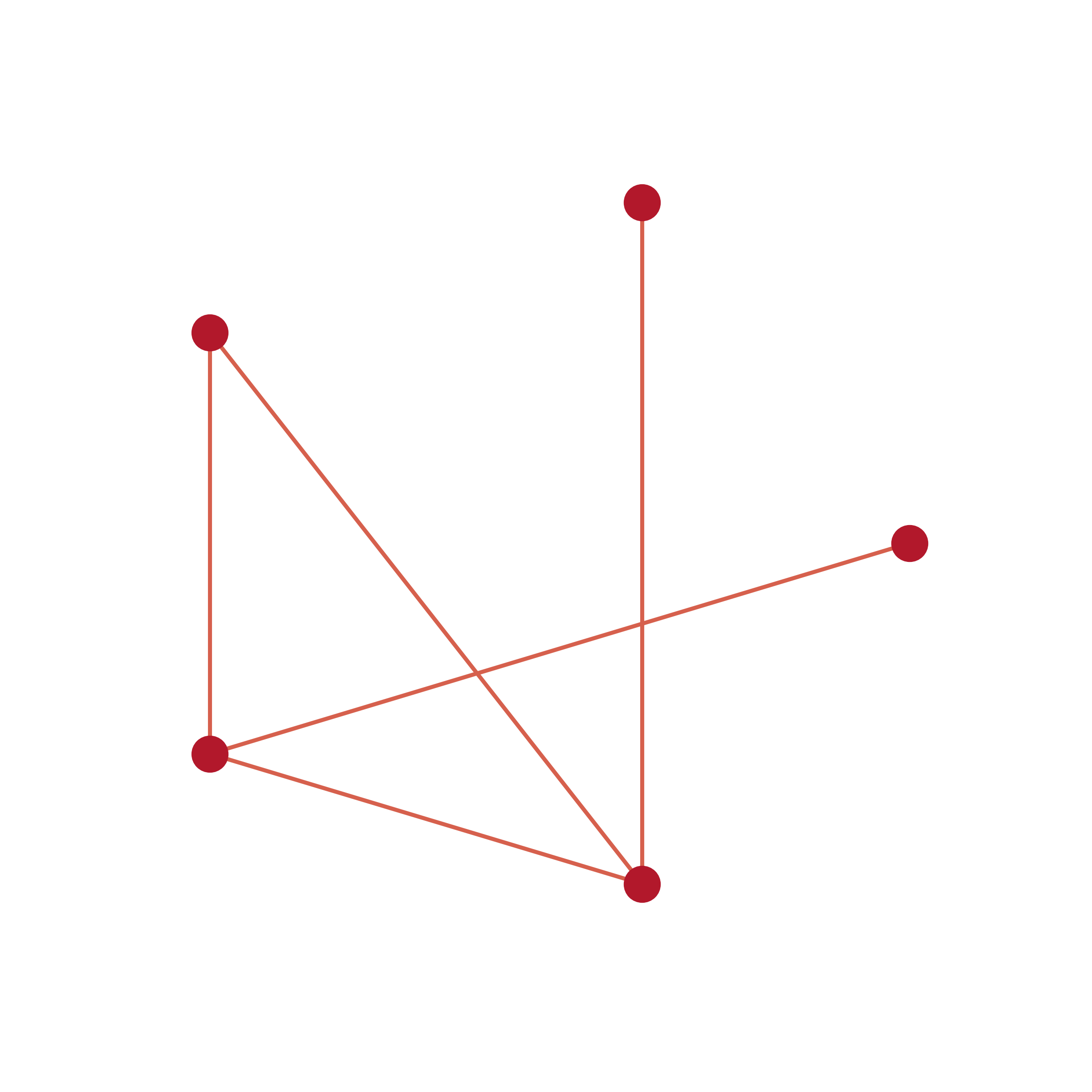}}  & \includegraphics[width=0.45\linewidth]{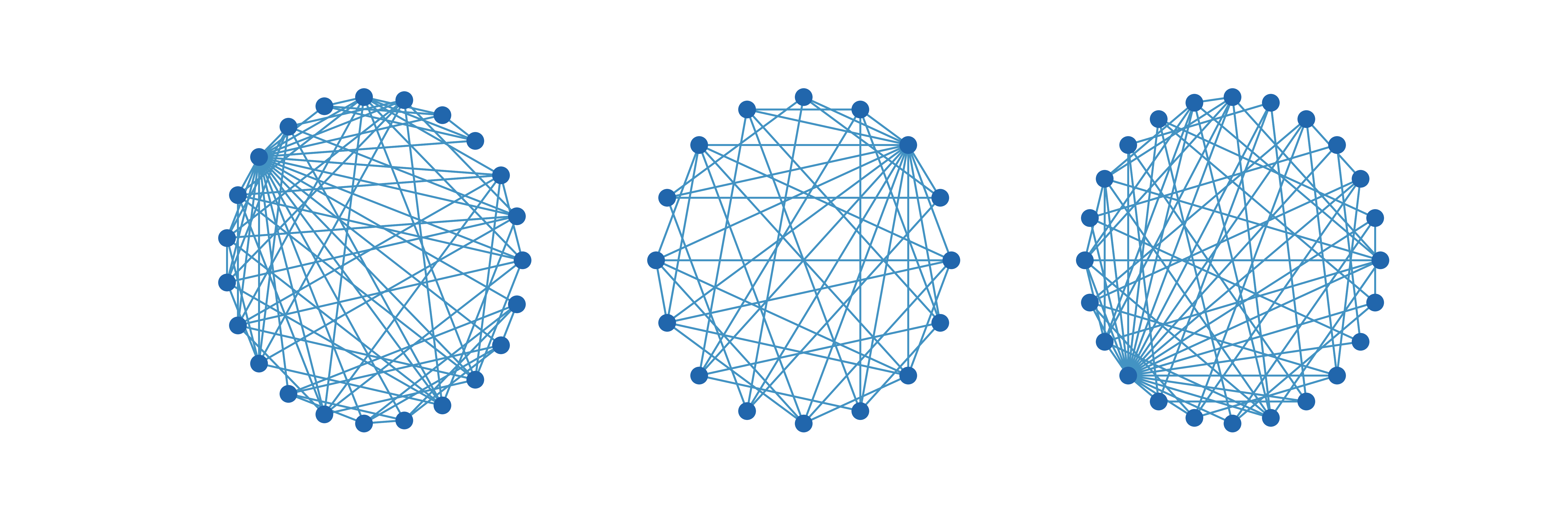}       \\ \cline{2-2} \cline{4-4}
                   & Connection Cleared & & Connection Cleared \\
                   & \includegraphics[width=0.45\linewidth]{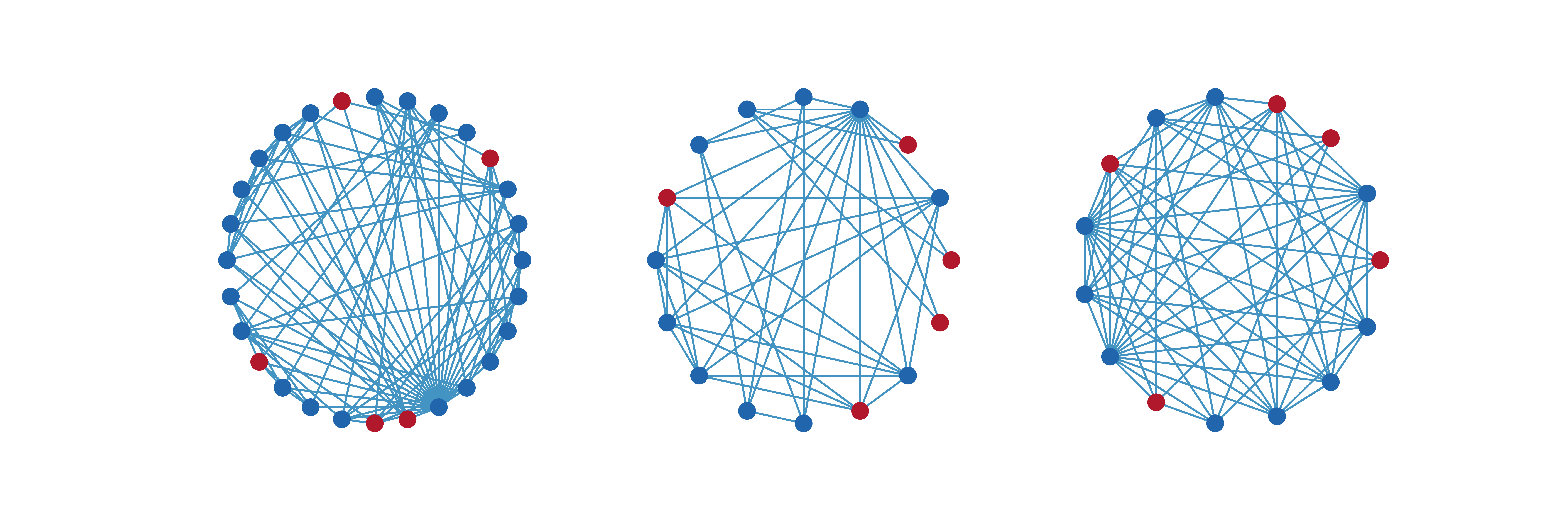}        &                    & \includegraphics[width=0.45\linewidth]{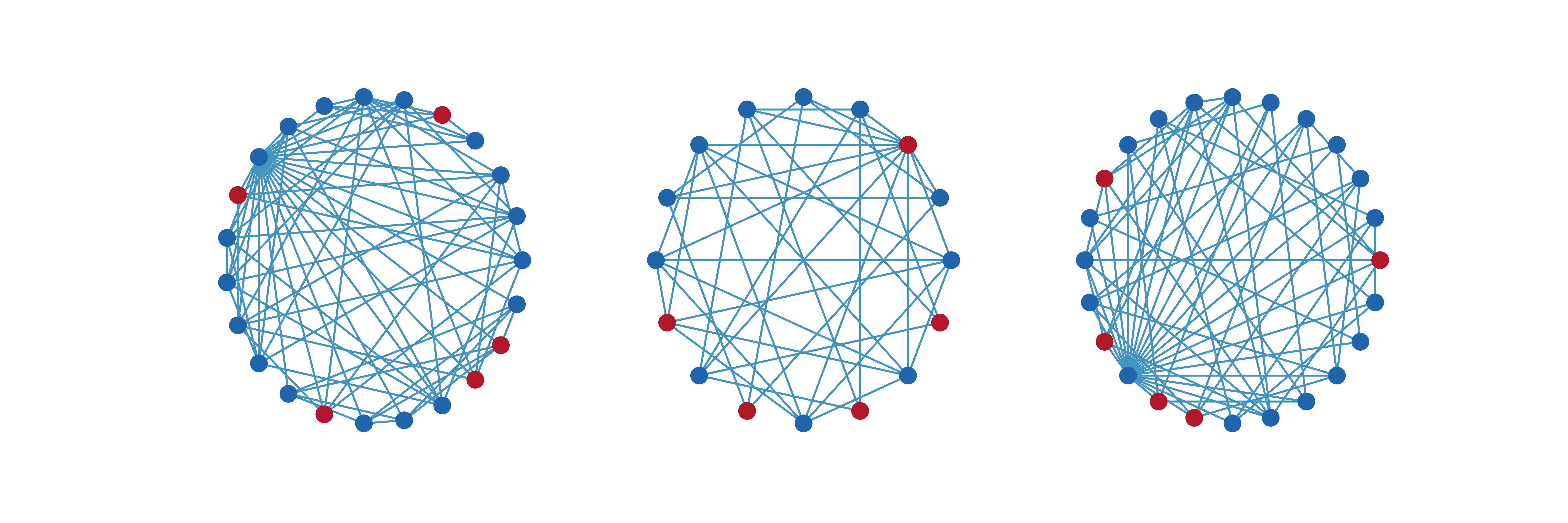}        \\ \cline{2-2} \cline{4-4}
                   & Subgraph Injected & & Subgraph Injected \\
                   & \includegraphics[width=0.45\linewidth]{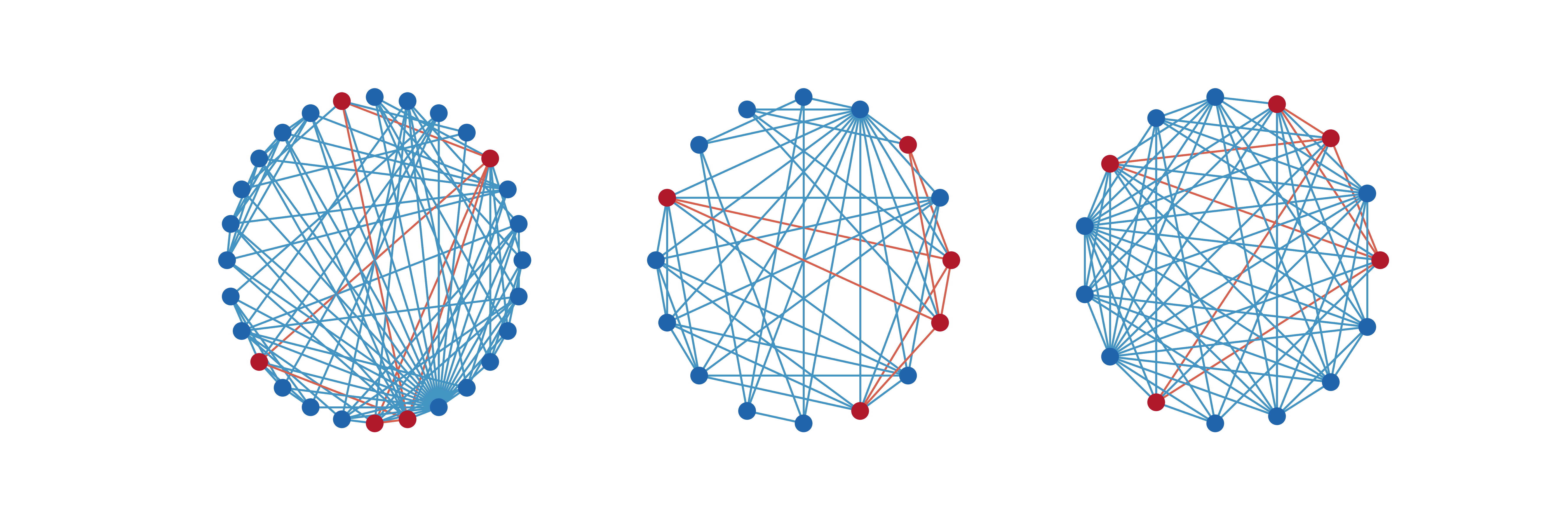}        &                    & \includegraphics[width=0.45\linewidth]{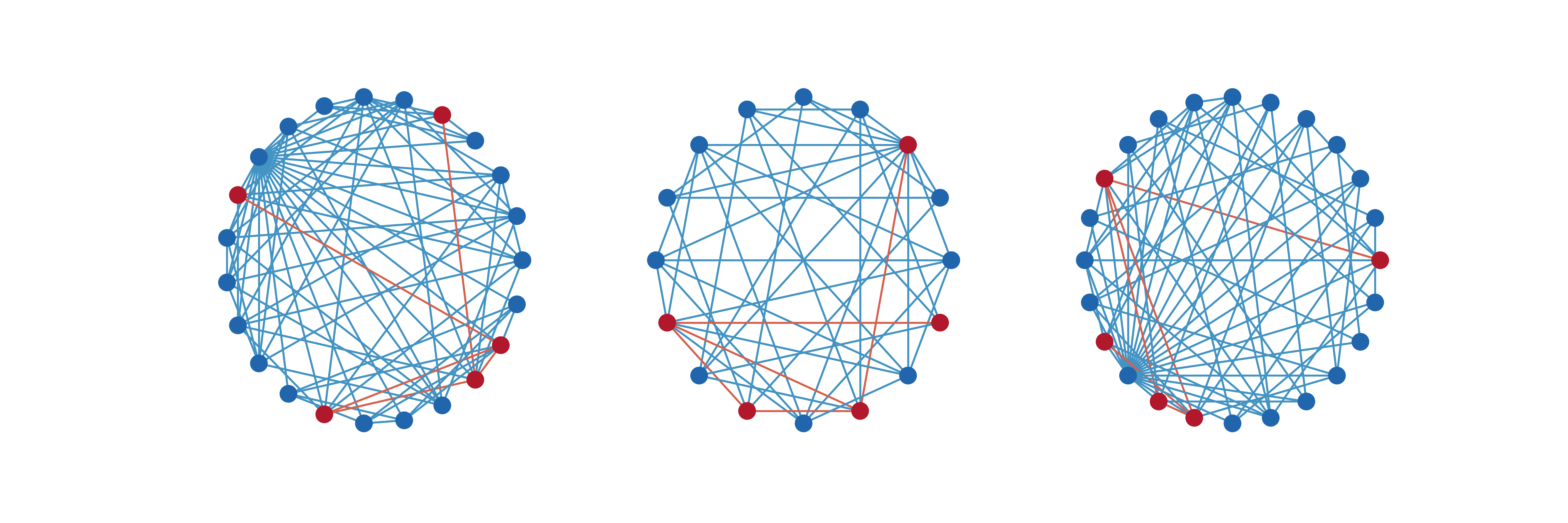}        \\ \bottomrule
\end{tabular}
}
\label{tab. suj. how}
\end{table*}

\subsection{Why \system work? }

\noindent Fundamentally, \system aims to craft ``invisible'' but delusive perturbation $\delta_i$ for obtaining cloaked graph $\hat{G}_i=G_i \oplus \delta_i$. Trained on these cloaked graphs, GNNs are likely to learn brittle noise-label correlations $\delta_i \rightarrow y_i$ instead of exploiting those truly robust informative features, that is, to learn $G_i  \rightarrow y_i$. As a consequence, we can protect the data since the trained models fail to generalize well on the clean graph. Our work is mainly inspired by previous works on adversarial attacks that demonstrate the vulnerability nature of DNNs in relying on some brittle features for classification. In our paper, we formulate the problem as a bi-level optimization problem and leverage two heuristic methods to solve it. 

\header{Why/How does our EMinF/EMinS work?}
Inspired by the error-minimizing approaches in the vision domain \cite{huang2021unlearnable}, we propose to craft delusive noises that greatly facilitate the training and seek to cause the trained model to overfit the noise. The EMinF and EMinS operate on the feature and structure space of graph data, respectively. With an iteratively trained surrogate model, EMinS and EMinF craft perturbation using gradient information with the purpose algorithms GradArgMax and PGD with softmax sampling, respectively. To verify that this kind of method can successfully decrease the training loss, we further provide the training loss curve of GNNs trained on clean data and cloaked data, respectively. 

\header{Why/How does our SubInj work? }
Inspired by studies on backdoor attacks, we also design manual-craft sub-graph triggers to create false correlations. Specifically, we inject class-wise Erdos-Renyi sub-graphs into the clean graphs to obtain the cloaked graphs and want to trick the models to learn the trigger-label correlation. Please see the illustration in the Tab. \ref{tab. suj. how}, which gives a more intuitive sense of how this method works.

\subsection{Discussion on future directions of \system}
To further enhance \system, we preliminarily propose the following promising directions to overcome the emerging challenges:
\begin{enumerate}
    \item \textbf{Apative-budget \system to migrate the cost of budget searching}: Recent studies suggest that training sample-dependent budget generator $c=f(x)$ is promising for conducting adaptive perturbation. Moreover, replacing the simple PGD algorithm in our paper with more advanced parameter-free variants (e.g., AutoAttack \cite{croce2020reliable}) might also further strengthen \system in generating more powerful adversarial noise. 
    \item \textbf{Target poisoning attack for low poison rate scenario}: A low poisoning rate scenario can be really challenging since there are some examples exposed that guide the model to learn the true correlation between the informative feature and the label. Motivated by the previous study that unearths the shortcut learning preference of DNNs, we hypothesize that injecting noise with target-poisoning attacks leveraging clean models may help us to tackle this challenge. Further studies are needed to demonstrate it. 
    
    \item \textbf{Model-ensemble EMinF/EMinS for improving transferability}: Currently, EMinF/EminS leverages a single fixed surrogate model for crafting delusive noise, which might be sub-optimal under a wide range of model settings. To improve the transferability of \system, one of the promising directions is to leverage model ensemble techniques for better data protection. 
    
\end{enumerate}

\begin{table*}[h]
    \caption{Graph visualization of the EMinS perturbed graphs on PROTEINS, IMDB-M, ENZYMES, and IMDB-BINARY datasets (the first row are clean graphs, and the following rows are the graphs we generated with different budgets).}
    \label{fig:morevisual}
    \centering
    \resizebox{0.92\linewidth}{!}{
    \begin{tabular}{c|cc}
        % \toprule
        &PROTEINS&IMDB-MULTI\\
        \midrule
        \rotatebox{90}{Clean}&  \includegraphics[width=0.4\linewidth]{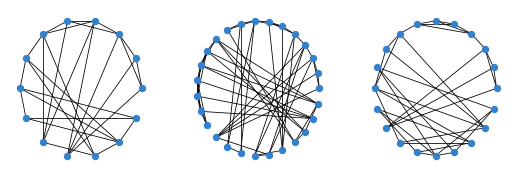}&  \includegraphics[width=0.4\linewidth]{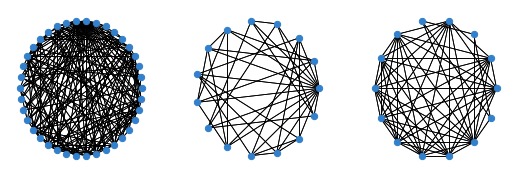}\\
        \rotatebox{90}{$\beta=0.03$} &  \includegraphics[width=0.4\linewidth]{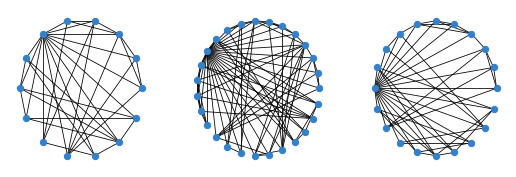}&  \includegraphics[width=0.4\linewidth]{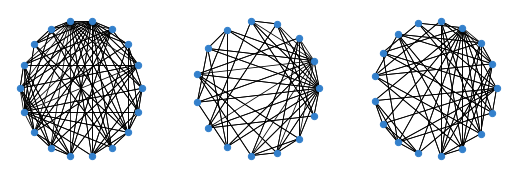}\\
        \rotatebox{90}{$\beta=0.05$} &  \includegraphics[width=0.4\linewidth]{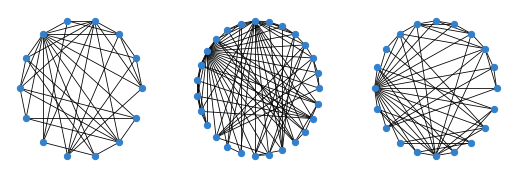}&  \includegraphics[width=0.4\linewidth]{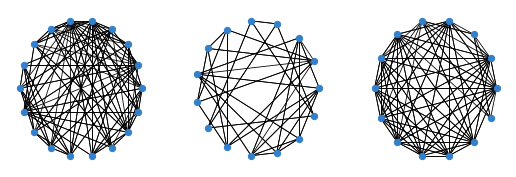}\\
        \rotatebox{90}{$\beta=0.08$} &  \includegraphics[width=0.4\linewidth]{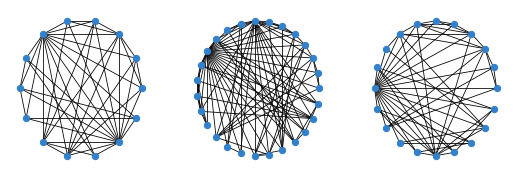}&  \includegraphics[width=0.4\linewidth]{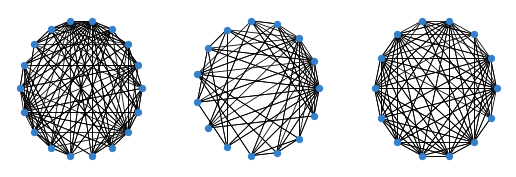}\\
        &ENZYMES&IMDB-BINARY\\
        \midrule
        \rotatebox{90}{Clean}&  \includegraphics[width=0.4\linewidth]{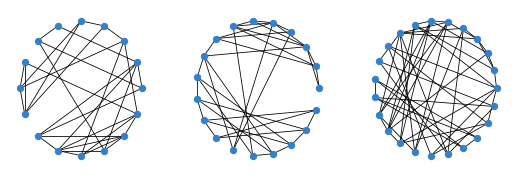}&  \includegraphics[width=0.4\linewidth]{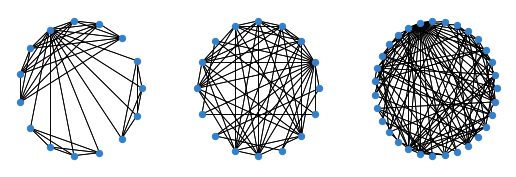}\\
        \rotatebox{90}{$\beta=0.03$} &  \includegraphics[width=0.4\linewidth]{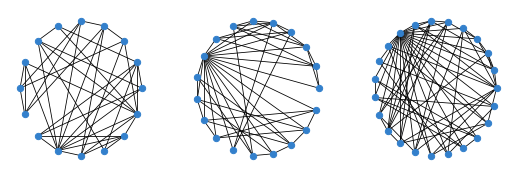}&  \includegraphics[width=0.4\linewidth]{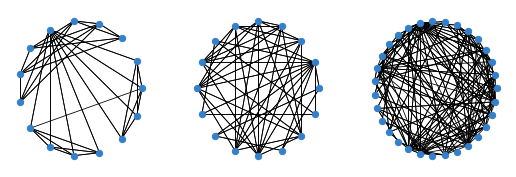}\\
        \rotatebox{90}{$\beta=0.05$} &  \includegraphics[width=0.4\linewidth]{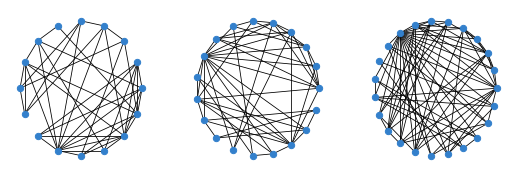}&  \includegraphics[width=0.4\linewidth]{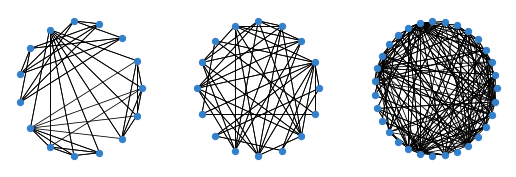}\\
        \rotatebox{90}{$=0.08$} &  \includegraphics[width=0.4\linewidth]{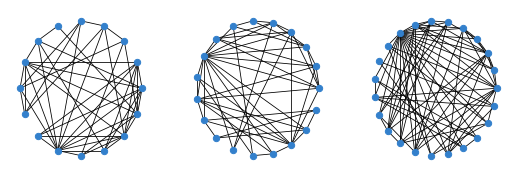}&  \includegraphics[width=0.4\linewidth]{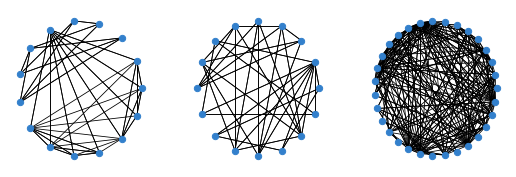}\\
    \bottomrule
    \end{tabular}
    }

\end{table*}

\begin{table*}[h]
\caption{Graph visualization of the EMinF perturbed graphs on PROTEINS, IMDB-M, ENZYMES, and IMDB-BINARY datasets (the first row are clean graphs, and the following rows are the graphs we generated with different budgets).}
    \label{fig:feat-visual}
    \centering
    \resizebox{0.92\linewidth}{!}{
    \begin{tabular}{c|cc}
        % \toprule
        &PROTEINS&IMDB-MULTI\\
        \midrule
        \rotatebox{90}{Clean}&  \includegraphics[width=0.4\linewidth]{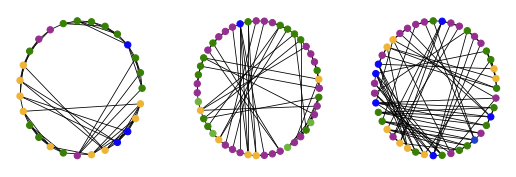}&  \includegraphics[width=0.4\linewidth]{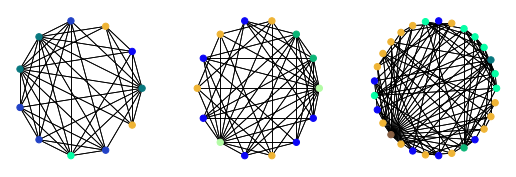}\\
        \rotatebox{90}{$\alpha=0.010$} &  \includegraphics[width=0.4\linewidth]{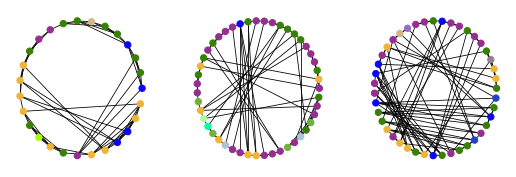}&  \includegraphics[width=0.4\linewidth]{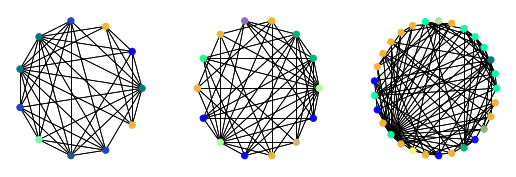}\\
        \rotatebox{90}{$\alpha=0.025$} &  \includegraphics[width=0.4\linewidth]{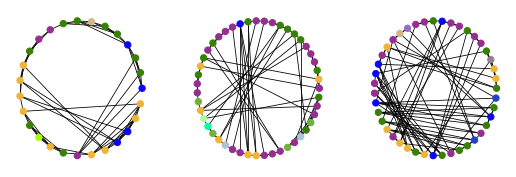}&  \includegraphics[width=0.4\linewidth]{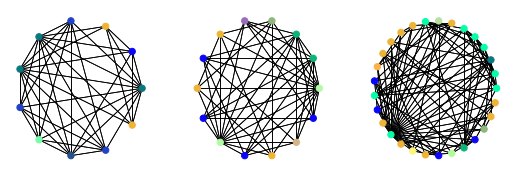}\\
        \rotatebox{90}{$\alpha=0.040$} &  \includegraphics[width=0.4\linewidth]{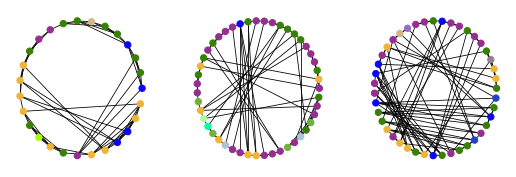}&  \includegraphics[width=0.4\linewidth]{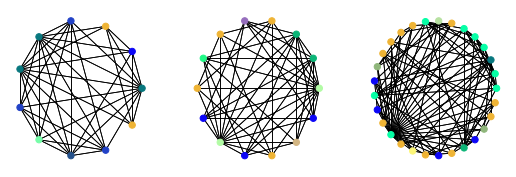}\\
        &ENZYMES&IMDB-BINARY\\
        \midrule
        \rotatebox{90}{Clean}&  \includegraphics[width=0.4\linewidth]{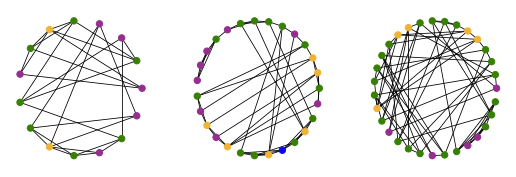}&  \includegraphics[width=0.4\linewidth]{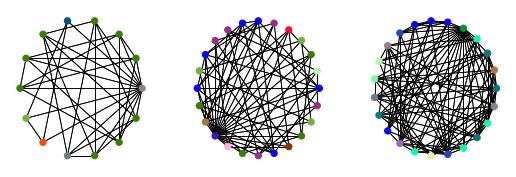}\\
        \rotatebox{90}{$\alpha=0.010$} &  \includegraphics[width=0.4\linewidth]{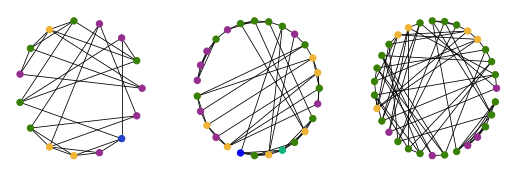}&  \includegraphics[width=0.4\linewidth]{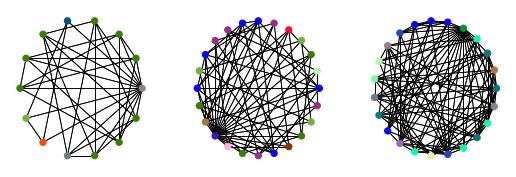}\\
        \rotatebox{90}{$\alpha=0.025$} &  \includegraphics[width=0.4\linewidth]{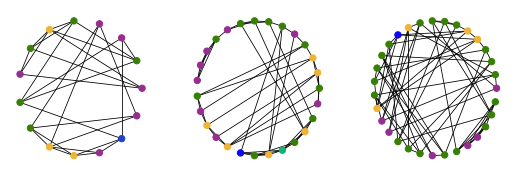}&  \includegraphics[width=0.4\linewidth]{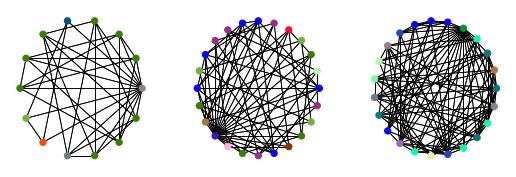}\\
        \rotatebox{90}{$\alpha=0.040$} &  \includegraphics[width=0.4\linewidth]{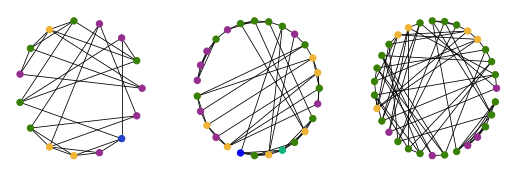}&  \includegraphics[width=0.4\linewidth]{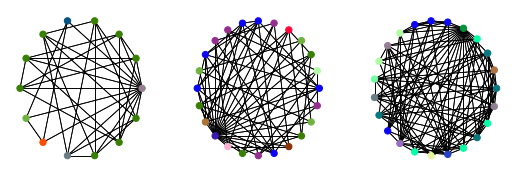}\\
    \bottomrule
    \end{tabular}
    }
    
\end{table*}

% \subsection{Stability Analysis }

% \begin{figure}[h]
% \centering  
%     % \subfigure[]{
%     % \label{fig.sub.1}
%     % }
%     \includegraphics[width=0.48\linewidth]{figs/imdb-m-heat.png}
%     \includegraphics[width=0.48\linewidth]{figs/proteins-heat.png}
%     % \subfigure[]{
%     % \label{fig.sub.2}
%     % }
%     \caption{The transferability test across models on IMDB-MULTI and PROTEINS datasets. The unlearnable examples are generated by the source model on which victim models are trained. The numbers in the matrix report the test accuracy and the ones in the brackets report the accuracy loss from models trained on clean examples. \rtodo{move to app.}}
%     \label{fig:trans}
% \end{figure}

\end{document}